\documentclass[10pt,twocolumn,aps,prb,balancelastpage,longbibliography]{revtex4-1}
\usepackage[latin9]{inputenc}
\usepackage{color}
\usepackage{verbatim}
\usepackage{amsmath}
\usepackage{amssymb,ulem,manfnt}
\usepackage{graphicx,cellspace,booktabs}
\usepackage{esint}
\usepackage[unicode=true,pdfusetitle,
 bookmarks=true,bookmarksnumbered=false,bookmarksopen=false,
 breaklinks=false,pdfborder={0 0 1},backref=false,colorlinks=true]{hyperref}
\usepackage{breakurl}
\usepackage{epstopdf}

\makeatletter

\@ifundefined{textcolor}{}
{%
 \definecolor{BLACK}{gray}{0}
 \definecolor{WHITE}{gray}{1}
 \definecolor{RED}{rgb}{1,0,0}
 \definecolor{GREEN}{rgb}{0,1,0}
 \definecolor{BLUE}{rgb}{0,0,1}
 \definecolor{CYAN}{cmyk}{1,0,0,0}
 \definecolor{MAGENTA}{cmyk}{0,1,0,0}
 \definecolor{YELLOW}{cmyk}{0,0,1,0}
}

\usepackage[caption=false]{subfig}
\usepackage{bm}

\newcommand{\bra}[1]{\langle #1 |}
\newcommand{\ket}[1]{|#1\rangle}
\newcommand{\braket}[2]{\langle #1 | #2 \rangle}

\newcommand{\bv}[1]{\mathbf{#1}}
\newcommand{\sx}{\sigma_x}
\newcommand{\sz}{\sigma_z}
\newcommand{\SWAP}{\text{SWAP}}

\newcommand{\ua}{\mathord{\uparrow}}
\newcommand{\da}{\mathord{\downarrow}}

\def\l@subsubsection#1#2{}
\makeatother

\begin{document}

\title{Gauging fractons: immobile non-Abelian quasiparticles, fractals, \\ and position-dependent degeneracies}

\author{Daniel Bulmash}
\author{Maissam Barkeshli}

\affiliation{Condensed Matter Theory Center and Joint Quantum Institute, Department of Physics, University of Maryland, College Park, Maryland 20472 USA}
\begin{abstract}
  The study of gapped quantum many-body systems in three spatial dimensions has uncovered the existence of quantum states hosting quasiparticles
  that are confined, not by energetics but by the structure of local operators, to move along lower dimensional submanifolds. These so-called ``fracton'' phases
  are beyond the usual topological quantum field theory description, and thus require new theoretical frameworks to describe them. Here we consider
  coupling fracton models to topological quantum field theories in (3+1) dimensions by starting with two copies of a known fracton model and
  gauging the $\mathbb{Z}_2$ symmetry that exchanges the two copies. This yields a class of exactly solvable lattice models that we study in detail
  for the case of the X-cube model and Haah's cubic code. The resulting phases host finite-energy non-Abelian immobile quasiparticles with
  robust degeneracies that depend on their relative positions. The phases also host non-Abelian string excitations with robust degeneracies
  that depend on the string geometry. Applying the construction to Haah's cubic code in particular provides an exactly solvable model with finite energy
  yet immobile non-Abelian quasiparticles that can only be created at the corners of operators with fractal support.
\end{abstract}

\maketitle

\section{Introduction}

The last several decades in condensed matter physics has seen substantial progress in our understanding of gapped quantum many-body systems, ranging from
topologically ordered systems\cite{WenBook}, such as fractional quantum Hall states and quantum spin liquids, to symmetry protected topological states, such as topological insulators. While
such phases can be described through the framework of topological quantum field theory (TQFT), it has more recently been understood
that in three and higher spatial dimensions there are also possible gapped states that are beyond the usual TQFT description. These states\cite{ChamonGlass,CastelnovoFirstXCubePaper, BravyiChamonModel, HaahsCode, BravyiHaahSelfCorrection, HaahModules, YoshidaFractal, VijayFractons, VijayGaugedSubsystem, NandkishoreFractonReview}, which
are now referred to as having ``fracton order,'' share a number of features in common with more conventional topological orders, including topologically
non-trivial quasiparticle excitations and robust topology-dependent ground state degeneracies.

In contrast to conventional topological orders, fracton phases possess a strong geometry dependence, most notably in that local operators can create or destroy excitations only in certain geometric patterns, which implies that the mobility of excitations is highly constrained. For example, quasiparticles may be confined to move along certain lower dimensional submanifolds, or perhaps may even be
completely immobile, despite the energy of a state with well-separated quasiparticles being finite. Importantly, these restrictions
do not arise from energetics, but rather from the structure of local operators in the Hilbert space. Particles that are fully immobile 
are referred to as fractons. Such mobility restrictions can also occur for electric and magnetic charges in gapless $U(1)$ gauge
theories\cite{PretkoSubdimensional,BulmashGeneralizedGauge,GromovMultipole}, though we focus on the gapped case.

Since the original fracton model by Chamon\cite{ChamonGlass, BravyiChamonModel} was presented, an intriguing development was
the discovery that there are states, such as the ground states of Haah's cubic code\cite{HaahsCode} and several others\cite{HsiehFractonsPartons, TianGeneralizedHaah,HaahBifurcation}, that do not support any
mobile topologically non-trivial quasi-particles; in other words, the system does not support any
string operator that creates a pair of quasiparticles at its ends. Instead, in Haah's code the well-separated non-trivial quasiparticles can
only be created at the corners of ``fractal operators,'' which are operators with support on a fractal subsystem.

These developments raise the question of how to develop a general theoretical framework to describe the set of allowed
gapped states in (3+1) dimensions. Given the rich mathematical structure of topological quantum field theory and in particular
the understanding of unitary modular tensor categories for (2+1) dimensional states, it is possible that the understanding
of gapped fracton orders may similarly uncover a rich mathematical structure.

The Hamiltonians for Chamon's model and Haah's code both consist entirely of Pauli operators. Ref.~\onlinecite{HaahModules} has
since provided a classification of commuting Pauli Hamiltonians, and there has been a flurry of recent activity\cite{SlagleFieldTheory, SlagleGenericLattices, ShirleyXCubeFoliations, SlagleGaugeTheoryCurvedSpace,FractonCorrFunctions, ShiFractonEntanglement, FractonEntanglement, ShirleyEntanglement, RecoverableInformation, HalaszFractons, EmergentPhasesFractons, SivaBravyiMemory, PremDynamics, BulmashHiggs, MaHiggs, ShirleyFoliatedCheckerboard, ShirleyFoliatedFractional, BulmashGappedBoundaries, YouLego, SlagleFoliatedFieldTheory, WangFoliatedCheckerboard} studying various
aspects of the possible fracton phases of matter.  Since commuting Pauli Hamiltonians can only give rise to Abelian topological quasiparticles, which have a unique fusion outcome when
fused with themselves, a natural challenge is to understand models that can give rise to non-Abelian fractons. Such phases would arise
from Hamiltonians that are necessarily beyond the commuting Pauli Hamiltonian classification.

To date, there have been two approaches to developing fracton models with non-Abelian quasiparticles. The first is through a layer construction\cite{MaLayer,VijayLayer,VijayNonAbelianFractons,PremCageNet},
where layers of non-Abelian (2+1)D topological orders are stacked and subsequently strings of particles from different layers are condensed
in various geometrical patterns. However in this approach, in all known cases the non-Abelian quasiparticles are mobile in at least one direction. The second approach\cite{SongTwisted}
proceeds by considering generalized gauge theories with Abelian gauge group, but combined with an analog of a Dijkgraaf-Witten cohomological twist,
which renders the excitations non-Abelian. In this latter approach, immobile non-Abelian quasiparticles can be created at the corners of membrane operators. To date, phases containing immobile non-Abelian quasiparticles that can only be created by fractal operators have not been found.

A powerful way of obtaining conventional non-Abelian topological order from Abelian topological order is to gauge a global symmetry
that permutes quasiparticle types\cite{BarkeshliOrbifoldFQH, BarkeshliCSParafermion, BarkeshliTwistedZN, GCrossed,TeoTwists},
such as the permutation symmetry of multiple copies of a topological phase. After gauging, any quasiparticle
with a non-trivial orbit under the symmetry becomes non-Abelian, as are the fluxes of the symmetry (the twist defects). It is well-known that gauging subsystem symmetries of more conventional phases can produce fracton phases\cite{VijayGaugedSubsystem, Williamson2016, KubicaUngauging, ShirleyFoliatedGauge}, and fracton phases in the presence of additional symmetries\cite{PretkoSupersolidDuality, YouSymmetricFracton}
 and symmetry defects\cite{YouFractonDefects} are an active area of current research. It is a natural question, then, to ask whether gauging a global symmetry of an Abelian fracton phase produces non-Abelian fractons.

In this paper, we consider coupling fracton phases to topological quantum field theories by gauging the permutation symmetry of multiple
copies of a given fracton phase. Such a construction allows us to obtain an exactly solvable model where we couple $N$ copies of
a fracton phase to any discrete $G$ gauge theory, where $G$ is a subgroup of the permutation group on
$N$ copies, although we specialize to $N=2$. The interplay of fracton order and topological order is an interesting topic, as it is not \it a priori \rm clear whether,
for example, the subdimensional excitations of the fracton system would survive when coupled in a non-trivial way to a system with
topologically non-trivial mobile, deconfined excitations. We find that gauging global symmetries of Abelian fracton models
indeed leads to models with non-Abelian subdimensional excitations. Notably, the gauging procedure does not affect the mobility
of the excitations of the ungauged theory. Gauging symmetries in models with immobile excitations, including fractons created at the corners of
fractal operators, leads to non-Abelian immobile excitations created by operators with similar support. We show explicitly that the
robust degeneracy associated with these excitations depends in a highly nontrivial way on the relative positions of those excitations. The models we
construct, which are based on the X-cube model\cite{CastelnovoFirstXCubePaper,VijayGaugedSubsystem} and Haah's code,
have another unusual feature: flexible, dynamical string-like excitations which carry topological degeneracy which depends on the
geometry of the string. This is in contrast to conventional topological order, where the degeneracy associated to string excitations
is independent of their geometry. 

Our work, in particular a non-Abelian extension of Haah's code, may also be of interest for quantum information applications.
Our model allows the encoding of topologically protected degeneracy in fully immobile excitations (a benefit unique to fractons),
with some logical operations on the degenerate manifold implemented via a fractonic analog of braiding. This is in contrast to Haah's original code,
where qubits would have to be encoded in the ground state subspace of the system on a topologically nontrivial manifold such as a three-torus. 

The structure of this paper is as follows. In Sec. \ref{sec:ToricCode}, we explain the gauging procedure in a more familiar context by gauging the layer-swap symmetry in the bilayer (2+1)D toric code. Although the results are already well-understood from the perspective of topological order, we go into considerable detail because the technical framework is precisely the same as the one which we use in Sec. \ref{sec:X-Cube} to gauge the layer-swap symmetry of the bilayer X-Cube model. We then use Sec. \ref{sec:X-CubePheno} to discuss some of the phenomenology of the gauged bilayer X-Cube model, including computing topological degeneracies and non-Abelian fracton braiding-like procedures. In Sec. \ref{sec:Haah}, we apply the gauging procedure to gauge the layer-swap symmetry of the bilayer Haah's code model. Finally, Sec. \ref{sec:Discussion} contains discussion and conclusions. Exhaustive lists of properties of our models and some technical calculations are relegated to several appendices.


\section{Warmup: Gauging the (2+1)D Bilayer Toric Code}
\label{sec:ToricCode}

In this section, we will gauge the layer-swap symmetry of the bilayer toric code to obtain a model with the same topological order as the
$\left[\mathbb{Z}_2 \times \mathbb{Z}_2\right] \rtimes \mathbb{Z}_2 \simeq \mathbb{D}_4$ quantum double model (here $\mathbb{D}_4$ is the symmetry group of a square).
Although the resulting topological order is well-understood, at least at an abstract level, understanding the procedure in detail and at the level of
explicit exactly solvable Hamiltonians generalizes directly to the fracton models. As such, it is useful to study this example in considerable depth
before considering the fracton case. 

Before proceeding, we comment that, at the level of exactly solvable models, there are several ways in which the gauging procedure can be performed.
One method explicitly reproduces the $\mathbb{D}_4$ quantum double model\cite{KitaevQC03}. Although such a result is desirable and natural for (2+1)D topological order,
the required procedure does not appear to generalize easily to the (higher dimensional) fracton models, so we relegate that choice of gauging procedure
to Appendix \ref{app:alternateGauge} and instead implement a gauging procedure that leads to a more convenient generalization.

\subsection{Building the model}

\begin{figure}
\centering
\includegraphics[width=0.4\columnwidth]{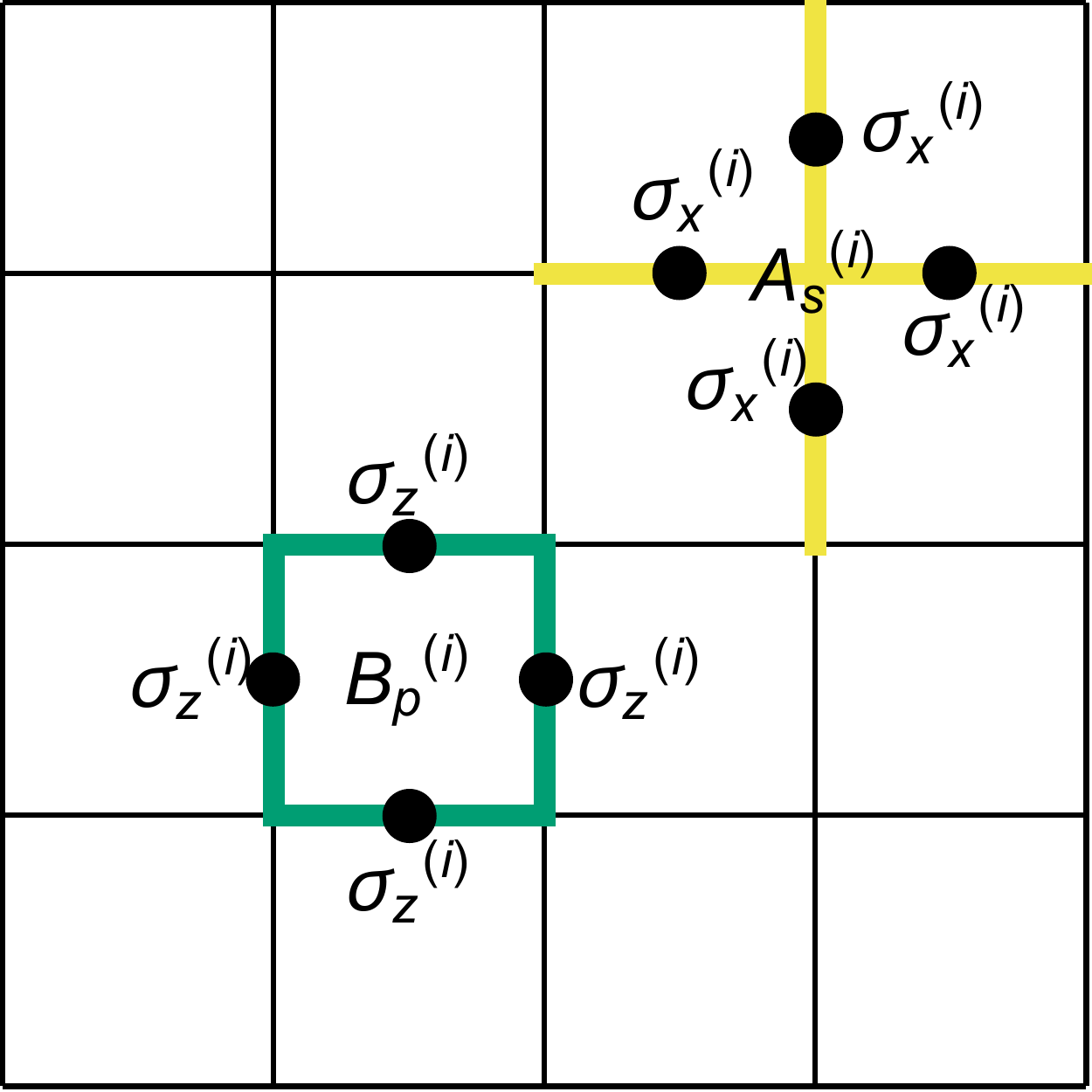}
\caption{Hamiltonian terms for each layer of the bilayer (2+1)D toric code. There are two spins per link of the lattice, and the superscripts refer to which layer the operator acts on.}
\label{fig:toricCodeH}
\end{figure}

We start from two copies of the usual toric code. The Hilbert space for the ungauged model is two qubits per link of the square lattice, with ungauged Hamiltonian
\begin{align}
H_0 &= H_1 + H_2 ,\label{eqn:ungaugedBilayerTC}\\
H_i &= -\sum_{s} A_s^{(i)} - \sum_p B_p^{(i)} .
\end{align}
Here $A_s^{(i)} = \otimes_{+} \sx^{(i)}$ and $B_p^{(i)} = \otimes_{\square} \sz^{(i)}$ are the usual star and plaquette operators shown in
Fig.~\ref{fig:toricCodeH}. As usual, $\sx$ and $\sz$ are Pauli operators. The ungauged Hamiltonian $H_0$ has the global symmetry
$\bigotimes \SWAP$, where $\SWAP$ is the local two-spin operator which exchanges the state of a single spin in layer 1 with the
spin on the same link in layer 2, and the tensor product is over all links. In order to gauge the symmetry, we
introduce extra ``gauge" spins and modify the Hamiltonian so that some operator involving SWAP on a single link commutes with
every term in the new Hamiltonian. 

\begin{figure*}
\centering
\subfloat[\label{fig:plaquetteRewrite}]{
\includegraphics[width=1.5\columnwidth]{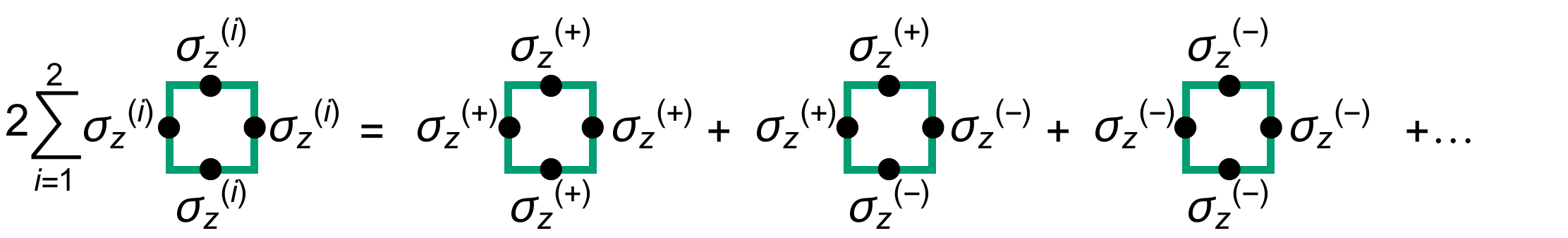}}\\
\subfloat[\label{fig:gaugedTCHilbertSpace}]{
\includegraphics[width=0.4\columnwidth]{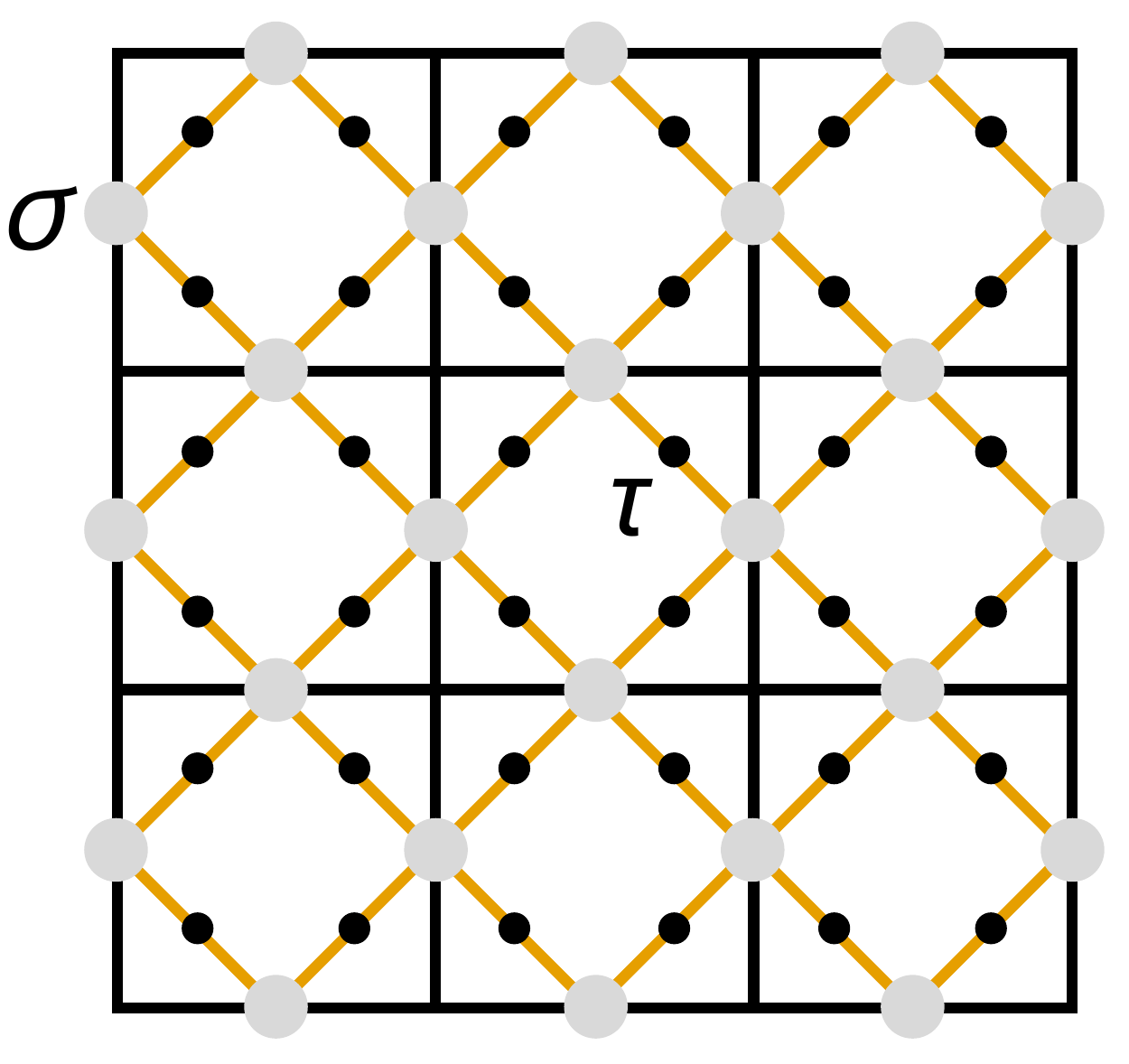}} \hspace{0.1\columnwidth}
\subfloat[\label{fig:TCgaugeGenerator}]{
\includegraphics[width=0.4\columnwidth]{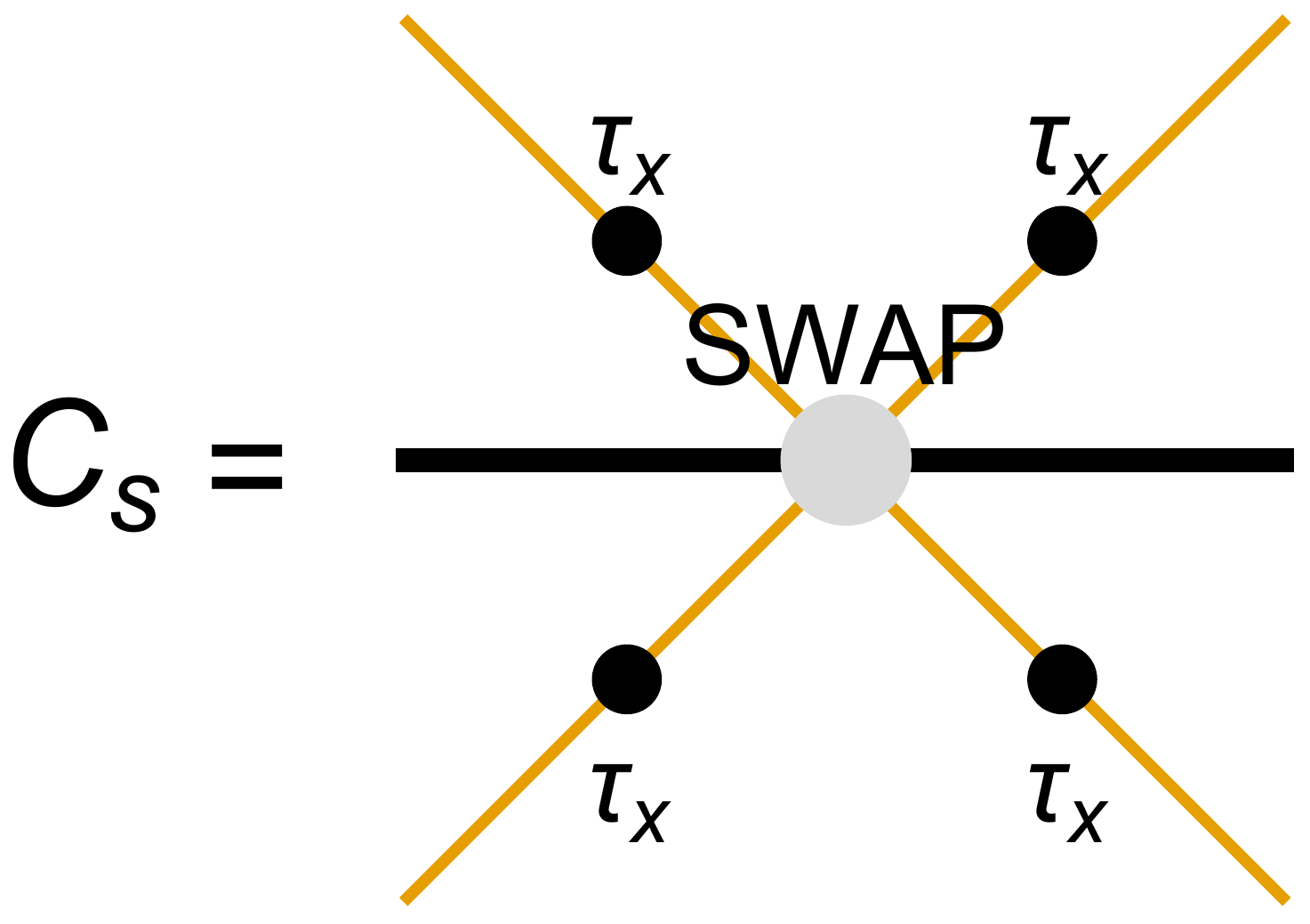}}\\
\subfloat[\label{fig:gaugedPlaquette}]{
\includegraphics[width=1.5\columnwidth]{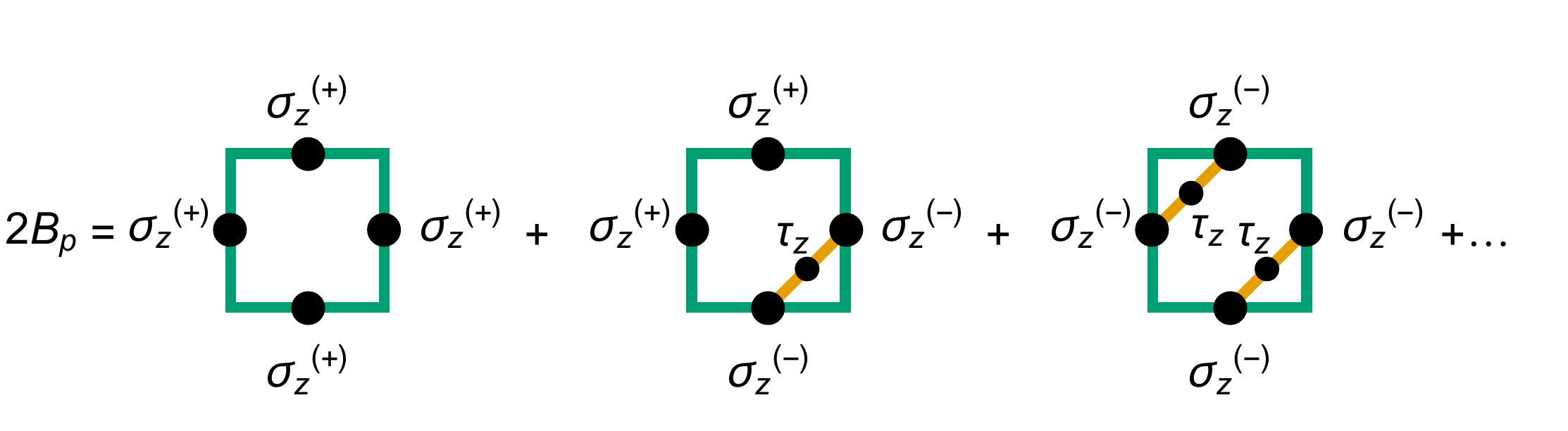}}
\caption{Gauging procedure for the bilayer toric code. (a) Rewriting of the ungauged sum of plaquette operators in Eq. \eqref{eqn:plaquetteRewrite}. The sum on the right-hand side is over all choices of signs with an even number of $\sz^{(-)}$ operators. (b) Hilbert space for the gauged bilayer toric code. There are two of the original ``matter" qubits per link (light grey circles) and four additional ``gauge" qubits per site (dark circles), one per orange link. (c) Generator of the gauge symmetry in the gauged bilayer toric code. (d) Gauge-invariant version $B_p$ of (a). Distinct pairs of $\sz^{(-)}$ operators are connected by $\tau_z$ operators.}
\end{figure*}

First we decompose the Hamiltonian into terms of definite parity under local SWAPs. The decomposition is done using the operators $\sigma_a^{(\pm)} = (\sigma_a^{(1)} \pm \sigma_a^{(2)})/\sqrt{2}$ for $a=x,z$. In what follows, we will need the following algebraic facts about $\sigma_{a}^{(\pm)}$ with $a=x,z$:
\begin{align}
\SWAP \sigma_a^{(\pm)}\SWAP &= \pm \sigma_a^{(\pm)} \nonumber \\
\sigma_a^{(\pm)}\sigma_a^{(\pm)} &= 1 \pm \sigma_a^{(1)}\sigma_a^{(2)} \nonumber \\
\sigma_a^{(\pm)}\sigma_a^{(\mp)} &= 0 \nonumber \\
\sx^{(\pm)}\sz^{(+)} &= -\sz^{(-)}\sx^{(\mp)} \nonumber \\
\sx^{(\pm)}\sz^{(-)} &= -\sz^{(+)}\sx^{(\mp)} \label{eqn:sigmaAlgebra}
\end{align}
One can check straightforwardly that
\begin{equation}
B_p^{(1)} + B_p^{(2)} = \frac{1}{2}\sum_{\lbrace s_{\bv{r}} | \prod s_\bv{r} = +1\rbrace} \bigotimes_{\bv{r} \in \square} \sx^{(s_{\bv{r}})}
\label{eqn:plaquetteRewrite}
\end{equation}
where $s_{\bv{r}}=\pm$. The product is over the four different edges in a plaquette, and the condition in the sum enforces an even number of minus signs; see Fig.~\ref{fig:plaquetteRewrite}. We have suppressed the explicit position index on the $\sx^{(\pm)}$ operators. An even number of minus signs is required to maintain invariance under the global SWAP operation.

We implement the gauging procedure by adding ``gauge" qubits $\tau$ which live halfway between nearest neighbor ``matter" qubits $\sigma$ as shown in Fig.~\ref{fig:gaugedTCHilbertSpace}. We then define the $\mathbb{Z}_2$ gauge symmetry generator 
\begin{equation}
C_e = \SWAP_e \bigotimes_{\bv{r}\in +} \tau_{x,\bv{r}}
\end{equation}
Here $e$ is an edge of the lattice of $\sigma$ spins and the product of $\tau_x$ is taken over the four gauge spins surrounding that edge. This operator is shown in Fig.~\ref{fig:TCgaugeGenerator}. We also define the $\mathbb{Z}_2$ flux operator on both the sites and plaquettes of the original (matter) lattice by
\begin{equation}
D_{\Diamond} = \bigotimes_{\bv{r} \in \Diamond} \tau_{z,\bv{r}}
\end{equation}
where the product is over the four gauge spins surrounding a site or within a plaquette. We demand that $C_e$ generate a gauge symmetry by modifying every term in the Hamiltonian (using $\tau_z$s) so that they commute with all $C_e$. For example, the term with four $+$ signs in Fig.~\ref{fig:plaquetteRewrite} commutes with all the $C_e$ and is unmodified, while the other terms in Fig.~\ref{fig:gaugedPlaquette} require $\tau_z$s next to every $\sigma^{(-)}$ (since $\sigma^{(-)}$ anticommutes with SWAP). The gauged Hamiltonian $H_{\text{gauged}}$ has the form
\begin{equation}
H_{\text{gauged}} = -\sum_s A_s -\sum_p B_p - \sum_e C_e - \sum_{\Diamond} D_{\Diamond}
\label{eqn:gaugedTCHamiltonian}
\end{equation}
Here $A_s$ and $B_p$ are the gauge-invariant versions of the star and plaquette terms from the original bilayer model and are each the sum of eight terms, three of which are shown in Fig.~\ref{fig:gaugedPlaquette}. As before, $C_e$ is the local symmetry generator shown in Fig.~\ref{fig:TCgaugeGenerator}; we enforce this gauge symmetry energetically. The flux term $D_{\Diamond}$ energetically penalizes the presence of $\mathbb{Z}_2$ flux. 

\subsection{Algebraic properties of $H_{\text{gauged}}$}
The Hamiltonian $H_{\text{gauged}}$ is not a stabilizer model. Nevertheless, it has considerable
structure. First, it is straightforward to check that
\begin{equation}
(A_s)^2 = 2\left(1 + \otimes_{+}\sx^{(1)}\sx^{(2)}\right)
\end{equation}
This operator has eigenvalues $0$ and $4$, so $A_s$ has eigenvalues $\pm 2$ and 0. The same sort of computation holds for $B_p$.

\begin{figure*}
\includegraphics[width=1.5\columnwidth]{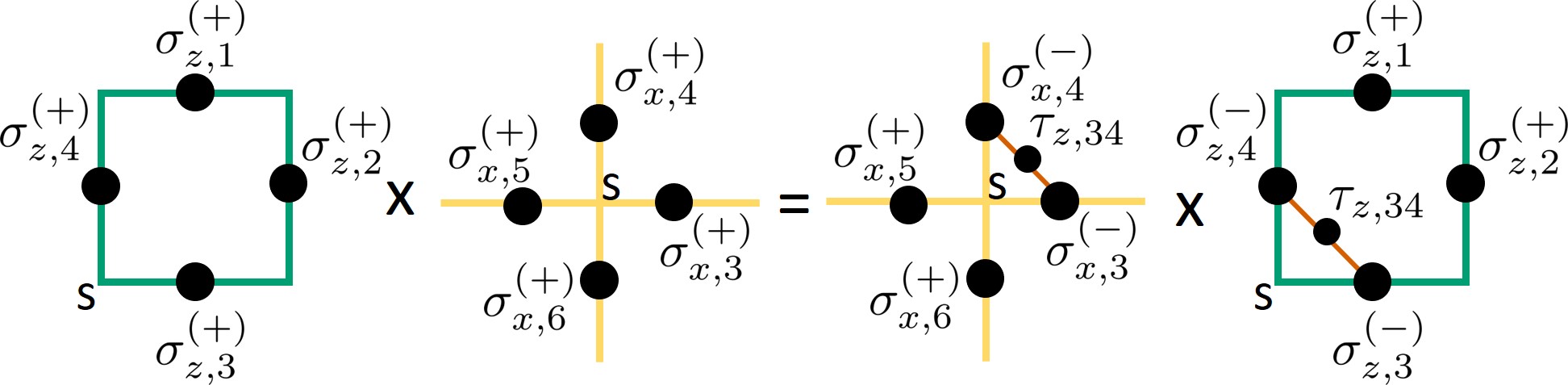}
\caption{Commutation properties of example terms in the gauged bilayer toric code's $A_s$ and $B_p$ operators. Here $s$ labels the same lattice site.}
\label{fig:gaugedCommRelations}
\end{figure*}

Furthermore,
\begin{align}
 [A_s,B_p]=0.
  \end{align}
To understand this fact when $A_s$ and $B_p$ have overlapping support, we isolate single terms in each of $A_s$ and $B_p$; consider, for example, the product of one plaquette term, which we call $B_p^{(+)}$, and one star term, which we call $A_s^{(+)}$, which have only $\sigma_{x,z}^{(+)}$ operators, as shown in Fig.~\ref{fig:gaugedCommRelations}. Using the algebraic properties in Eq. \eqref{eqn:sigmaAlgebra}, it is straightforward to check that
\begin{align}
B_p^{(+)}A_s^{(+)} &= \sigma_{z,1}^{(+)} \sigma_{z,2}^{(+)}\sigma_{z,3}^{(+)}\sigma_{z,4}^{(+)}\sigma_{x,3}^{(+)}\sigma_{x,4}^{(+)}\sigma_{x,5}^{(+)}\sigma_{x,6}^{(+)}\\
&= \sigma_{z,3}^{(-)}\sigma_{z,4}^{(-)}\sigma_{z,5}^{(+)}\sigma_{z,6}^{(+)}\sigma_{x,1}^{(+)} \sigma_{x,2}^{(+)}\sigma_{x,3}^{(-)}\sigma_{x,4}^{(-)}\\
&=\sigma_{z,3}^{(-)}\tau_{34}^z\sigma_{z,4}^{(-)}\sigma_{z,5}^{(+)}\sigma_{z,6}^{(+)}\sigma_{x,1}^{(+)} \sigma_{x,2}^{(+)}\sigma_{x,3}^{(-)}\tau_{34}^z\sigma_{x,4}^{(-)}
\end{align}
where in the last line we simply inserted $\tau_{34}^z\tau_{34}^z=1$. This equation is shown pictorially in Fig.~\ref{fig:gaugedCommRelations}. Here the numbers in the subscript indicate labels for links. This is a product of a different term in $A_s$ with a different term in $B_p$; in commuting these operators past each other, the terms in the Hamiltonian have been \textit{permuted}. Using a similar computation for each term, it is easy to check that indeed $[A_s,B_p]=0$.

By construction, both $C_e$ and $D_{\Diamond}$ commute with every individual summand in $A_s$ and $B_p$ and with each other. Therefore, although $A_s$ and $B_p$ are not projectors, all four types of terms in the Hamiltonian commute.

\subsection{String-net wavefunction picture}

We now describe a simple string-net wavefunction picture for the ground states of the gauged bilayer toric code which will be convenient to
use both for computing the ground state degeneracy and for understanding its excitations.

The ground states of the toric code can be understood as condensates of closed strings on (say) the links of the lattice; that is, a superposition
of all possible closed strings on the lattice. In our convention, presence of a string means $\sx$ has eigenvalue $-1$ and absence means $\sx$
has eigenvalue $+1$. These strings can be thought of as $e$-strings, because an open string induces violations of the star operators, which correspond
to the $e$ particles ($\mathbb{Z}_2$ charges) of the toric code. The topologically degenerate ground states can thus be labeled by the
parity of the number of closed strings wrapping the various non-contractible cycles of the space. The (ungauged) bilayer toric code is
understood similarly; the only difference is that the strings have a color which corresponds to their layer, and strings of different colors
can occupy the same link. We will use orange and blue to denote strings in layer 1 and layer 2 respectively.

\begin{figure}
\centering
\subfloat[\label{fig:twistDefectStringNet}]{\includegraphics[width=0.6\columnwidth]{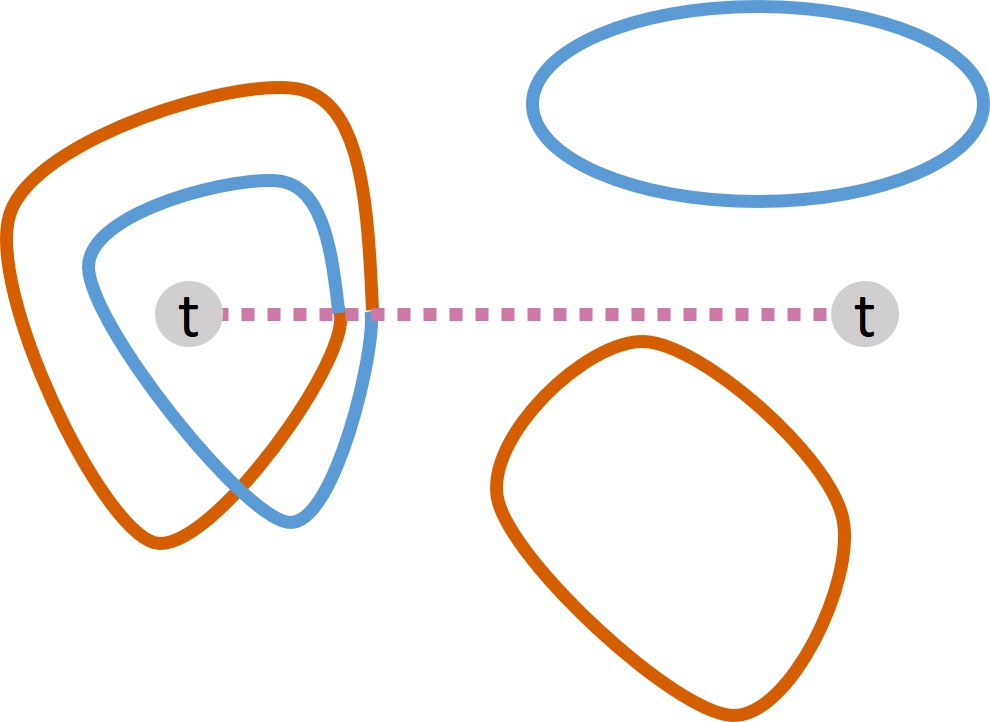}}\\
\subfloat[\label{fig:TCstringNet}]{\includegraphics[width=0.6\columnwidth]{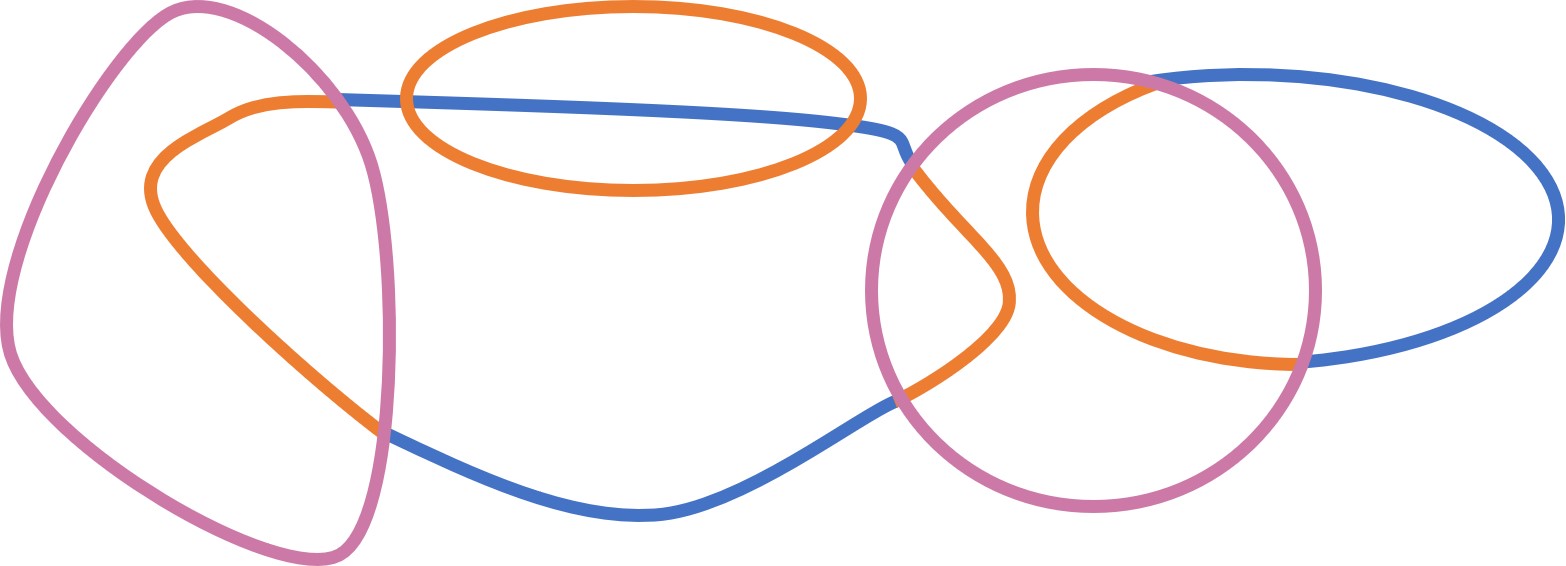}}
\caption{String-net configurations before and after gauging the SWAP symmetry of the bilayer toric code. Orange and blue strings correspond to the original toric code string-net wavefunctions. (a) Example string-net configuration in the presence of twist defects (grey circles labeled $t$). Dashed purple line is a branch cut for the layer-swap symmetry. (b) Example ground state string-net configuration for the gauged bilayer toric code. Solid purple strings are the proliferated branch cuts.}
\end{figure}

Before gauging the SWAP symmetry, consider a pair of extrinsic twist defects, corresponding to the end-points of a branch cut across which
the two layers are glued to each other. Whenever a string crosses the branch cut, it changes colors. The twist defects correspond to the
$\mathbb{Z}_2$ flux of the global $\mathbb{Z}_2$ SWAP symmetry. A string which makes a full loop around the twist defect must change
colors since it gets transformed by the action of the symmetry, as in Fig.~\ref{fig:twistDefectStringNet}. 

Gauging the SWAP symmetry means that the ground state is now also a condensate of closed branch cuts. Therefore, the ground state for the gauged model
corresponds to a string-net condensate with three different colored strings - the strings in the original two layers of toric code (orange and blue),
corresponding to the $\sigma^{(1)}$ and $\sigma^{(2)}$ spins, and the branch cut or ``twist defect" strings (which we denote with purple),
corresponding to the $\tau$ spins. Note that since the open twist defect strings corresponded to $\mathbb{Z}_2$ fluxes, the purple strings
should be thought of as $m$ strings associated with the $\tau$ spins; thus here the presence or absence of a string corresponds to whether $\tau_z = \pm 1$. 

Given a particular configuration of the purple string condensate, we can still write down a string-net
wavefunction for the orange and blue strings, but with the rule that whenever an orange or blue string crosses a purple string, it changes colors.
An example string configuration is given in Fig.~\ref{fig:TCstringNet}. Note in particular that we may start with a configuration
with no purple strings, create a closed purple string from the vacuum, wrap it all the way around the system (or bring it out to
infinity if the system is on a plane), and then annihilate it. This implements a global SWAP operation; therefore, all
ground states must be invariant under a global SWAP.

\subsection{Ground state degeneracy}

The string-net wave function picture is helpful to understand the ground state degeneracy of the gauged model on a torus
by considering topologically nontrivial strings. 

Before gauging, there are $2^4$ ground states labeled by the parity of the number of nontrivial strings winding around each
handle of the torus in each layer.

\begin{figure}
\centering
\subfloat[\label{fig:torusDoubleWrap}]{\includegraphics[width=.4\columnwidth]{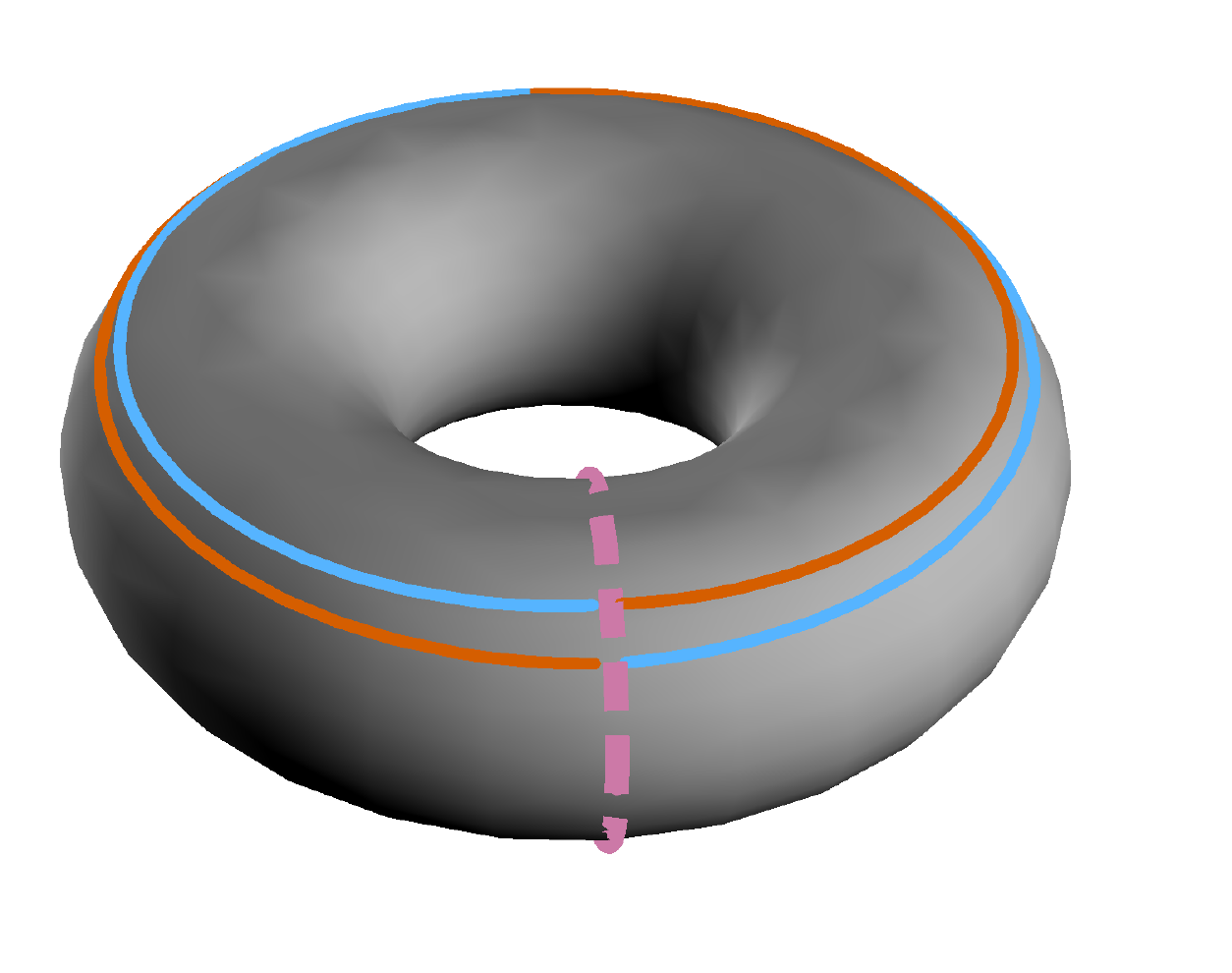}} \hspace{.1\columnwidth}
\subfloat[\label{fig:torusMoveLoop}]{\includegraphics[width=.4\columnwidth]{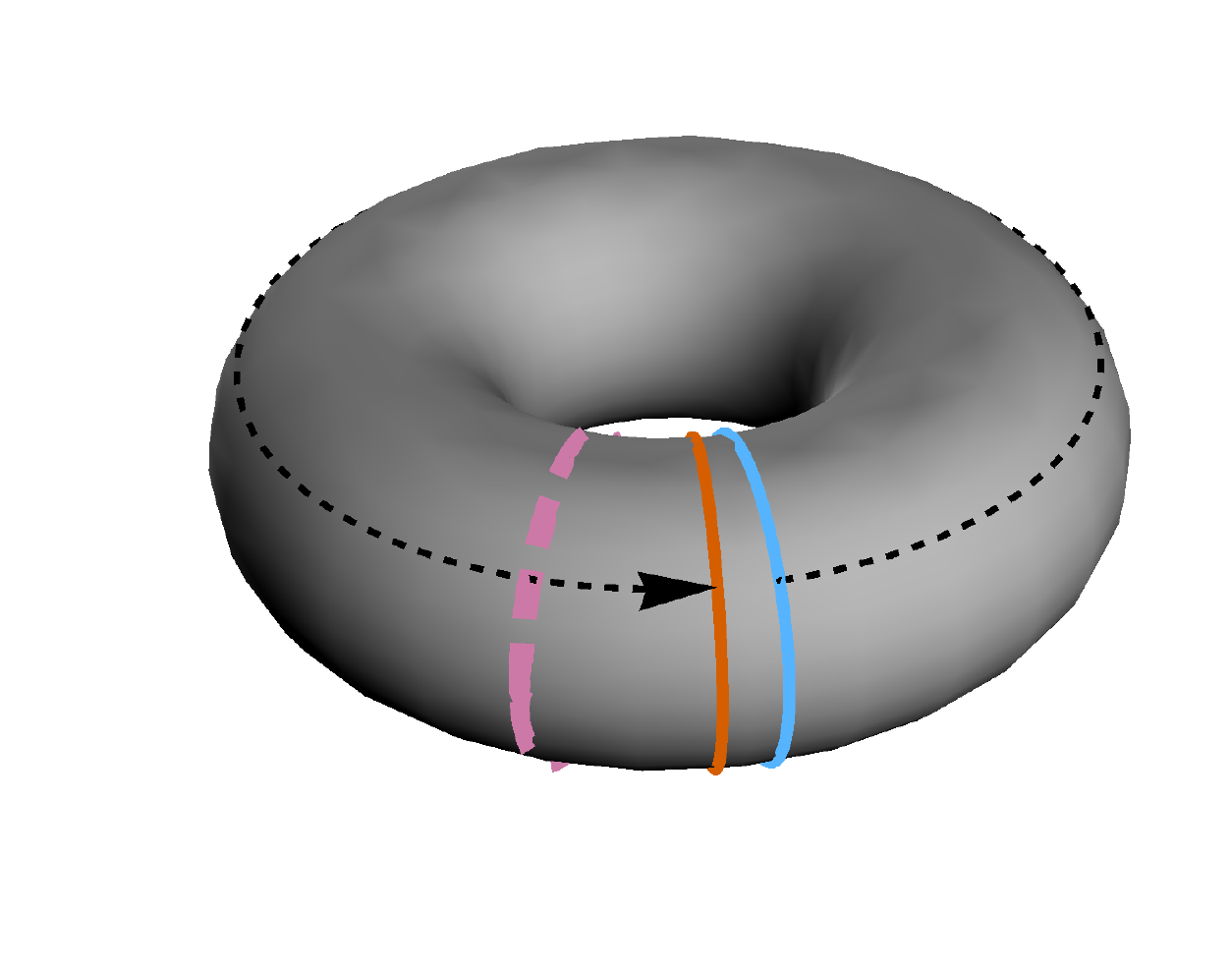}}
\caption{String-net configurations in the gauged bilayer toric code in the presence of a symmetry flux through one handle of the torus. Orange (resp. blue) is a string in the first (resp. second) layer of toric code and the purple dashed line is the branch cut implementing the symmetry twist due to the symmetry flux. All topologically trivial strings are omitted. (a) Nontrivial Wilson loop around the $x$ handle of the torus. (b) Nontrivial Wilson loop around the $y$ handle of the torus. A nontrivial red string is equivalent to a nontrivial blue string by the process of moving the string all the way around the torus (black arrow).}
\end{figure}

After gauging, all states must be invariant under the global SWAP symmetry since the twist defects proliferate. Moreover,
we are now allowed configurations where there are an odd number of twist defect strings (purple loops) wrapping around
any handle of the torus. In the presence of an odd number of topologically nontrivial twist defect strings, the original strings acquire
twisted boundary conditions; they must wind \textit{twice} around the torus (since a layer 1 string becomes a layer 2 string
and vice-versa after winding once around the torus), as shown in Fig.~\ref{fig:torusDoubleWrap}. It is therefore most
convenient to compute the ground state degeneracy by separately considering each topological sector for the purple strings
(i.e. boundary conditions for the orange/blue strings), a task to which we now turn.

In the untwisted sector, we simply select the states of the ungauged theory which are symmetric under the global SWAP symmetry.
It is straightforward to check that there are 10 such states.

In the sector with twisted boundary conditions around the $x$ handle only, a state can either have no string around the $x$ handle or have one string which winds twice around the $x$ handle, as in Fig.~\ref{fig:torusDoubleWrap}. We also note that a string which winds around the $y$ (untwisted) handle in layer 1 can be translated all the way around the $x$ handle, in which case it becomes a string in layer 2, as in Fig.~\ref{fig:torusMoveLoop}. This also implies that the state with both colors of strings winding around the $y$ handle is equivalent to the state with no string around that handle. The superposition $\ket{\text{layer }1 \text{ string}}+\ket{\text{layer }2 \text{ string}}$ of strings winding around the $x$ handle is, however, topologically distinct and invariant under a global SWAP. In total, there are thus 4 states in this sector, and another 4 when the twist is instead only around the $y$ handle.

Finally, in the sector with twists around both handles of the torus, there are 4 states (each handle either has no string or a doubly-wound string).

In total, this means the ground state degeneracy is $10+4+4+4=22$. This calculation can be straightforwardly extended to closed surfaces of any genus $g$, which then
allows extraction of the quantum dimensions of each of the topologically non-trivial quasi-particles through the formula
\begin{align}
\text{dim } \mathcal{H}(\Sigma_g) = \frac{1}{\mathcal{D}} \sum_a d_a^{2-2g} ,
\end{align}
$\mathcal{H}(\Sigma_g)$ is the ground state subspace on a closed genus $g$ surface, $\Sigma_g$, $\mathcal{D}$ is the total quantum dimension, and $d_a$ is the quantum dimension of
the quasiparticle labeled $a$. 

\subsection{Excitations and string operators}

The elementary excitations of the $\mathbb{D}_4$ quantum double model are labeled by conjugacy classes $C$ and irreducible representations (irreps) of the centralizer $Z(C)$, where the trivial conjugacy class corresponds to pure charges and the trivial irrep corresponds to pure fluxes. For labeling purposes, we give a convenient presentation of $\mathbb{D}_4$ as $\langle a,b,c| a^2=b^2=c^2=1, cac=b\rangle$. The nontrivial pure charges and pure fluxes in this topological order are listed in Table \ref{tab:D4Excitations}; all other anyons are dyons obtained from various fusions of the pure charges and pure fluxes. In total, there are $22$ topologically distinct quasiparticles. 

\begin{table*}
\centering
\renewcommand{\arraystretch}{1.3}
\begin{tabular}{@{}lllllllllllllllllll}
\toprule[2pt]
\textbf{Label} && \textbf{Conjugacy Class $C$}&&\textbf{Irrep of $Z(C)$}&& $d_a$ && $\mathcal{U}_+$ && $\mathcal{U}_-$ && $\mathcal{V}_+$ && $\mathcal{V}_-$ && $\mathcal{W}_+$ && $\mathcal{W}_-$ \\ \hline
$\phi$ && $\lbrace 1\rbrace$ && $(a,c)\rightarrow (1,-1)$ && 1 && 1 && 1 && 1 && 1 && $\tau_z$ && $\tau_z$ \\
$e^{(1)}e^{(2)}$ && $\lbrace 1\rbrace$ && $(a,c) \rightarrow (-1,1)$ && 1 &&  $\sz^{(1)}\sz^{(2)}$ && $\sz^{(1)}\sz^{(2)}$ && 1 && 1 && 1 &&1\\
$\phi e^{(1)}e^{(2)}$ && $\lbrace 1 \rbrace$ && $(a,c) \rightarrow (-1,-1)$ &&1 && $\sz^{(1)}\sz^{(2)}$ && $\sz^{(1)}\sz^{(2)}$ && 1 && 1 && $\tau_z$ && $\tau_z$ \\
$[e]$ && $\lbrace 1\rbrace$ && 2D irrep of $\mathbb{D}_4$ && 2 && $\sz^{(+)}$ && $\sz^{(-)}$ && 1 && 1 && 1 && $\tau_z$  \\
$m^{(1)}m^{(2)}$ && $\lbrace ab \rbrace$ && Trivial && 1 && 1 && 1 && $\sx^{(1)}\sx^{(2)}$ &&  $\sx^{(1)}\sx^{(2)}$ && 1 && 1 \\
$[m]$ && $\lbrace a, b \rbrace$ && Trivial && 2 && 1 && 1 && $\sx^{(+)}$ && $\sx^{(-)}$ && 1 && $\tau_z$ \\
$\sigma_1$ && $\lbrace c, abc \rbrace$ && Trivial && 2 && $1$ && $\sz^{(1)}\sz^{(2)}$ && $1+\sx^{(1)}\sx^{(2)}$ && $1-\sx^{(1)}\sx^{(2)}$ && $\tau_x$ && $\tau_x$  \\
$\sigma_m$  && $\lbrace ac, bc \rbrace$ && Trivial && 2 && 1 && -$\sz^{(1)}\sz^{(2)}$ && $\sx^{(+)}$ && $\sx^{(-)}$ && $\tau_x$ && $\tau_y$ \\	\bottomrule[2pt]	
\end{tabular}
\caption{Nontrivial pure charges and pure fluxes of $\mathbb{D}_4$ quantum double topological order, which correspond to irreps and conjugacy classes of $\mathbb{D}_4$ respectively, their quantum dimensions $d_a$, and a specification $\mathcal{U}_{\pm},\mathcal{V}_{\pm},\mathcal{W}_{\pm}$ of the string operators that create those excitations. See Sec.~\ref{subsubsec:TCstringOps} for the construction of the string operators.}
\label{tab:D4Excitations}
\end{table*}

Their properties can be understood
systematically using the theoretical framework developed in Ref. \onlinecite{GCrossed}. Here we briefly mention how to think of some of the quasiparticles.
We can label the quasiparticles of the bilayer toric code model as $a^{(1)} b^{(2)}$, where $a, b= 1,e,m, \psi$ are $\mathbb{Z}_2$ charge, flux, and fermion of the toric code.
After gauging, the remaining quasiparticles are as follows. We have the invariant Abelian particles $1$, $e^{(1)} e^{(2)}$, $m^{(1)} m^{(2)}$ and
$\psi^{(1)} \psi^{(2)}$, which retain their braiding and fusion properties even after gauging. The particles that are not invariant
under the SWAP symmetry are grouped into orbits, $[ab] \sim a^{(1)}b^{(2)} + b^{(1)}a^{(2)}$, with $a \neq b$,
leading to the particles $[e] \sim e^{(1)} + e^{(2)}$, $[m] \sim m^{(1)} + m^{(2)}$, etc. Each of these particles is non-Abelian, with
quantum dimension $2$, and there are 6 such particles in total. Here the $\sim$ means that these particles can be roughly thought of as
superpositions of the original particles, as seen in more detail in the next subsection. 

In addition, we have the $\mathbb{Z}_2$ charge associated with the $\tau$ degrees of freedom, which is an Abelian particle labeled by $\phi$.
We also have a $\mathbb{Z}_2$ flux $\sigma_1$ that corresponds to the twist defect after gauging, which also has quantum dimension $2$.
In total we can obtain $8$ twist defects $\sigma_a$ and $\phi \sigma_a$, for $a = 1,e,m,\psi$, which arise in the fusion outcomes
of $\sigma_1$ with the other particles. The $22$ particles thus correspond to $\phi^s a^{(1)} a^{(2)}$, $[ab]$, and $\phi^s \sigma_a$,
with $a,b = 1,e,m,\psi$ and $s = 0,1$, which gives a total of 22 particles. The particles $\phi^s a^{(1)} a^{(2)}$ are all Abelian, with quantum dimension $1$,
while $[ab]$, and $\phi^s \sigma_a$ are non-Abelian, with quantum dimension $2$. In Table \ref{tab:D4Excitations}, we list how
the pure charges and pure fluxes of the $\mathbb{D}_4$ quantum double model can be related to this labeling.

In the next subsection, we will describe how these excitations can be understood in terms of gauging two copies of the toric code model, and we
will explicitly construct the string operators that create the pure charge and pure flux excitations. The techniques used to construct the
string operators in this model will carry over nicely to the fractonic case. In particular, we will
develop an understanding of why some of the particles, such as $[e]$, are non-Abelian with quantum dimension $2$. We will explicitly
see how the fusion rule
\begin{align}
  \label{eFusion}
[e] \times [e] = 1 + \phi + e^{(1)} e^{(2)} + \phi e^{(1)} e^{(2)} 
\end{align}
arises, both at the level of operators and also from understanding the string-net wave function picture. 

Due to the structure of our model, there is no obvious mapping to a system whose degrees of freedom are elements of $\mathbb{D}_4$
(the degrees of freedom do not quite match - there are two extra qubits per site), so we identify the excitations in the lattice
model using the following intuition. Since conjugation by SWAP exchanges operators in the two layers, we will think of the Pauli operators
$\sigma_i^{(1)}$ (resp. $\sigma_i^{(2)}$) as being related to the group element $a$ (resp. $b$), while SWAP is related to the group element $c$.
Since $C_e$ anticommutes both with anything odd under SWAP and with $\tau_z$, we also think of $\tau_z$ as being related
to the group element $c$. This intuition can be confirmed by examining the fusion rules.

\subsubsection{Wave function picture of excitations}
\label{subsubsec:TCstringOps}

\begin{figure*}
\centering
\subfloat[\label{fig:TCe1e2StringNet}]{\includegraphics[width=0.7\columnwidth]{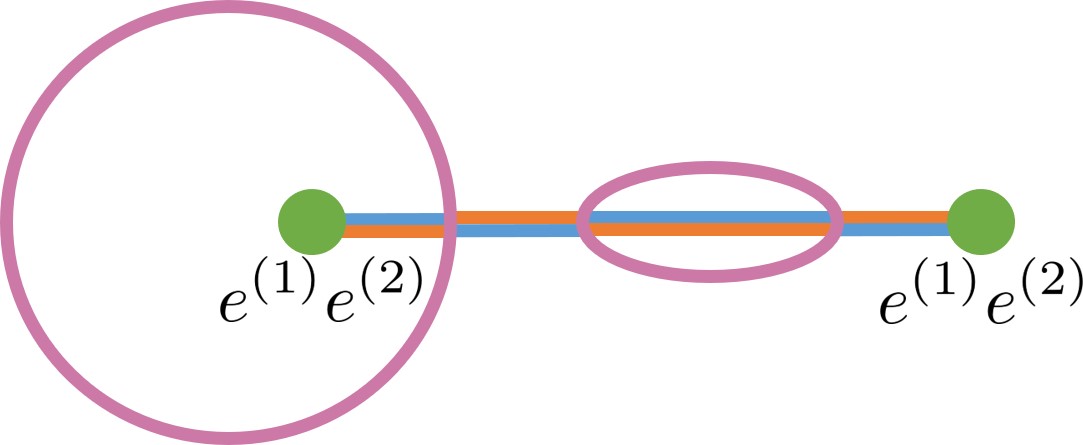}}\\
\subfloat[\label{fig:TCabStringNet}]{\includegraphics[width=1.5\columnwidth]{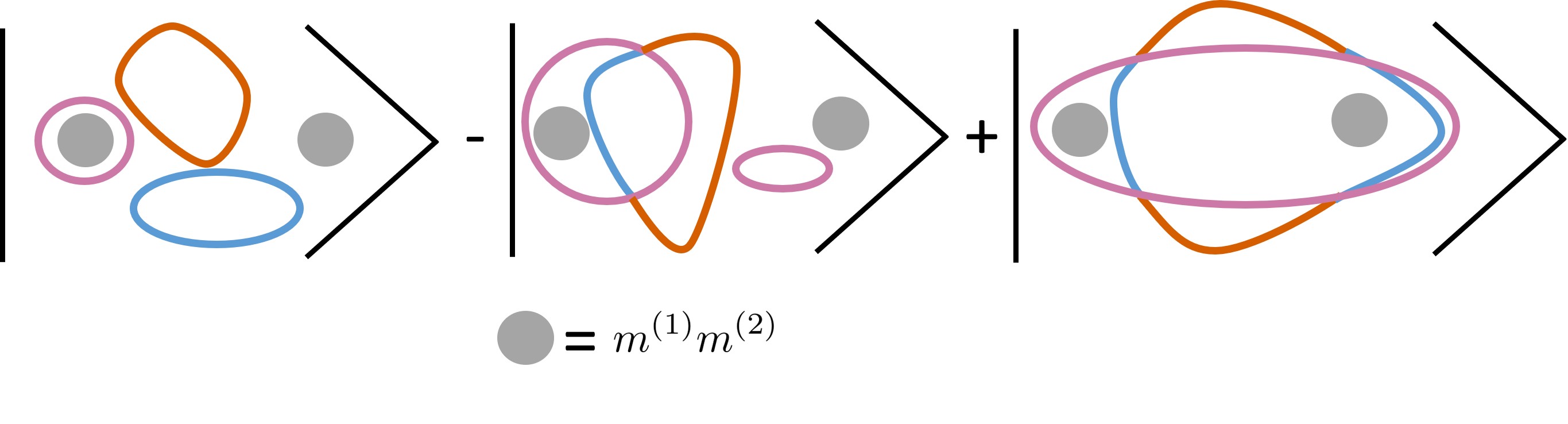}}\\
\subfloat[\label{fig:TCphiStringNet}]{\includegraphics[width=1.5\columnwidth]{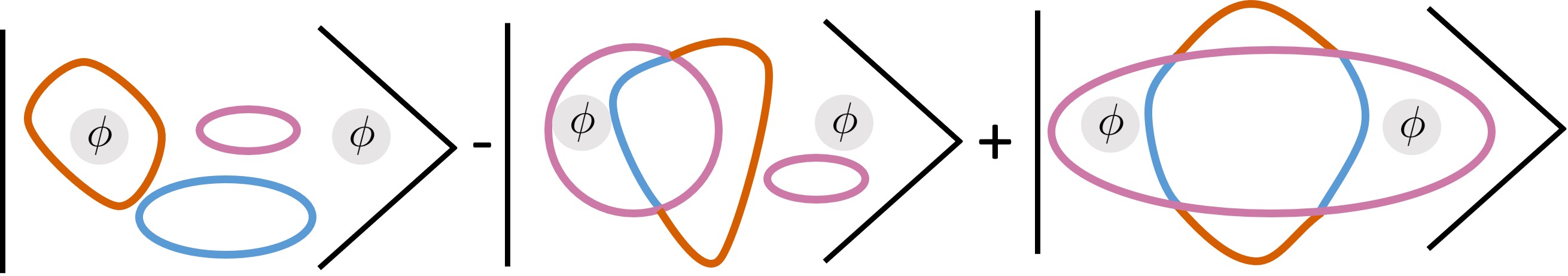}}
\caption{Abelian excitations in the string-net picture of the gauged bilayer toric code. Orange strings are $\sz^{(1)}=-1$, blue strings are $\sz^{(2)}=-1$, and purple strings are $\tau_z=-1$. (a) One term in the superposition for a pair of $e^{(1)}e^{(2)}$ excitations (green). A blue string and an orange string both end at the excitation. (b) Some terms in the superposition for a pair of $m^{(1)}m^{(2)}$ excitations (dark grey). (c) Some terms in the superposition for a pair of $\mathbb{Z}_2$ charges $\phi$ (light grey).}
\end{figure*}

All of the Abelian excitations have simple string operators, several of which arise from the original bilayer toric code.
In particular, string operators in the ungauged model which create SWAP-invariant excitations are unaffected by the
gauging procedure. For example, the Abelian charge $(a,c) \rightarrow (-1,1)$ should be identified with the bound state
$e^{(1)}e^{(2)}$ of $e$ excitations in both original toric code layers; it is straightforward to check that a string of
$\sz^{(1)}\sz^{(2)}$ operators creates this charge, just as it would in the ungauged model. In the string-net picture,
this excitation just looks like a bound state of an open orange string and an open blue string, as in Fig.~\ref{fig:TCe1e2StringNet}.

Likewise, the flux $m^{(1)}m^{(2)}$ is simply the bound state of $m$ excitations in both of the original toric code layers, and is
created by a string of $\sx^{(1)}\sx^{(2)}$ operators on the dual lattice. In our choice of string-net basis, whenever an odd number
of blue and orange strings surround an $m^{(1)}m^{(2)}$ excitation, the configuration enters the ground state superposition
with a minus sign, as shown in Fig.~\ref{fig:TCabStringNet}.

The other Abelian charges are also simple; a string of $\tau_z$ operators on the lattice of gauge qubits anticommutes with $C_e$
and commutes with everything else, so it should be identified as the $\mathbb{Z}_2$ gauge charge $(a,c)\rightarrow (1,-1)$,
which we denote as $\phi$. In the string-net picture, configurations where an odd number of purple strings surround the
$\phi$ excitation enter the ground state superposition with a minus sign, as in Fig.~\ref{fig:TCphiStringNet}. The other pure charge
is simply the fusion of $\phi$ and $e^{(1)}e^{(2)}$.

The operators which create non-Abelian anyons are more interesting and quite instructive. Intuitively, the non-Abelian charge
(which we shall dub $[e]$) can be thought of schematically as superposition of $e^{(1)}$ and $e^{(2)}$ from the ungauged model, since it arises from the
symmetry orbit of the ungauged $e$ particles. It should therefore be constructed using $\sz^{(1)}$ and $\sz^{(2)}$. In the
subsequent section, we explicitly construct string operators for a variety of non-Abelian excitations and verify their fusion rules.

\begin{figure*}
\centering
\subfloat[\label{fig:TCeStringNet}]{\includegraphics[width=1.1\columnwidth]{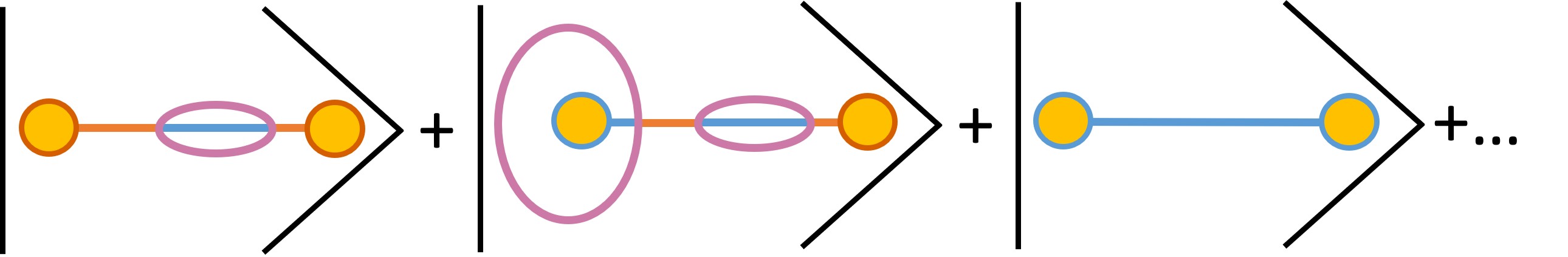}}\\
\subfloat[\label{fig:fourEBasis}]{\includegraphics[width=1.3\columnwidth]{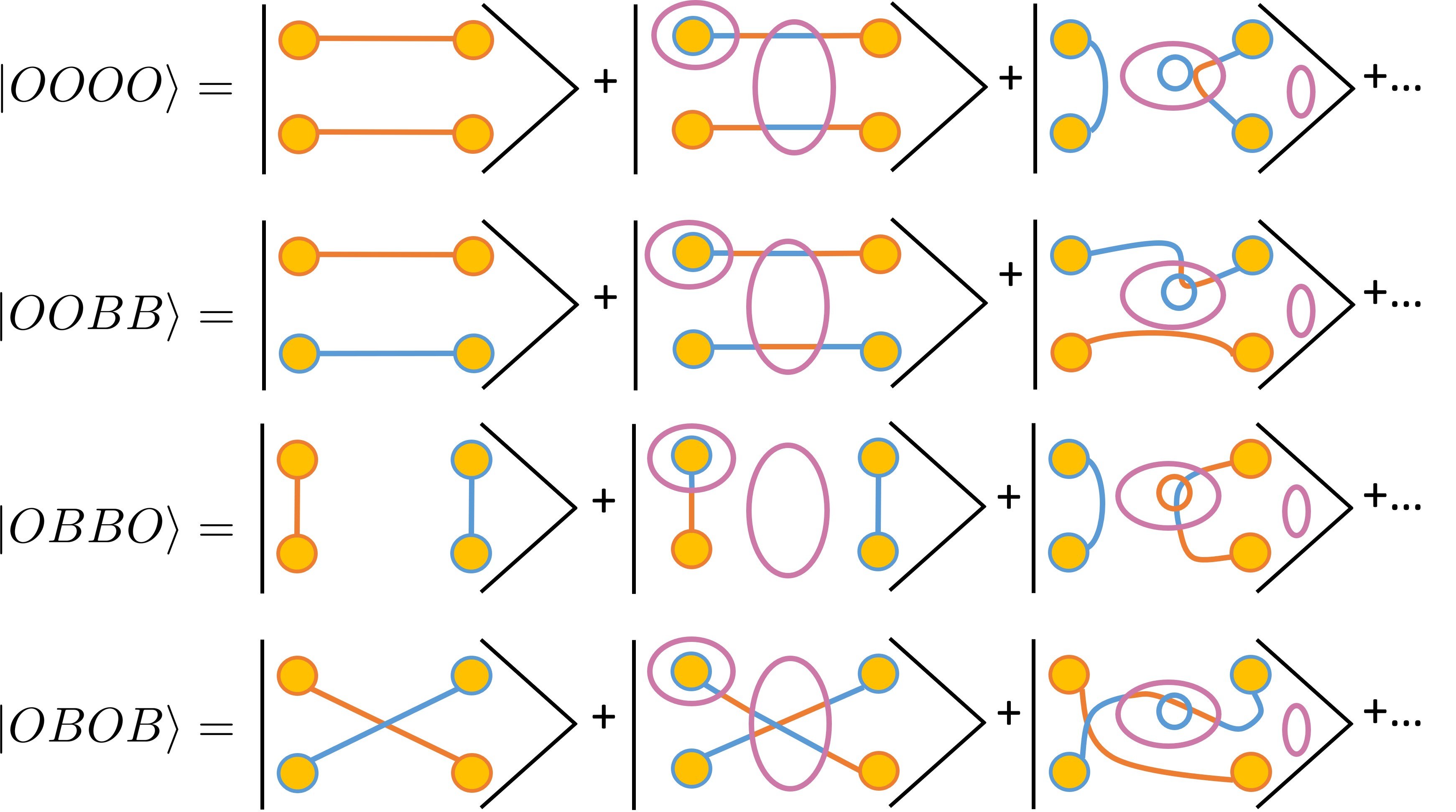}}
\caption{Example string-net configurations associated with $[e]$ excitations (yellow). The outline on $[e]$ indicates whether it is the end $e^{(1)}$ of a layer-1 string (orange) or the end $e^{(2)}$ of a layer-2 string (blue). The first term in each picture is the ``reference" configuration with no branch cuts (purple). (a) Two $[e]$ excitations, demonstrating how $[e]$ can be thought of as a superposition of $e^{(1)}$ and $e^{(2)}$. (b) A basis for the set of four degenerate states associated to a set of four $[e]$ excitations with overall fusion channel equal to the identity.}
\end{figure*}

In the string-net picture, a pair of $[e]$ excitations (created from vacuum) are shown in Fig.~\ref{fig:TCeStringNet}. The key point is that due to the presence of the proliferated branch cuts, the color of the string connecting the excitations is not definite, as whenever the string crosses a branch cut (purple string), it must change color. In this way, the excitation is, roughly speaking, a superposition of an $e^{(1)}$ (the end of an orange string) and an $e^{(2)}$ (the end of a blue string). 

We next provide a pictorial understanding from the string-net wave function of why the non-Abelian $[e]$ excitations have the four fusion channels
corresponding to the fusion rules in Eq.~\eqref{eFusion}. Let us consider four $[e]$ excitations, corresponding to
a pair of open $[e]$ strings, such that the overall fusion channel of all four is the identity. There is a remaining $4$-fold
topological degeneracy in this case: depending on the state, fusing the two open strings together leaves behind an open string associated with
$1, \phi, e^{(1)}e^{(2)}, $ or $\phi e^{(1)}e^{(2)}$. We can understand this four-fold degeneracy as follows. 

Since no open branch cuts (open purple strings) are present, any string-net configuration can be deformed to a ``reference" configuration where no branch cut are present at all. Such a reference configuration, which is unique up to a global SWAP, must have an even number of anyons associated with each layer to ensure that the configuration can be created from vacuum. It is immediately clear that there are four inequivalent (modulo a global SWAP) reference configurations consistent with these rules; they are shown as the first terms in Fig.~\ref{fig:fourEBasis}, along with several other terms in their string-net superpositions. There are therefore four states, each labeled by one of these inequivalent reference configurations, which can easily be checked to be orthogonal. This is exactly the desired topological degeneracy. To relate the degeneracy to fusion, consider the particular superpositions
\begin{align}
\ket{1}&=\ket{OOOO}+\ket{OOBB}\\
\ket{\phi} &= \ket{OOOO}-\ket{OOBB}\\
\ket{e^{(1)}e^{(2)}} &= \ket{OBBO} + \ket{OBOB}\\
\ket{\phi e^{(1)}e^{(2)}} &= \ket{OBBO} - \ket{OBOB}
\end{align}
where normalization is ignored and the states are labeled on the right-hand side as they are in Fig.~\ref{fig:fourEBasis}, that is, O and B refer to the color (orange or blue) of the excitations in the reference configuration in clockwise order starting from the top left. We claim that the labeling on the left-hand side corresponds to the fusion channel for the top two and bottom two particles, that is, 
\begin{equation}
\ket{a} = \raisebox{-.5\height}{\includegraphics[width=.3\columnwidth]{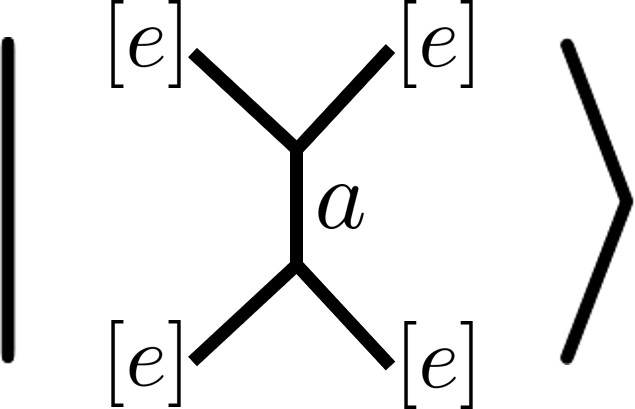}}
\label{eqn:fusionDiagram}
\end{equation}
where $a$ is an anyon label and the right-hand side is a fusion diagram.

To check that these fusion channels are correct, note that the two $[e]$ particles being fused have the same color in the reference configurations for $\ket{OOOO}$ and $\ket{OOBB}$. As we allow the purple strings to fluctuate, the layer labels of the $[e]$ excitations change, but the color correlation remains. That is, every time that both excitations are surrounded by the same parity of purple strings, then the $[e]$ particles carry the same color label, and every time they are each surrounded by a different parity of purple strings,
they carry different color labels. In particular, in \textit{every} configuration in the superposition, moving these two excitations together (possibly crossing branch cuts along the way) will always lead to an
$e$ meeting an $e$ from the \textit{same} layer, so they can only fuse to the identity or $\phi$. If instead we considered $\ket{OBBO}$ and $\ket{OBOB}$, bringing the endpoints together always causes the $[e]$s to carry \textit{opposite} layers, meaning that only $e^{(1)}e^{(2)}$ and $\phi e^{(1)}e^{(2)}$ are allowed fusion
channels.

\begin{figure*}
\includegraphics[width=1.3\columnwidth]{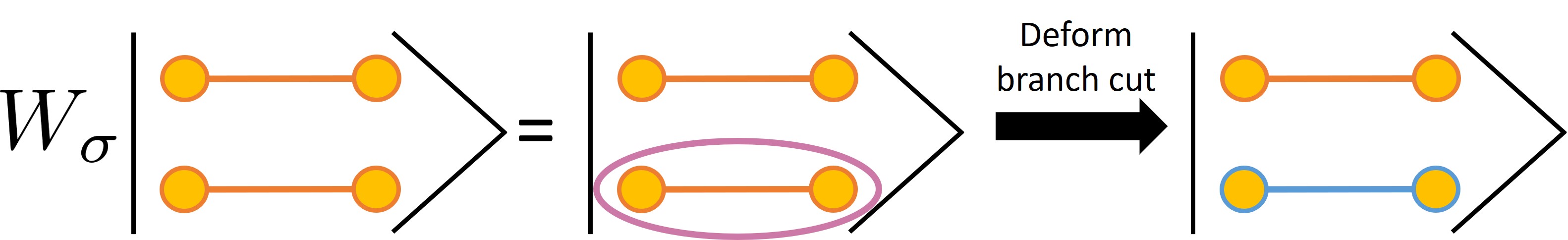}
\caption{Action of $W_{\sigma}$ on the reference configuration for $\ket{OOOO}$.}
\label{fig:WAction}
\end{figure*}

The presence or absence of a $\mathbb{Z}_2$ charge in a fusion product is measured with an operator $W_{\sigma}$ which creates a pair of twist defects $\sigma$, braids them around the two excitations being fused, and re-annihilates said defects, leaving behind a closed branch cut. This operator's support is far from the excitations, so it inserts this branch cut (modifying strings in the condensate as appropriate) \textit{without} changing the color labels of the excitations as shown in Fig.~\ref{fig:WAction}. Deforming the resulting configuration to the reference configuration, also shown in Fig.~\ref{fig:WAction} we find that $W_{\sigma}\ket{OOOO}=\ket{OOBB}$ when $W_{\sigma}$ encloses the bottom two anyons. A more delicate examination shows that indeed $W_{\sigma}\ket{OBBO}=\ket{OBOB}$ as well. Therefore, $W_{\sigma}$ has eigenvalue $+1$ in the states $\ket{1}$ and $\ket{e^{(1)}e^{(2)}}$ and eigenvalue $-1$ in the states $\ket{\phi}$ and $\ket{\phi e^{(1)}e^{(2)}}$, demonstrating that our claimed fusion outcomes are indeed correct.

\subsubsection{Explicit non-Abelian string operators}
\label{subsubsec:TCnonAbStrings}

\begin{figure*}
\subfloat[\label{fig:stringStarComm}]{\includegraphics[width=1.5\columnwidth]{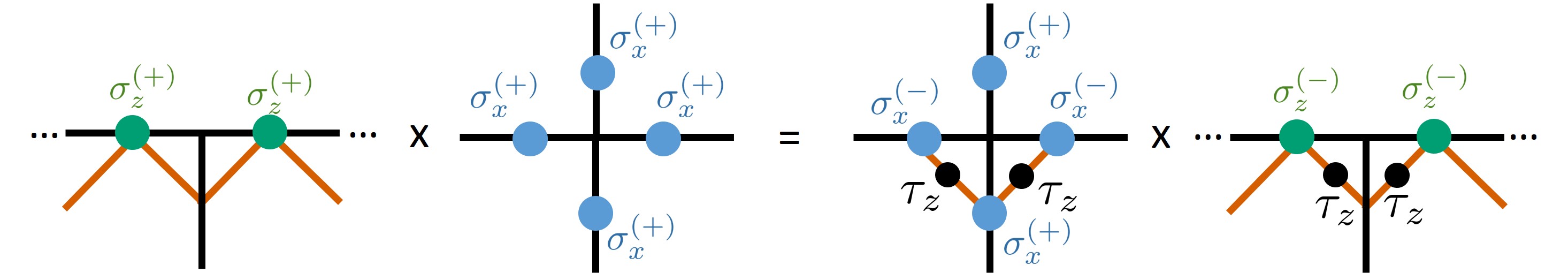}}\\
\subfloat[\label{fig:TCNonAbCharge}]{\includegraphics[width=1.5\columnwidth]{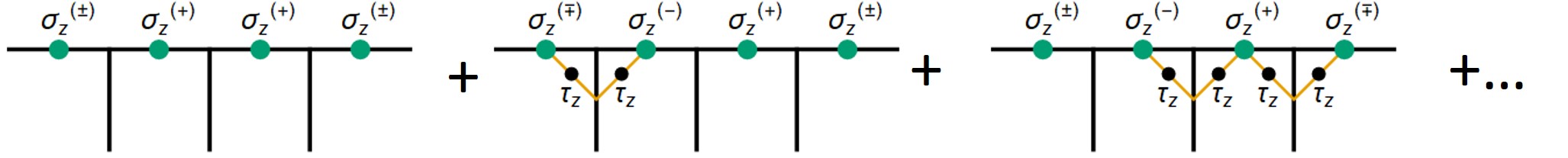}}\\
\caption{Constructing string operators for the non-Abelian $[e]$ particle in the gauged bilayer toric code. (a) Commutation relations between one term in the superposition of string operators for $[e]$ commuting with one star term in superposition making up $A_s$ in the Hamiltonian Eq.~\eqref{eqn:gaugedTCHamiltonian}. Moving the plaquette operator past the string operator turns the star term into a different term in the superposition forming $A_s$. The new string operator which results should be superposed with the original one. (b) Superposition making up the full string operator. There are four such operators, obtained by choosing either the upper or lower sign on each end of the string. These operators commute with all terms in the Hamiltonian except for $A_s$ (and possibly $C_e$) at their endpoints.}
\end{figure*}

Here we explicitly construct string operators for the non-Abelian excitations that correspond to pure charges and pure fluxes of the
$\mathbb{D}_4$ gauge theory. After constructing the operators, we then verify the resulting fusion rules of the non-Abelian quasiparticles. 

We begin by considering the string operator for the $[e]$ excitation. As remarked in the previous section, this should be constructed
using $\sz^{(1)}$ and $\sz^{(2)}$. To ensure gauge invariance, we should start with a string of $\sz^{(\pm)}$ operators, where disjoint pairs of $\sz^{(-)}$ operators are connected pairwise
by a string of $\tau_z$ operators. Such operators commute with $B_p$ and $D$ everywhere, but not with any $A_s$; this is a problem because the fact that the quasiparticles are deconfined requires that our string commute with all the $A_s$ away from the end of the string. To ensure that this occurs, we use equations like the one in Fig.~\ref{fig:stringStarComm}, which arise directly from the commutation relations Eq.~\eqref{eqn:sigmaAlgebra}. In particular, commuting a string of $\sz^{(\pm)}$ operators permutes the various summands in $A_s$ at the cost of interchanging $\sz^{(+)}\leftrightarrow\sz^{(-)}$ on the edges touching site $s$ (and dressing with $\tau_z$ operators). Superposing over all operators obtained by these interchanges, as in Fig.~\ref{fig:TCNonAbCharge}, will therefore produce a string operator which commutes with the Hamiltonian everywhere except at its endpoints; the resulting excitation is the $[e]$ particle.

If the local unitary $\sx^{(i)}$ acts on the end of the string operator, which interchanges $\sz^{(+)}$ and $\sz^{(-)}$ on that link only, we obtain a new string operator which no longer commutes with $C_e$ on the last link of the string, but still commutes with all other terms along the length of the string. This new operator therefore also creates $[e]$ at its endpoints. There are four such operators, specified by the four choices of sign at the endpoints of the operator in Fig.~\ref{fig:TCNonAbCharge}, which we call $\mathcal{S}_{[e]}^{\pm,\pm}$ in accordance with Fig.~\ref{fig:TCNonAbCharge}. The two $\pm$ signs refer to the left and right endpoints of the string respectively.

The string-net picture also gives us another way to understand the structure of the operator $\mathcal{S}^{++}_{[e]}$. Suppose that we consider configurations where the left end of the string in Fig.~\ref{fig:TCeStringNet} is orange, that is, we act with $\sigma_{z,1}^{(1)}$ where the lower index indicates that it is the first link in the string. Then to continue the string operator, we should act with $\sigma_{z,2}^{(1)}$ if an odd number of branch cuts pass between links 1 and 2, and act with $\sigma_{z,2}^{(2)}$ if an even number do. This is expressed by acting next with the operator
\begin{equation}
\sigma_{z,2}^{(1)}\left(1+\tau_{z,12}\right) + \sigma_{z,2}^{(2)}\left(1-\tau_{z,12}\right) \propto \sigma_{z,2}^{(+)} + \tau_{z,12}\sigma_{z,2}^{(-)}
\end{equation}
where $\tau_{z,12}$ is the product of $\tau_z$ on a string of gauge spins connecting links 1 and 2, because $\tau_{z,12}$ measures the parity of the number of branch cuts passing between links 1 and 2. Continuing, we obtain a string operator
\begin{equation}
S^{(1)} = \sigma_{z,1}^{(1)}\left(\sigma_{z,2}^{(+)} + \tau_{z,12}\sigma_{z,2}^{(-)}\right)\left(\sigma_{z,3}^{(+)} + \tau_{z,12}\tau_{13}\sigma_{z,3}^{(-)}\right)\cdots
\end{equation}
The same thing can be done starting from a blue link, i.e. beginning with $\sigma^{(2)}_{z,1}$, which, following the same logic, leads to a string operator
\begin{equation}
S^{(2)} = \sigma_{z,1}^{(2)}\left(\sigma_{z,2}^{(+)}- \tau_{z,12}\sigma_{z,2}^{(-)}\right)\left(\sigma_{z,3}^{(+)} - \tau_{z,12}\tau_{13}\sigma_{z,3}^{(-)}\right)\cdots
\end{equation}
Adding these together (to obtain a globally SWAP-invariant operator), one can check that
\begin{equation}
S^{(1)}+S^{(2)}=\mathcal{S}_{[e]}^{++}
\end{equation}

The same idea of choosing a particular gauge-invariant string operator and then interchanging local operators to ensure that the
string has no tension can be used to find (largely by inspection) the rest of the string operators in the theory. Below we formalize
this structure and give a list of the string operators for all the pure charges and pure fluxes in the theory.
(Dyon string operators can of course be constructed by taking products of the charge and flux operators.)

\begin{figure}
\centering
\subfloat[\label{fig:generalString_reference}]{\includegraphics[width=0.9\columnwidth]{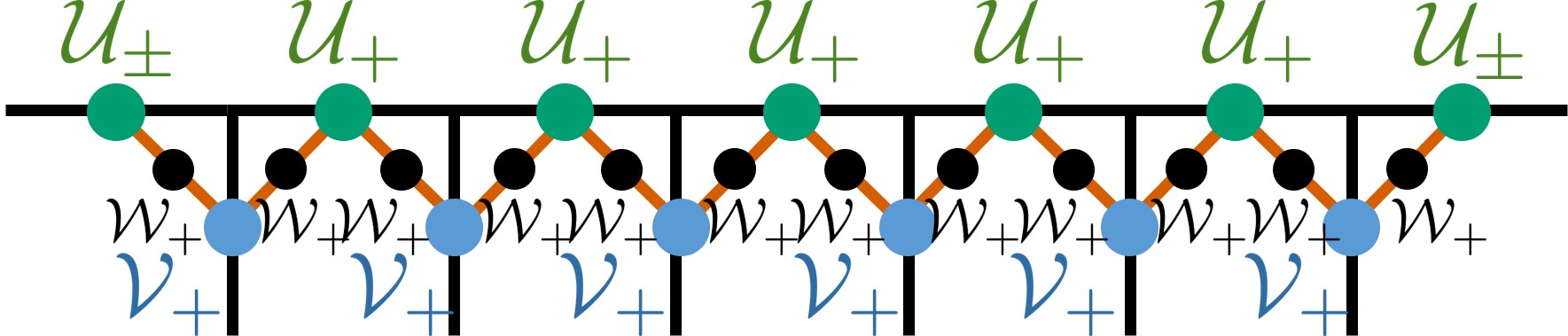}}\\
\subfloat[\label{fig:generalString_charge}]{\includegraphics[width=0.9\columnwidth]{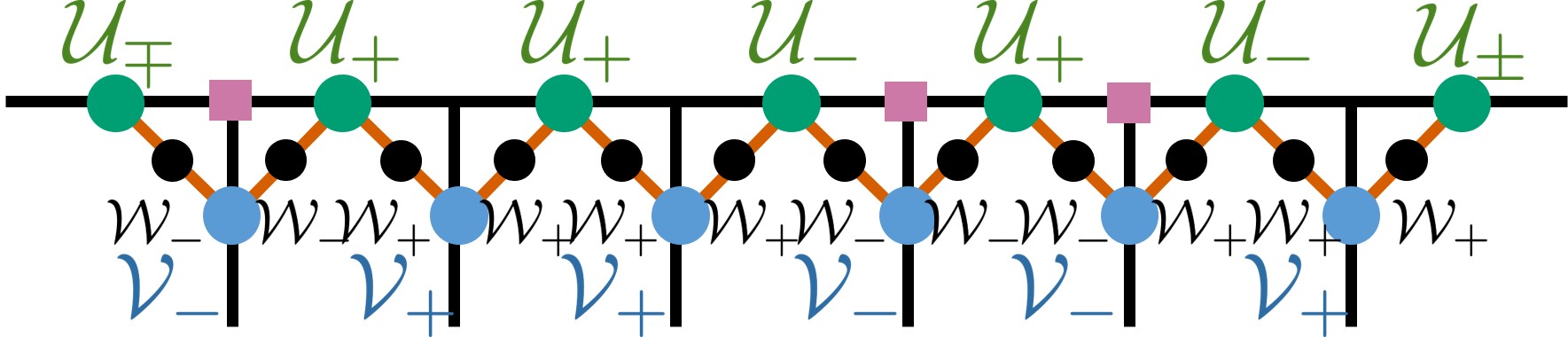}}\\
\subfloat[\label{fig:generalString_flux}]{\includegraphics[width=0.9\columnwidth]{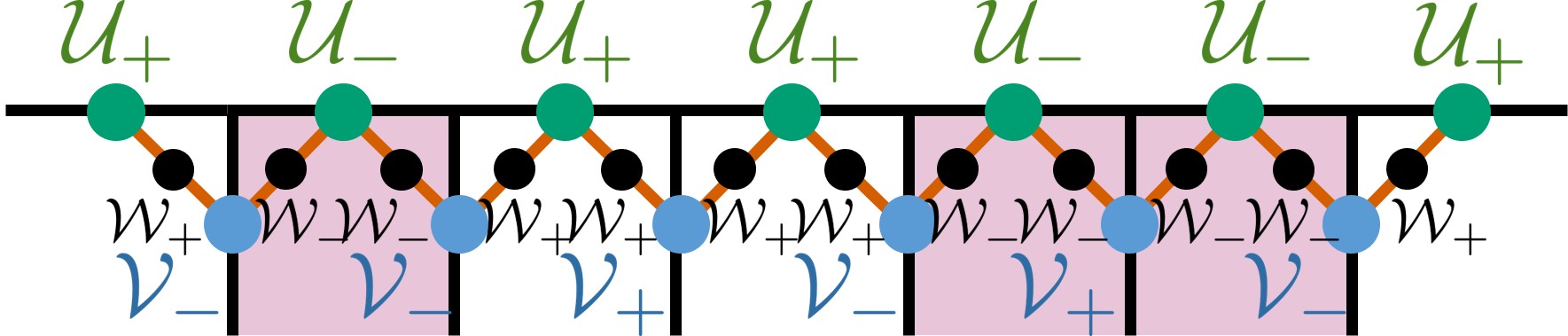}}
\caption{Terms in the expression of string operators in the gauged bilayer toric code as a superposition of products of local operators. (a) Reference operator for a pure charge; each end can have either $\mathcal{U}_+$ or $\mathcal{U}_-$. This freedom would move to the outermost $\mathcal{V}$ operators for a pure flux. (b) Another term in the superposition for a string operator that creates pure charges. At each site marked with a purple square, we have interchanged all surrounding subscripts $+\leftrightarrow -$. (c) Another term in the superposition for a string operator that creates pure fluxes. At each plaquette which is shaded purple, we have interchanged all surrounding subscripts $+ \leftrightarrow -$. This operator has $\mathcal{V}_+$ on both ends in its reference configuration.}
\label{fig:generalString}
\end{figure}

We specify the string operators as a superposition of products of local operators. For an Abelian anyon, there will only be one term in the superposition, whereas for a non-Abelian anyon, the superposition will contain $2^{\ell-1}$ terms, where $\ell$ is the length of the string. As we will explain shortly, the strings all have a common structure which is specified by six local operators. Their locations are shown in Fig.~\ref{fig:generalString}. Two, which we shall call $\mathcal{U}_+$ and $\mathcal{U}_-$, live on the square lattice links along the length of the string. Two, which we shall call $\mathcal{V}_+$ and $\mathcal{V}_-$, live on the square lattice links jutting out from the string. The last two, which we shall call $\mathcal{W}_-$ and $\mathcal{W}_-$, live on the $\tau$ links. The choice of whether $\mathcal{U}_+$, $\mathcal{V}_+$, and $\mathcal{W}_+$ or $\mathcal{U}_-$, $\mathcal{V}_-$, and $\mathcal{W}_-$ appears at various positions on the string will specify the different terms in the superposition.

A convenient way to explain the structure of the operators is as follows. To form our superposition of local operators, start from a ``reference" operator where the body of the string consists entirely of $\mathcal{U}_+$, $\mathcal{V}_+$, and $\mathcal{W}_+$ operators, as in Fig.~\ref{fig:generalString_reference}. For non-Abelian quasiparticles, the operators with the $+$ signs should be chosen to commute with all gauge generators $C_e$ and the operators with the $-$ signs are chosen to anticommute with any $C_e$ with overlapping support. This ensures that the reference operator is gauge-invariant.

Next, for a length-$\ell$ string, construct $2^{\ell-1}$ string operators as follows. Given the reference string for a pure charge (resp. flux) excitation, proceed down the string from site to site (resp. plaquette to plaquette). At each site (resp. plaquette), choose whether or not to interchange $\mathcal{U}_+ \leftrightarrow \mathcal{U}_-$,$\mathcal{V}_+ \leftrightarrow \mathcal{V}_-$, and $\mathcal{W}_+ \leftrightarrow \mathcal{W}_-$ on the links surrounding that site (resp. within that plaquette). After making such a choice on every site (resp. plaquette), we have obtained a new product of local operators; our desired string operator, which we denote $\mathcal{S}_a^{\pm,\pm}$ (where $a$ labels the anyon type and the $\pm$ signs indicate whether $\mathcal{U}_{\pm}$ is chosen on the left and right ends of the string, respectively) is simply the superposition over all such choices. An example term is shown in Fig.~\ref{fig:generalString_charge} for the pure charge excitations and Fig.~\ref{fig:generalString_flux} for the pure flux excitations. Obviously, if $\mathcal{U}_+=\mathcal{U}_-$,$\mathcal{V}_+=\mathcal{V}_-$ and $\mathcal{W}_+ = \mathcal{W}_-$, all of these terms are the same and there is only one term in the superposition; such a string operator will create Abelian anyons. 

Given this structure, the $\mathcal{U}$,$\mathcal{V}$, and $\mathcal{W}$ operators are chosen such that this superposition procedure causes $\mathcal{S}_a^{\pm,\pm}$ to commute with the Hamiltonian everywhere except at its ends. In the non-Abelian case, this is done with equations similar to the one in Fig.~\ref{fig:stringStarComm} that we used for the $[e]$ anyon.

The full set of operators, along with their identifications, are listed in Table~\ref{tab:D4Excitations}. Note in particular that we have an explicit form for the string operator for the twist defect $\sigma_1$.

The string operator algebra can also be used to obtain (partial) information about the fusion rules of the theory. Because the string operators, when acting on the ground state, create excitations from the vacuum, every string operator obviously creates excitations which fuse to the vacuum sector. Furthermore, given two string operators $\mathcal{S}_{a}$ and $\mathcal{S}_{b}$ supported on the same region which create the anyons $a$ and $b$ respectively at their ends, their product $\mathcal{S}_{a}\mathcal{S}_{b}$ must create fusion products $a \times b$ at each endpoint (we make no distinction between particles and antiparticles here because in $\mathbb{D}_4$ topological order, all particles are their own antiparticles). The string operator algebra should thus contain operators $\mathcal{S}_a$ such that
\begin{equation}
\mathcal{S}_{a}\times \mathcal{S}_{b} = \sum_{c}N_{ab}^c \mathcal{S}_{c}
\label{eqn:stringFusion}
\end{equation}
for any $a,b$, where $N_{ab}^c$ are fusion multiplicities. Note that if $\mathcal{O}$ is a local operator, and defining 
\begin{equation}
\tilde{\mathcal{S}}_a = \mathcal{O}\mathcal{S}_a\mathcal{O}^{\dagger}
\end{equation}
then Eq.~\eqref{eqn:stringFusion} still holds with $\mathcal{S}$ replaced by $\tilde{\mathcal{S}}$. However, a product like $\tilde{\mathcal{S}}_a\mathcal{S}_b$ need not obey the same equation, so only certain elements of the string operator algebra produce the full list of fusion outcomes. 

We defer an exhaustive enumeration of the string operator algebra to Appendix \ref{app:bilayerTCStrings} and presently consider some representative examples to demonstrate the fusion rules. Consider first fusing any of the Abelian charges with themselves. Their string operators square to the identity, which means that, as expected, these charges are Abelian and are their own antiparticles.

Next, consider fusing the $e^{(1)}e^{(2)}$ charge with the non-Abelian charge. Since $\sz^{(1)}\sz^{(2)}\sz^{(\pm)} = \pm \sz^{(\pm)}$, multiplying the string operator in Fig.~\ref{fig:TCNonAbCharge} by the string operator which creates the $e^{(1)}e^{(2)}$ excitations produces the same string operator for the non-Abelian charge up to a possible minus sign (since the number of $\sz^{(-)}$ operators has the same parity for every term in the sum). Therefore, these charges fuse to the non-Abelian charge, verifying the fusion rule
\begin{align}
[e] \times e^{(1)}e^{(2)} = [e]
\end{align}
This is also easy to see in the string-net wave function picture: Fusing an open string of both colors onto the non-Abelian string clearly just permutes the terms in the ground state superposition for the non-Abelian string (e.g. it switches the terms with a purely orange string and a purely blue string), thus giving us back the same picture. 

Finally, consider fusing the non-Abelian charge $[e]$ with itself. Using the operator definitions in Fig.~\ref{fig:TCNonAbCharge},
a straightforward but tedious calculation shows that, up to normalization,
\begin{align}
\left(\mathcal{S}_{[e]}^{++}+\mathcal{S}_{[e]}^{--}\right)^2 &= 1 + \bigotimes \tau_z + \bigotimes \sz^{(1)}\sz^{(2)} + \bigotimes \tau_z\sz^{(1)}\sz^{(2)}\\
&= 1 + \mathcal{S}_{\phi} + \mathcal{S}_{e^{(1)}e^{(2)}}+\mathcal{S}_{\phi e^{(1)}e^{(2)}}
\end{align}
where the tensor products are over the full length of the original string operator. The second line, where we identify these operators
with string operators for Abelian charges, follows by inspection. That is, all of the Abelian charges are possible fusion outcomes of
two non-Abelian charges. By counting the quantum dimension, one can check that this exhausts all possible fusion outcomes.
The operator algebra for the individual $\mathcal{S}_{[e]}^{\pm \pm}$ is listed in the Appendix. 

\section{Gauged Bilayer X-Cube}
\label{sec:X-Cube}

The X-cube model has been discussed elsewhere in considerable detail~\cite{CastelnovoFirstXCubePaper,VijayGaugedSubsystem}. We give a brief overview of its properties in Appendix~\ref{app:XCubeReview}.
In this section, we construct the gauged bilayer X-Cube model, provide a cage-net picture for the ground state wavefunctions, and
sketch the calculation of the ground state degeneracy on a 3-torus, relegating technical details of the calculation to Appendix~\ref{app:GSD}.

\subsection{Construction of the model}

\begin{figure}
\centering
\includegraphics[width=0.9\columnwidth]{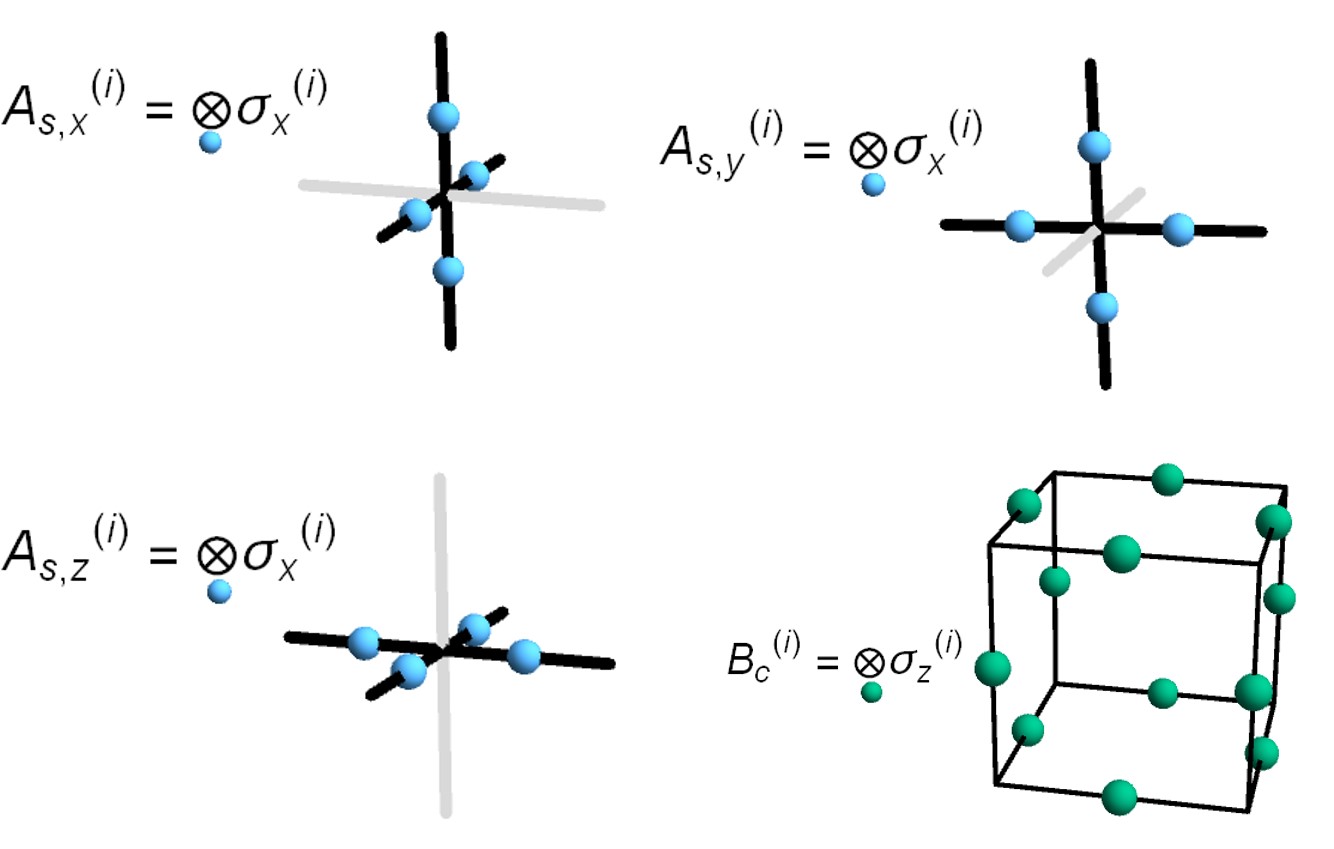}
\caption{Operators appearing in each layer of the bilayer X-Cube Hamiltonian.}
\label{fig:XCubeTerms}
\end{figure}

\begin{figure}
\centering
\subfloat[\label{fig:octahedronHilbertSpace}]{\includegraphics[width=0.4\columnwidth]{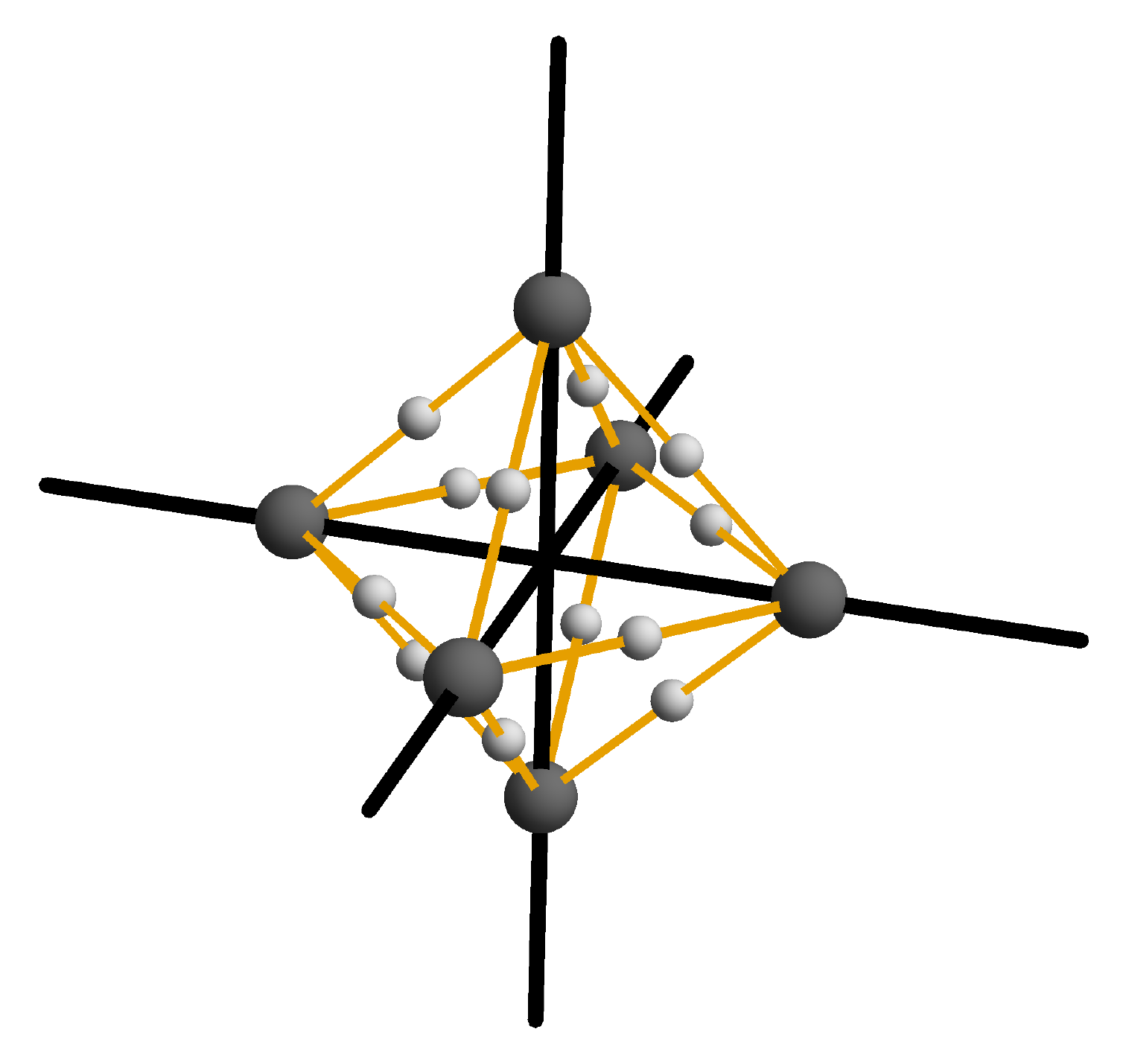}}\hspace{0.1\columnwidth}
\subfloat[\label{fig:XCubeGaugeGenerator}]{\includegraphics[width=0.45\columnwidth]{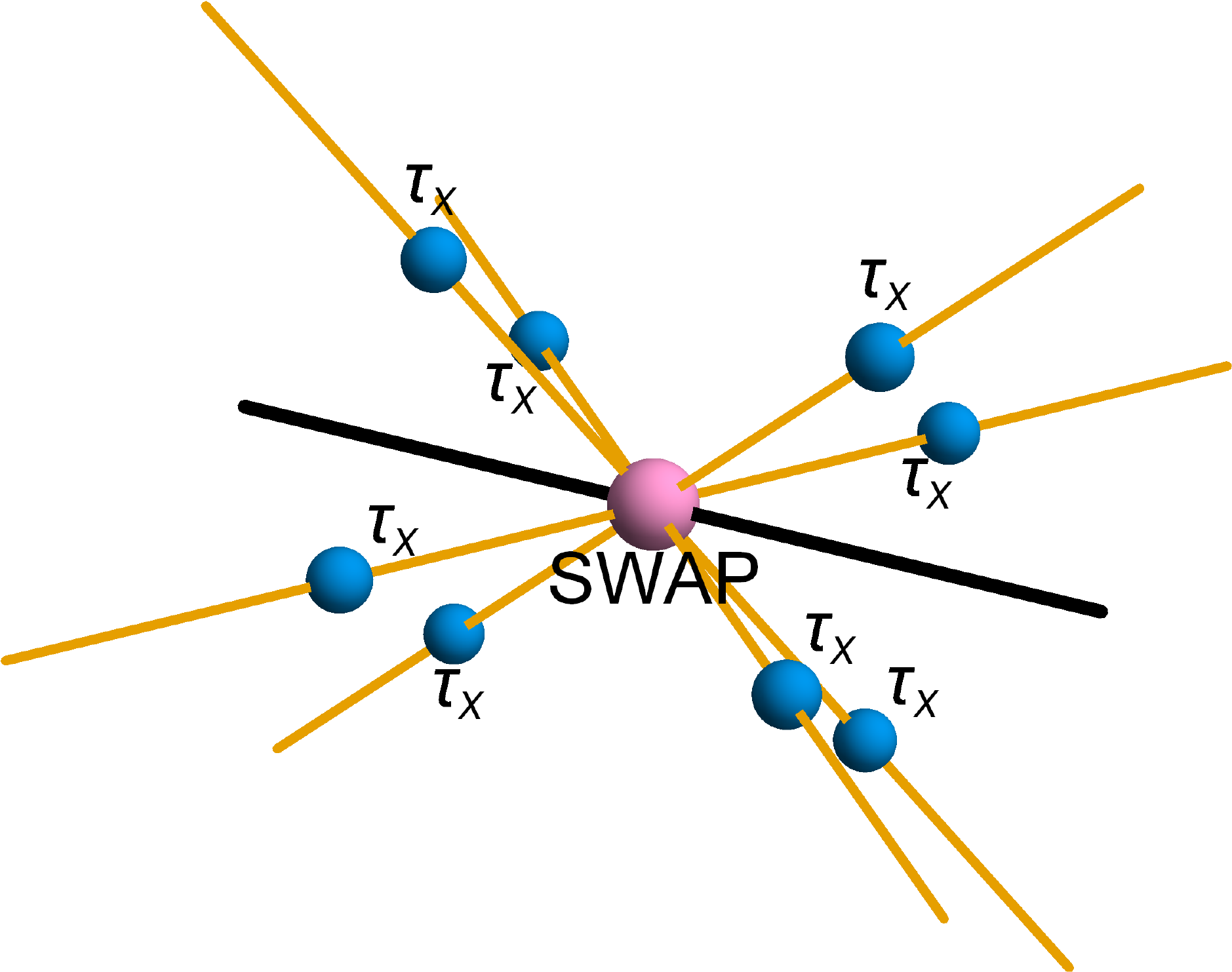}}\\
\subfloat[\label{fig:gaugedCubeTerms}]{\includegraphics[width=0.95\columnwidth]{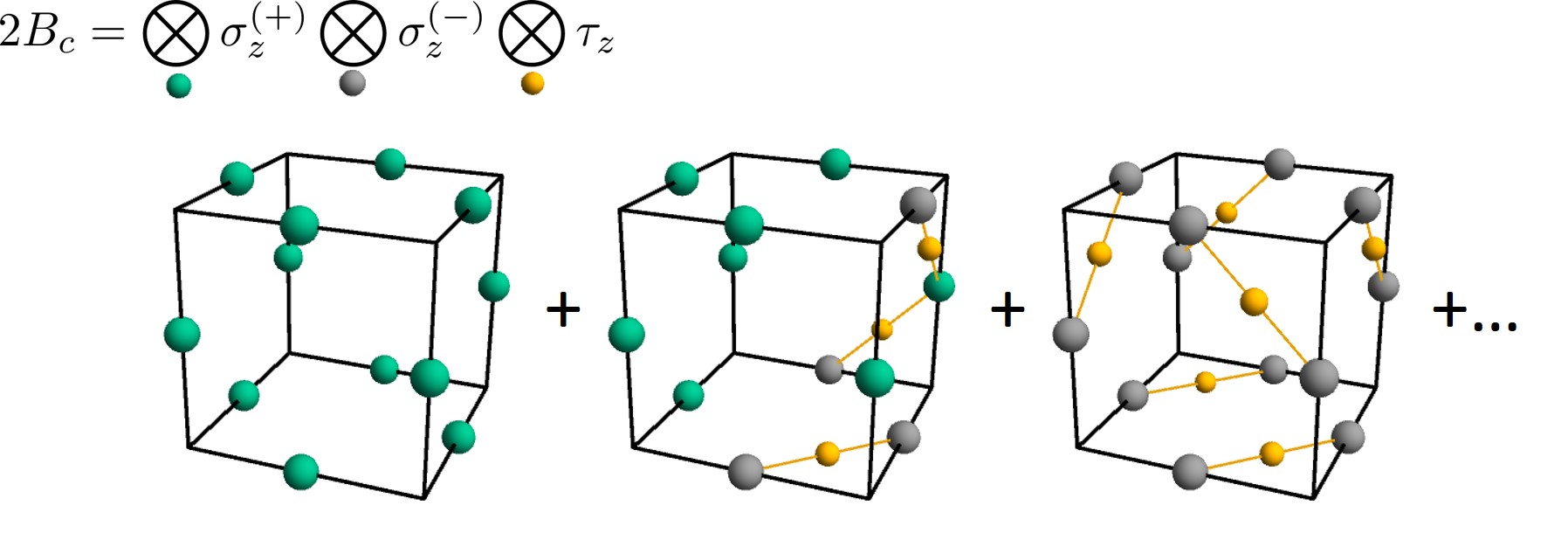}}\\
\subfloat[\label{fig:XCubeDOperators}]{\includegraphics[width=0.6\columnwidth]{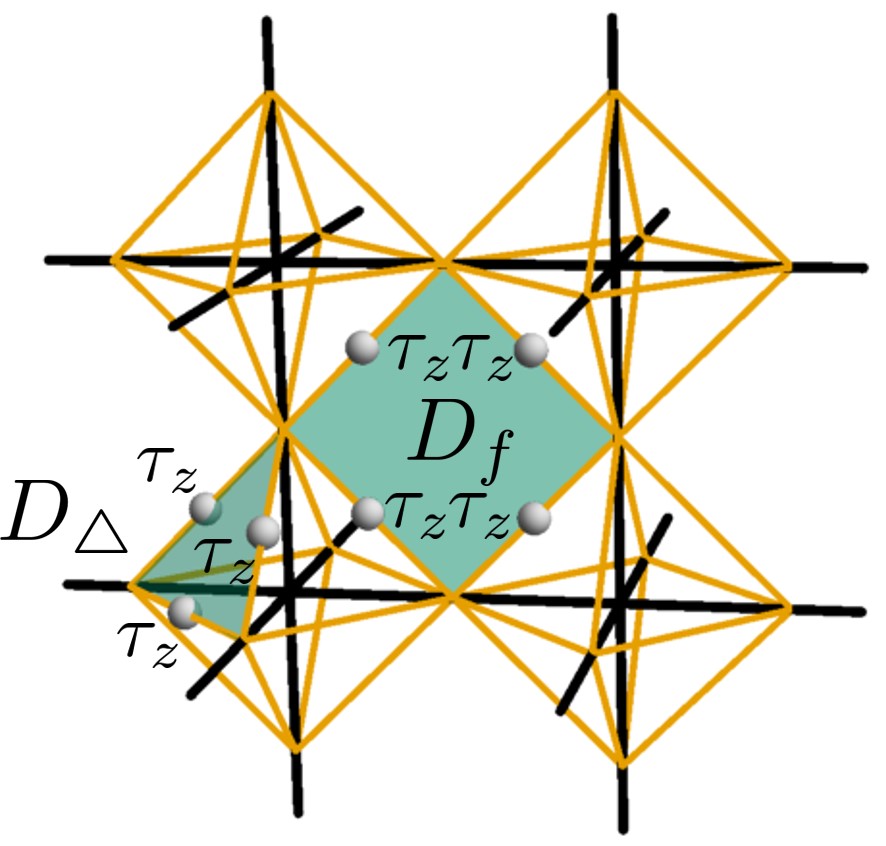}}
\caption{Gauging the bilayer X-Cube model. (a) Positions of spins for the Hilbert space of the gauged model. There are two ``matter" spins (light) per link of the lattice and one ``gauge" spin (dark) per orange line connecting nearest-neighbor matter spins, for a total of 18 spins per site. (b) Generator $C_e$ of local gauge transformations. The operator acts as SWAP on the matter spins $\sigma$ and acts with $\tau_x$ on the eight surrounding	gauge spins. (c) Gauged cube term $B_c$. The sphere color indicates whether $\sz^{(+)}$ (green) or $\sz^{(-)}$ (grey) acts on the matter spins, and $\tau_z$ acts on the orange gauge spins. Three of the 2048 possible terms (which all have an even number of $\sz^{(-)}$ operators) are shown. (d) Flux operators $D_{\triangle}$, which is a three-spin operator on the faces of the octahedra, and $D_f$, which is a four-spin operator within each face of the cubic lattice.}
\end{figure}

The procedure is analogous to the bilayer toric code case. The ungauged model consists of two layers of spins on the links of a cubic lattice with two copies of the X-Cube Hamiltonian:
\begin{align}
H_0 &= H_1 + H_2\\
H_i &= -\sum_{s}\sum_{p=x,y,z} A_{s,p}^{(i)} - \sum_c B_c^{(i)} \label{eqn:ungaugedXCube}
\end{align}
with $A_{s,p}^{(i)} = \otimes_{+_p} \sx^{(i)}$ and $B_c^{(i)} = \otimes_{\text{\mancube}}\text{ } \sz^{(i)}$ the usual star and cube operators shown in Fig.~\ref{fig:XCubeTerms}. Here $s$ labels a site, $p$ labels the orientation of the star operators, and $c$ labels elementary cubes. For each term in the ungauged model, we can rewrite the sum of the two layers' terms as a sum of products of $\SWAP$-even and $\SWAP$-odd operators; for example,
\begin{equation}
B_c^{(1)}+B_c^{(2)} = \frac{1}{2}\sum_{\lbrace s_{\bv{r}}| \prod s_{\bv{r}} = 1 \rbrace} \bigotimes_{\bv{r}\in \text{\mancube}} \sz^{(s_{\bv{r}})}
\end{equation}
where $s_{\bv{r}}=\pm$. As before, we have suppressed the explicit position index on the $\sz^{(s_{\bv{r}})}$ operators.
To gauge the symmetry, we add gauge qubits $\tau$ connecting nearest neighbor links. This corresponds to adding an octahedral ``cage" of gauge qubits surrounding each site, as in Fig.~\ref{fig:octahedronHilbertSpace}. Next, we define a $\mathbb{Z}_2$ gauge symmetry generator 
\begin{equation}
C_e = \SWAP_e \bigotimes_{\bv{r} \in \text{star}} \tau_{x,\bv{r}} 
\end{equation}
where the $\tau_x$  act on the eight gauge qubits which neighbor an edge $e$ of the original cubic lattice; this operator is shown in Fig.~\ref{fig:XCubeGaugeGenerator}. We also define $\mathbb{Z}_2$ gauge flux operators
\begin{align}
D_{\triangle} &= \bigotimes_{\bv{r} \in \triangle}\tau_{z,\bv{r}}\\
D_{f} &= \bigotimes_{\bv{r} \in \diamond}\tau_{z,\bv{r}}
\end{align}
Here $D_{\triangle}$ acts on each face of the octahedron of gauge spins surrounding each site and $D_p$ acts on the four $\tau$ spins within a face of the cubic lattice, as shown in Fig.~\ref{fig:XCubeDOperators}.

We now modify the Hamiltonian so that the model is invariant under the gauge symmetry generated by $C_e$. This invariance is obtained by putting $\tau_z$ operators into the Hamiltonian terms such that every $\sigma_a^{(-)}$ term has an odd number of $\tau_z$ operators acting on the qubits which surround it. This can always be achieved by choosing disjoint pairs of $\sigma^{(-)}$ operators and connecting them with a path of $\tau_z$ operators. Some examples of the 2048 terms are shown in Fig.~\ref{fig:gaugedCubeTerms}.

The gauged Hamiltonian is
\begin{equation}
H_{\text{gauged}} = -\sum_s \sum_{n=x,y,z} A_{s,n} - \sum_c B_c - \sum_e C_e - \sum_{\triangle} D_{\triangle} - \sum_f D_f
\label{eqn:gaugedXCubeH}
\end{equation}
where $A_{s,n}$ is the sum of the eight gauged star terms for each orientation of the star and $B_c$ is the sum of the 2048 gauged cube terms.

All of the algebraic computations for the gauged bilayer toric code model carry through for the gauged bilayer X-Cube model. That is,
$C_e$, $D_{\triangle}$, and $D_f$ square to 1, while the $A$ and $B$ operators square to $2(1+\otimes \sigma^{(1)}\sigma^{(2)})$.
Likewise, $A_{s,n}$, $B_c$, $C_e$, $D_{\triangle}$, and $D_f$ all mutually commute.

\subsection{Wavefunction picture of ground states}

The basic building blocks of our wavefunction picture for the ground states is the cage-net picture for a single layer of the X-cube model\cite{PremCageNet}, which we now briefly review. 

\begin{figure}
\centering
\subfloat[\label{fig:ungaugedXCubeCages}]{\includegraphics[width=0.4\columnwidth]{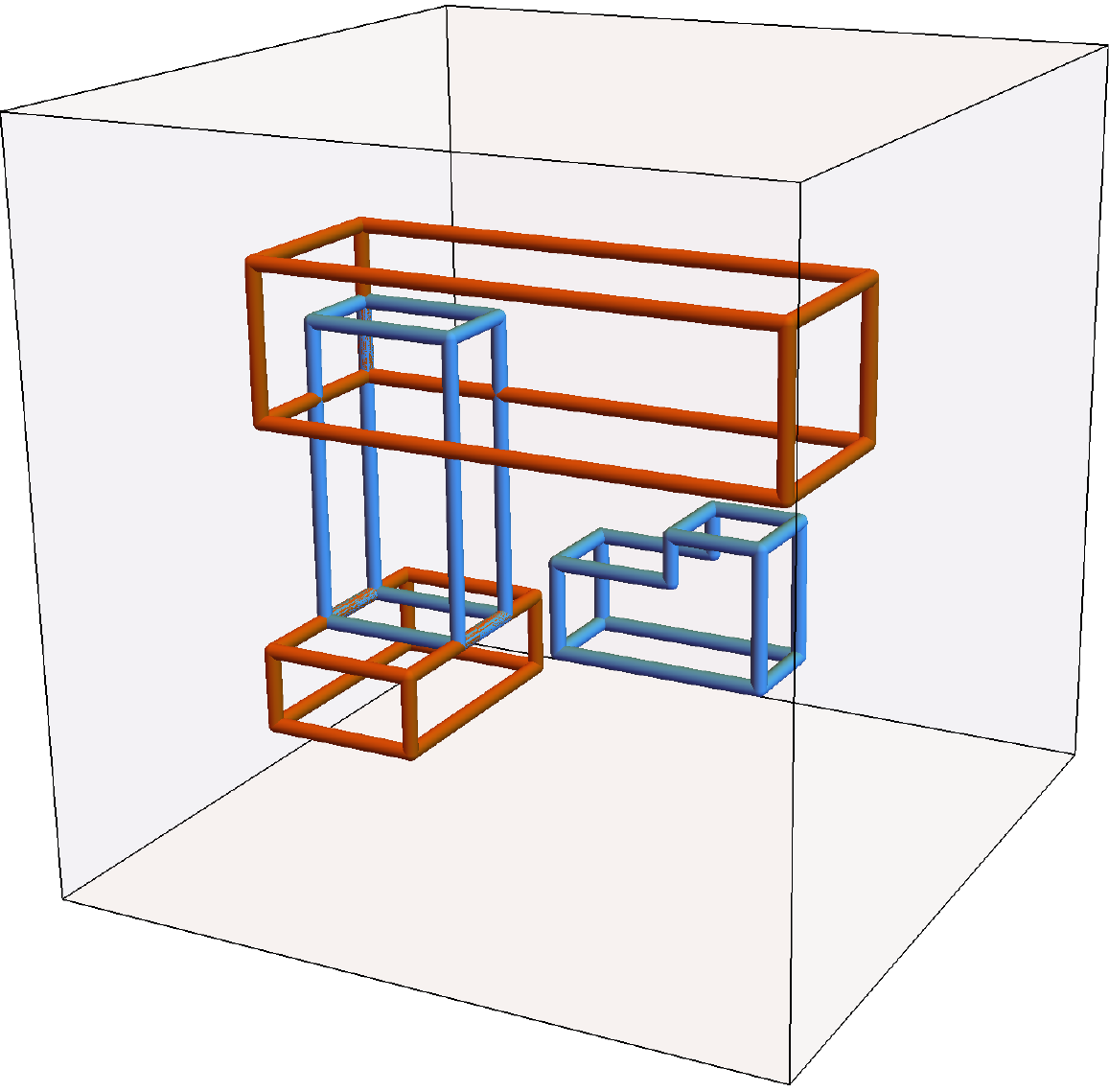}}\hspace{0.05\columnwidth}
\subfloat[\label{fig:cagesTwistDefect}]{\includegraphics[width=0.5\columnwidth]{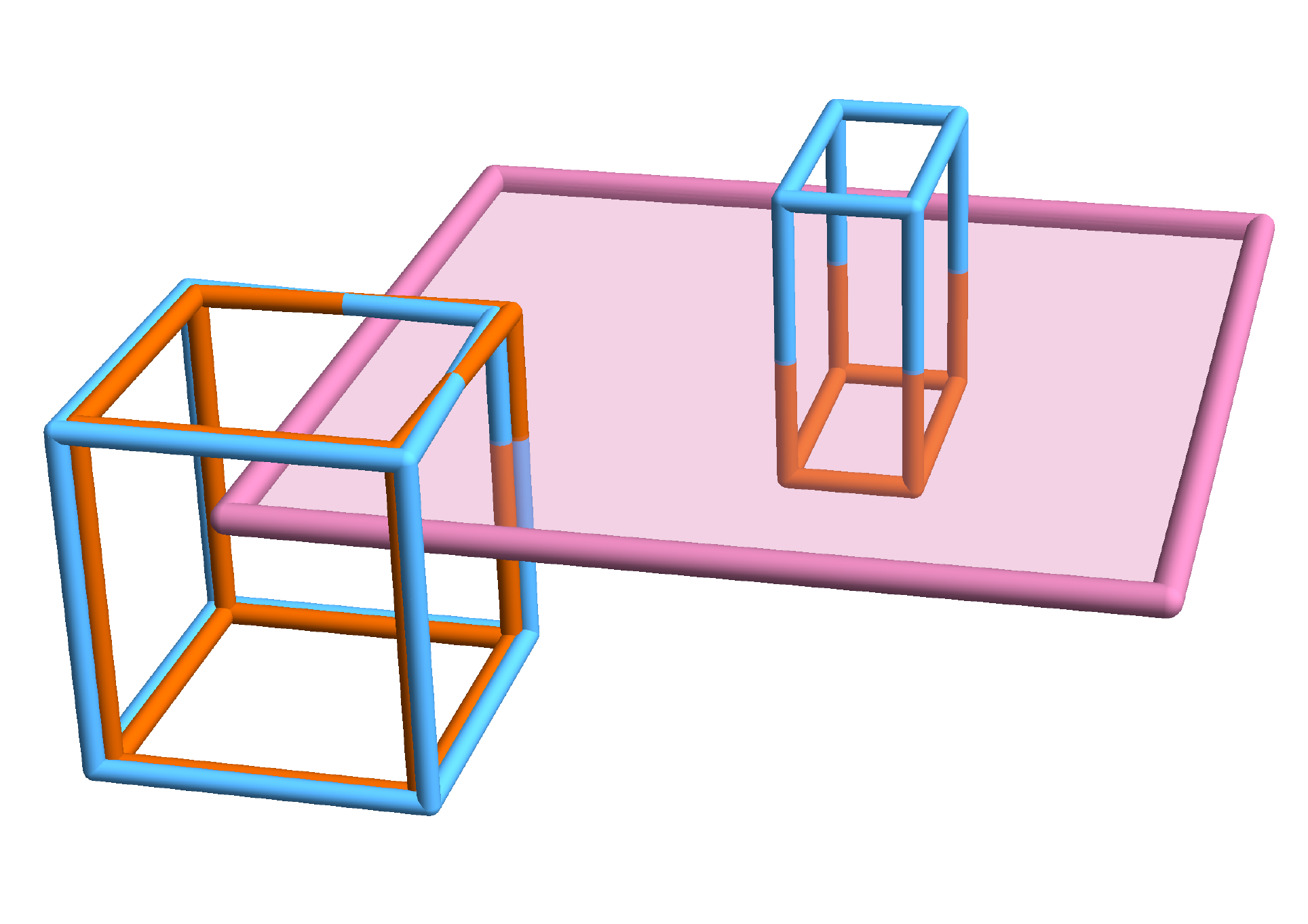}}\\
\subfloat[\label{fig:gaugedXCubeCages}]{\includegraphics[width=0.5\columnwidth]{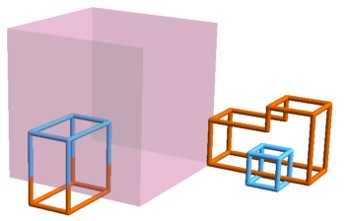}}
\caption{(a) Example ground state cage-net wavefunction configuration in the ungauged bilayer X-Cube model. Orange and blue are strings in different layers. (b) Cage-net configurations in the presence of a twist defect (purple string, branch cut membrane in light purple) Overlapping strings are displaced slightly for visibility. (c) Example ground state cage-net configuration in the gauged bilayer X-Cube model. The purple membrane is one of the proliferated closed branch cut membranes.}
\end{figure}

Configurations where $\sx = +1$ on a link are drawn with no string, whereas $\sx = -1$ is represented by the presence of a string on a link. In the
ungauged model, strings are colored to represent their layers (orange for layer 1, blue for layer 2) and both colors can live on the same link. A configuration
with no strings is an eigenstate of all the $A^{(i)}_{s,p}$ operators with eigenvalue $+1$, but not of the $B^{(i)}_c$ operators. Requiring that $B^{(i)}_c=+1$
amounts to superposing over all configurations of closed ``cages" of strings of each color, where an elementary (i.e. minimal size) ``cage" is the
wireframe outlining a unit cell cube. An example configuration in a ground state superposition for the ungauged bilayer model is shown in Fig.~\ref{fig:ungaugedXCubeCages}.

Distinct ground states on the 3-torus are labeled by the presence or absence of strings which wrap all the way around each handle of the torus.
Not all such strings are independent; we will discuss the constraints between them in Sec. \ref{subsec:GSDOutline}.

The system now supports extrinsic co-dimension 1 membrane defects, whose boundary is a twist defect string. The
membrane acts as a branch sheet, such that excitations that cross it go from one layer to the other. That is, any leg of a cage which passes through a branch membrane associated with the twist defect must change colors. Hence
any cage which links nontrivially with the twist defect must wrap the defect twice, as shown in Fig.~\ref{fig:cagesTwistDefect}.

Upon gauging, the ground state now also consists of superpositions of all possible closed branch membranes. We therefore obtain
two colors of cage-net condensates and a conventional condensate of closed membranes, generally shown in purple, which correspond
to the $\tau$ spins. The wavefunction for the membrane sector is the same as in conventional (3+1)D $\mathbb{Z}_2$ topological order.
However, the cage-nets are subjected to the rule that if a leg of a cage crosses a membrane, it changes colors. An example configuration
in the ground state superposition is shown in Fig.~\ref{fig:gaugedXCubeCages}.

As usual, a closed branch membrane can be nucleated from the vacuum, wrapped all the way around the system, and annihilated,
which implements a global SWAP. Hence all ground states should be invariant under a global SWAP.

\subsection{Ground state degeneracy on 3-torus}
\label{subsec:GSDOutline}

The detailed calculation of the GSD on an $L_x \times L_y \times L_z$ 3-torus is quite involved, requiring considerable constraint counting. The full details of this calculation are relegated
to Appendix \ref{app:GSD}; here we present an outline of the calculation and its result.

\begin{figure}
\centering
\subfloat[\label{fig:prismConstraint}]{\includegraphics[width=0.7\columnwidth]{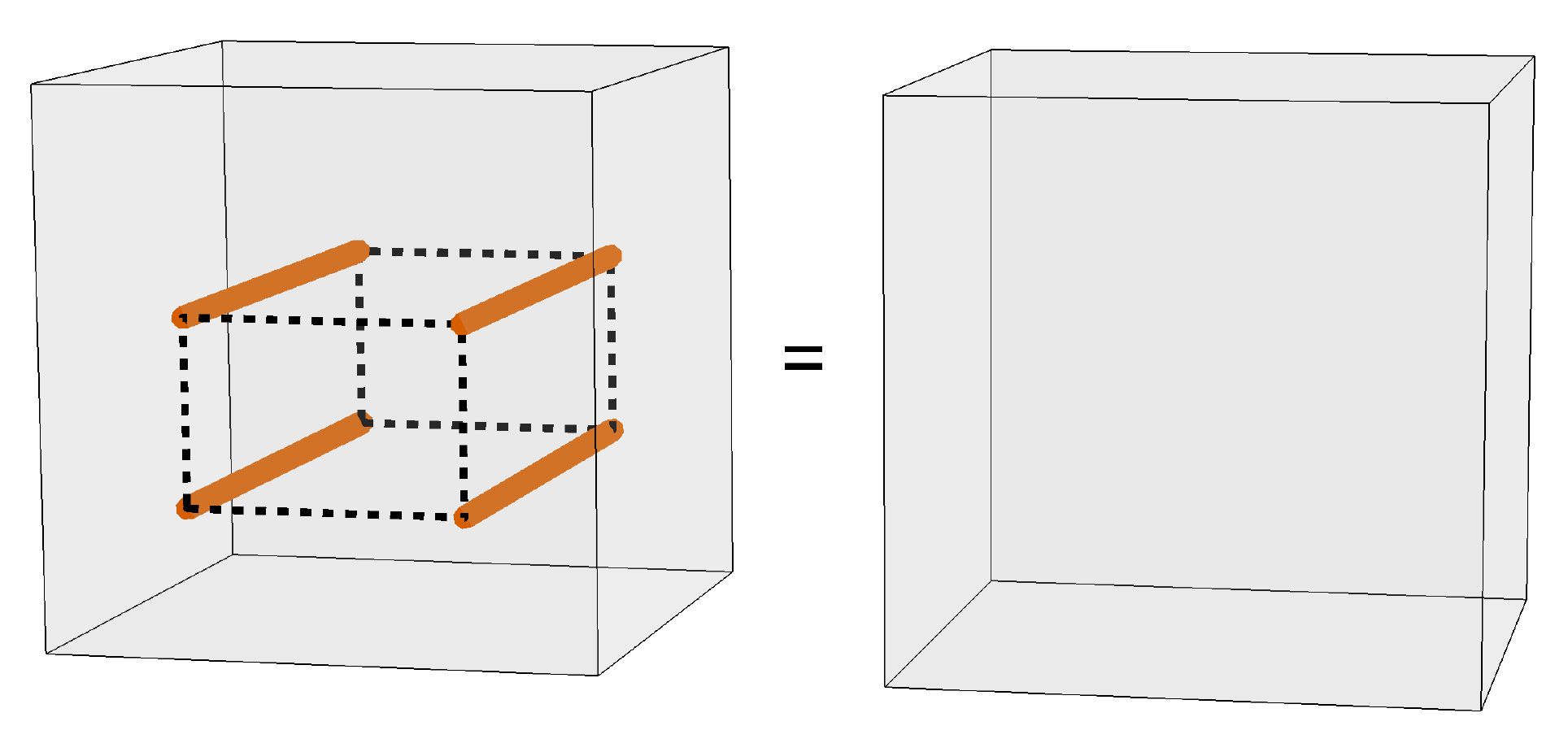}}\\
\subfloat[\label{fig:stringBasis}]{\includegraphics[width=0.5\columnwidth]{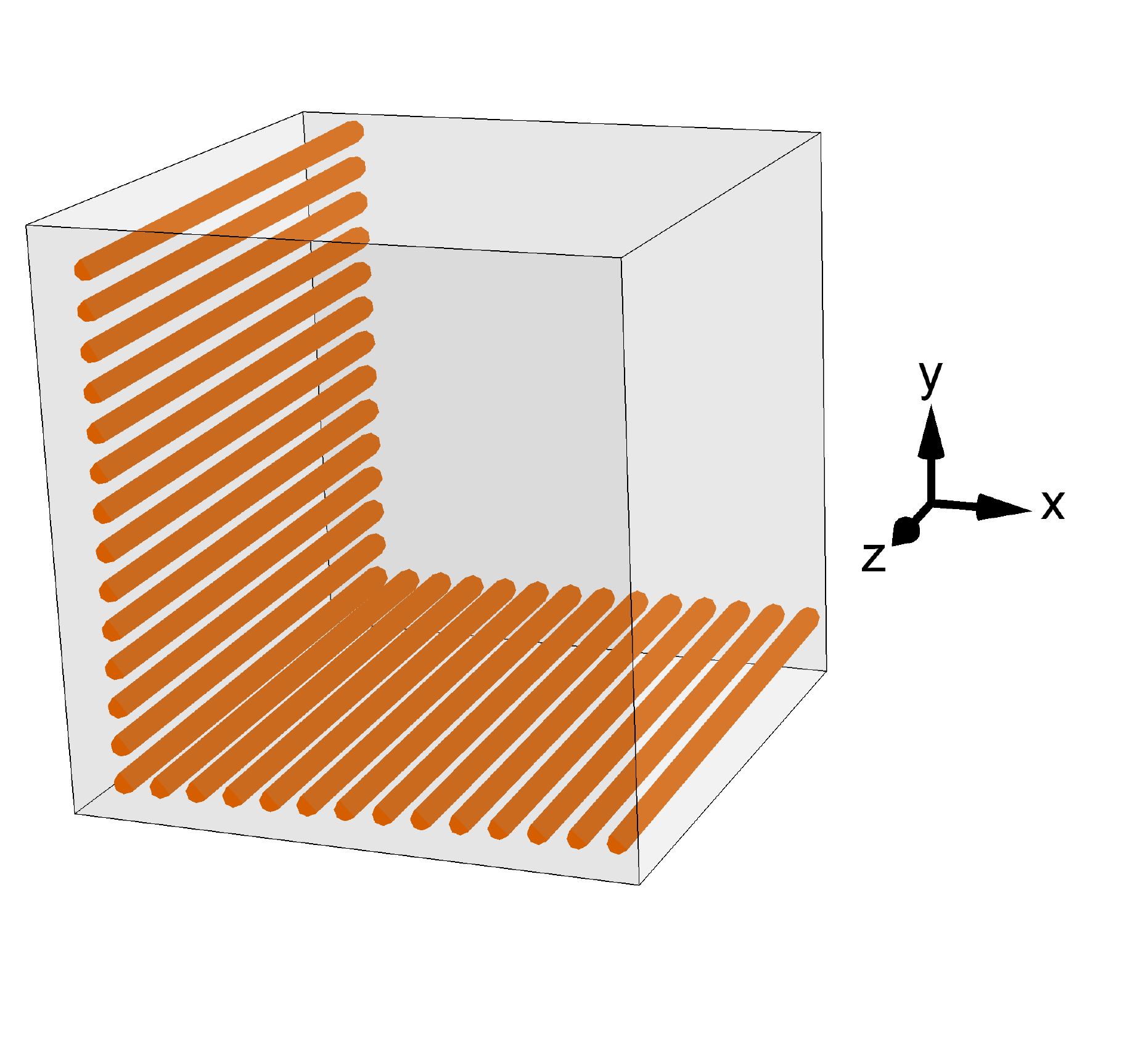}}
\caption{Properties of strings in the X-Cube model on the 3-torus. Opposite faces of the cube are identified in all subfigures. (a) Constraint on rigid, topologically nontrivial Wilson loops in the cage-net wavefunction picture of a single copy of the X-Cube model; four Wilson loops (orange) on the edges of a rectangular prism can disappear into the vacuum. The black dashed lines are guides to the eye. (b) Basis for topologically nontrivial Wilson loops oriented in the $z$ direction in a single copy of the X-Cube model. Each orange line is an independent Wilson loop.}
\end{figure}

In a copy of the ungauged X-Cube model, ground states are labeled by the eigenvalues of Wilson loop operators which physically correspond to
creating a particle-anti-particle pair for a particle constrained to move in one dimension, winding one around the torus, and then annihilating
the particles. These Wilson loops are products of $\sz$ operators. Since the quasiparticles involved are one-dimensional, these Wilson loops
are rigid, naively leading to $\sum_{i<j}L_iL_j$ distinct string operators with $i,j=x,y,z$. However, there is a constraint; the product of any four Wilson loop
operators which live on the edges of a rectangular prism, as shown in Fig.~\ref{fig:prismConstraint}, is equal to the identity in the ground
state (in the wavefunction picture, any four such strings can disappear). This follows from the energetic constraint that the product of
$\sz$ over the edges of a cube is equal to $+1$. Therefore, in the cage-net wavefunction picture, we may freely create or delete such a set of strings.

It is straightforward to show that a basis for the independent $z$-oriented Wilson loop operators is all such strings which are at $x=x_0$
and all such strings at $y=y_0$ for a fixed choice of $x_0,y_0$, as shown in Fig.~\ref{fig:stringBasis}. This leads to $L_x + L_y - 1$ independent
$z$-oriented Wilson loop operators. Including also the independent $x$ and $y$ oriented Wilson loop operators then leads to a total of
\begin{equation}
N=2L_x+2L_y+2L_z-3
\label{eqn:XCubeN}
\end{equation}
 Wilson loops in each copy of the X-Cube model. In the wavefunction picture, we represent a $+1$ eigenvalue
as the absence of a Wilson loop and a $-1$ eigenvalue as the presence of a Wilson loop. Two copies of the X-cube model thus have
$4^N$ ground states in total. 

For brevity, we call a particular string position and orientation its ``type" and refer to the X-Cube layer separately. 

Upon gauging the bilayer model, we first restrict to configurations of the ungauged model where all layer-twist membranes
are topologically trivial. The different states may then be labeled by ``reference configurations" where no topologically trivial
membranes or strings are present, as the other states in the superposition are obtained by allowing local fluctuations
of membranes and cages. However, we must project to ground states of the ungauged model which are invariant under a global SWAP operation. 

For each independent string type, if a configuration contains either no strings or both layers' strings, then the configuration is obviously symmetric under SWAP; there are $2^{N}$ such configurations. Given any of the other $4^N-2^N$ string configurations $\ket{\psi}$, a global SWAP acts nontrivially, so only the symmetric combination of $\ket{\psi}$ and $\SWAP\ket{\psi}$ invariant. 
Hence the total number of states in the untwisted sector is
\begin{equation}
\text{GSD}_{\text{untwisted}}=2^N + \frac{1}{2}\left(4^N-2^N\right) = 2^{2N-1}+2^{N-1}
\label{eqn:untwistedGSD}
\end{equation}

\begin{figure}
\centering
\subfloat[\label{fig:3TorusDoubleWrap}]{\includegraphics[width=0.5\columnwidth]{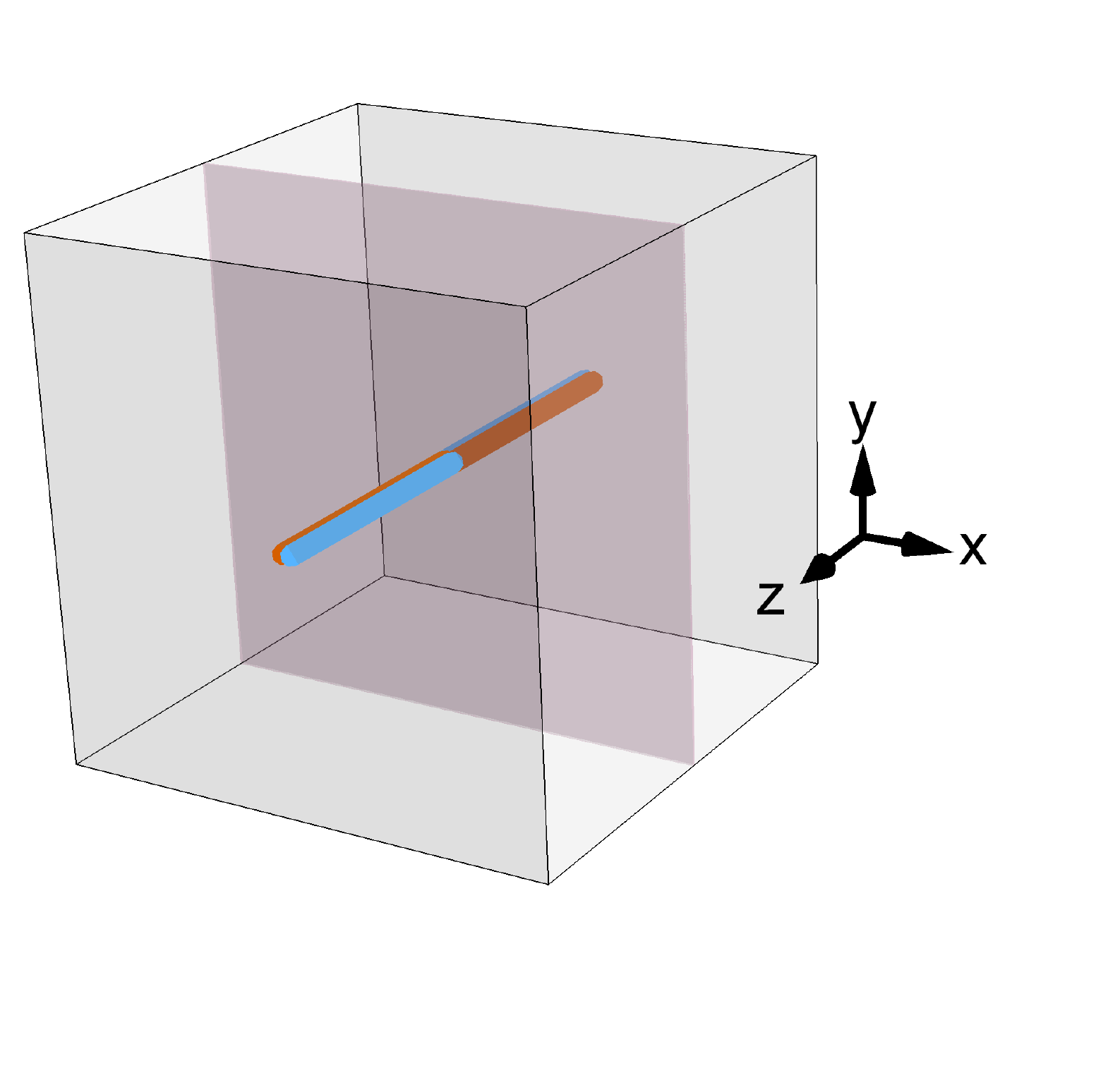}}\\
\subfloat[\label{fig:oneTwistPrismConstraint}]{\includegraphics[width=0.9\columnwidth]{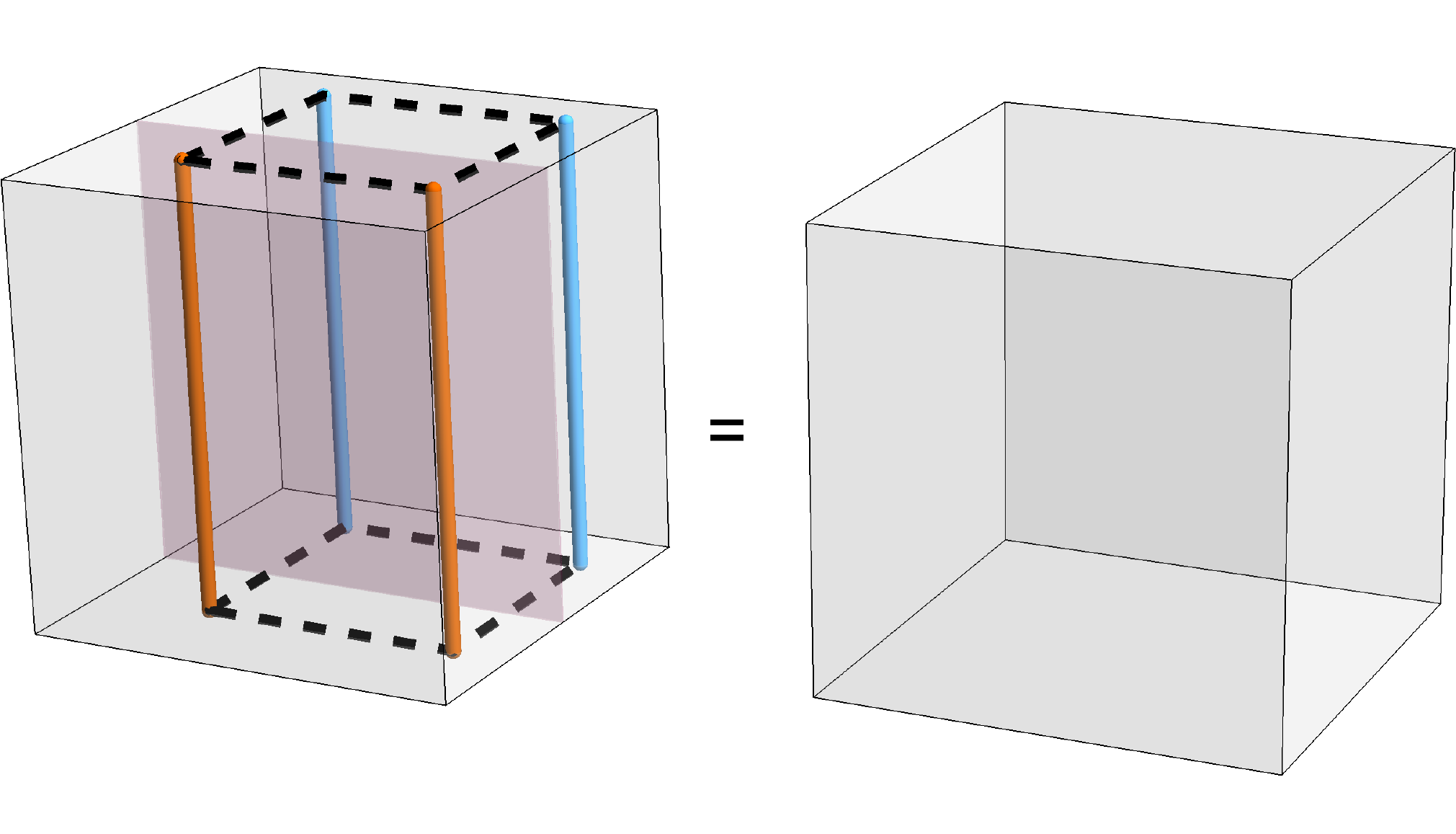}}\\
\subfloat[\label{fig:pairTranslation}]{\includegraphics[width=0.5\columnwidth]{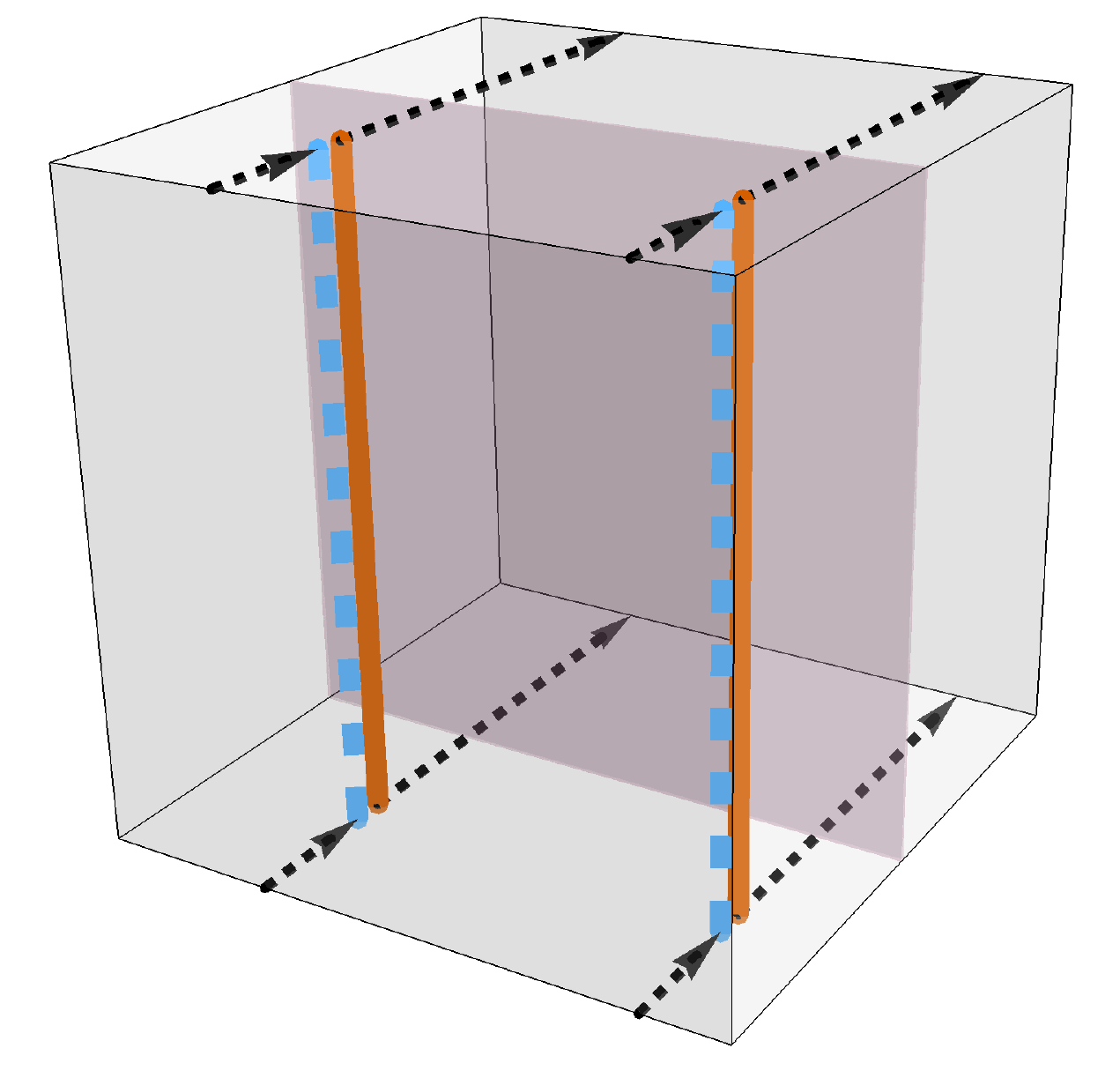}}
\caption{Topologically nontrivial cage-net wavefunction strings in the gauged bilayer X-Cube model in the presence of twisted boundary conditions (purple membrane is the branch cut) on the 3-torus. Opposite faces of the cubes are identified. (a)  Strings which pass through the branch cut must change layers (colors) and so must circle the 3-torus twice in order to close. (b) Modification of the prism constraint in Fig.~\ref{fig:prismConstraint} in the presence of the branch cut; strings on the opposite side of the branch cut have opposite colors. (c) Strings in layer 1 (orange) which are oriented in the $y$ direction at the same $x$ position can be moved around the torus using the constraint in (b) and end up in the other layer (dashed blue) after passing through the branch cut.}
\end{figure}

However, due to the proliferated layer-twist defects, additional ground states are available. In particular, the twist defects, which are created by flexible membrane operators, can extend all the way around the torus without costing energy. In a ground state on the 3-torus there are three independent possible twist defects corresponding to stretching the layer-twist membrane around the $x$ and $y$, $x$ and $z$, and $y$ and $z$ handles, an example of which is shown in Fig.~\ref{fig:3TorusDoubleWrap}.

These layer-twist defects are branch membranes for the strings and substantially change the ground states because a Wilson loop which passes through a layer-twist defect changes layers. For example, in the presence of a twist defect in the $xy-$plane, a $z$-oriented string must wrap around the torus twice in order to close, as shown in Fig.~\ref{fig:3TorusDoubleWrap}. Furthermore, the set of constraints on strings which do not pass through a twist defect also changes. The constraint that four strings on the edges of a prism are equal to the identity (see Fig.~\ref{fig:prismConstraint}) is modified in the presence of the twist defect to the one shown in Fig.~\ref{fig:oneTwistPrismConstraint}. Using these constraints, certain pairs of strings can be translated in certain directions, e.g. two $y$-oriented strings at the same $x$ position can be translated in the $z$ direction, as shown in Fig.~\ref{fig:pairTranslation}. If the strings are translated all the way around a handle of the torus, in the presence of an appropriate twist defect they change layers, as shown in Fig.~\ref{fig:pairTranslation}. This leads to additional constraints between the Wilson loops in different layers. This is similar to the gauged bilayer toric code; in that case, single strings around the untwisted handle can swap layers by being translated around the twisted handle. It is the same situation here, except only certain \textit{pairs} of strings can be translated around the twisted handle. 

In Appendix \ref{app:GSD}, we carefully account for the constraints in all sectors of the ground state subspace. The end result of the calculation is that the total ground state degeneracy on the 3-torus is
\begin{align}
\text{GSD} &= 2^{2N+1}+9\times 2^N + \nonumber \\
&+\sum_{i=x,y,z}\left(2^{N+2L_i-1}-3\times 2^{N-L_i}+2^{L_i + \frac{N+1}{2}}\right)
\label{eqn:gaugedGSD}
\end{align}
where $N$ is defined in Eq.~\eqref{eqn:XCubeN}.

\section{Excitations of the Gauged Bilayer X-Cube Model}
\label{sec:X-CubePheno}

In this section, we discuss at length the excitations of the gauged bilayer X-Cube model Eq.~\eqref{eqn:gaugedXCubeH}. In particular,
we show how the string operators for the non-Abelian excitations are constructed. We then compute the degeneracy associated to the
string-like twist defects and discuss some examples of bulk braiding analogs associated to the non-Abelian fractons. 

Similarly to the toric code case, the (ungauged) bilayer X-cube model has excitations which can be labeled as $a^{(1)}b^{(2)}$ where
$a,b=1,e_i,f_0,$ and bound states of those particles. (See Appendix~\ref{app:XCubeReview} for a review of the excitations of a single layer of the X-cube model using the labeling of Ref.~\onlinecite{BulmashGappedBoundaries}.) Here $e_i$, for $i=x,y,z$, is a one-dimensional particle which moves only in the $i$th direction,
 associated with exciting two star operators on the same site, and $f_0$ is the fracton associated with an excited cube operator.
Note that we are suppressing position indices since the true superselection sectors depend in a complicated way on the positions of various excitations.
We will also occasionally refer to a two-dimensional particle called $m_{ij}$, where $i,j= x,y,z$ are the directions of mobility; this excitation is a
bound state of two face-sharing cube excitations separated in the $k$ direction where $k\neq i \neq j$.

Upon gauging, any SWAP-invariant particle remains Abelian; in our case, these are $1$, $e^{(1)}_ie^{(2)}_i$, $f^{(1)}_0f_0^{(2)}$, and bound states thereof.
The non-invariant particles are grouped into orbits $[e_i]\sim e_i^{(1)}+e_i^{(2)}$ and $[f_0]\sim f_0^{(1)}+f_0^{(2)}$, which are non-Abelian. As in the toric code case, $\sim$ means a rough identification of the particle type with superpositions of the original particles. The non-Abelian quasiparticles have quantum dimension 2. Again, there is a $\mathbb{Z}_2$ charge associated with the $\tau$ degrees of freedom, which we shall call $\phi$ and will turn out to be fully mobile. Upon gauging, the string-like twist defect becomes a flexible string-like excitation $\sigma_1$, which can bind various point excitations. 

The pure charge excitations of the gauged model can be identified with representations of $\mathbb{D}_4$ in the same way as in the gauged
bilayer toric code case, but it is not clear that it is particularly meaningful or useful to describe the purely magnetic excitations in terms of conjugacy classes.
In general, monopoles (fractons) and fluxes are created by membrane operators. The nontrivial simple excitations are listed in Table \ref{tab:D4XCubeExcitations}.

\begin{table*}
\centering
\renewcommand{\arraystretch}{1.3}
\begin{tabular}{@{}lllllll}
\toprule[2pt]
\textbf{Label} && \textbf{Excitation}&& \textbf{Mobility} &&\textbf{$d_a$}\\ \hline
$\phi$ && Point $\mathbb{Z}_2$ charge $(a,c)\rightarrow (1,-1)$ && 3D && 1 \\
$e_i^{(1)}e_i^{(2)}$ && Point charge $(a,c) \rightarrow (-1,1)$ && 1D && 1\\
$\phi e_i^{(1)}e_i^{(2)}$ && Point charge $(a,c) \rightarrow (-1,-1)$ && 1D && 1 \\
$[e_i]$ && Point charge 2D irrep of $\mathbb{D}_4$ && 1D && 2\\
$f_0^{(1)}f_0^{(2)}$ && Abelian monopole && Fracton&& 1\\
$[f_0]$  && Non-Abelian monopole && Fracton && 2\\
$\sigma$&& Flux (twist) string && Flexible string excitation&& $2^{\ell_x + \ell_y +\ell_z-1}$
\\	\bottomrule[2pt]	
\end{tabular}
\caption{Nontrivial simple excitations in the gauged bilayer X-Cube model, their mobility, and their quantum dimensions $d_a$. Pure charges are labeled by representations of $\mathbb{D}_4$. Superscripts $(1)$ and $(2)$ label layers and subscripts $i=x,y,z$ label directions in which quasiparticles are mobile. Bound states of two point charges or point monopoles of the same type can have increased mobility, as in the ungauged X-cube model. The quantum dimension of the flux string depends on its linear extent $\ell_i$ in the $i$ direction and is computed in Sec. \ref{subsec:stringDegen}.}
\label{tab:D4XCubeExcitations}
\end{table*}

\subsection{Quasiparticle excitations}

\subsubsection{Abelian excitations}

\begin{figure*}
\centering
\subfloat[\label{fig:cageNetEE}]{\includegraphics[width=0.7\columnwidth]{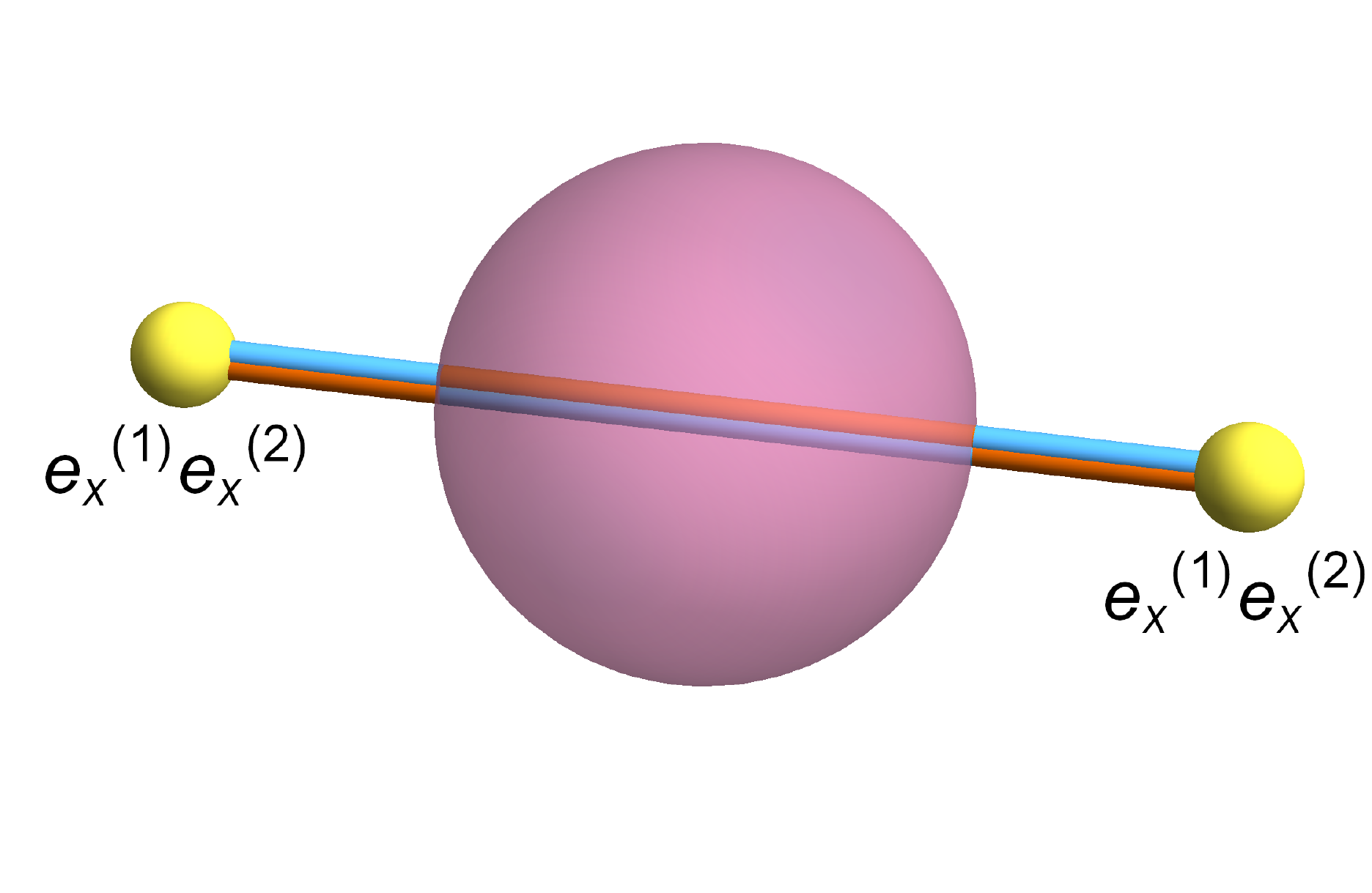}}\\
\subfloat[\label{fig:cageNetFF}]{\includegraphics[width=1.6\columnwidth]{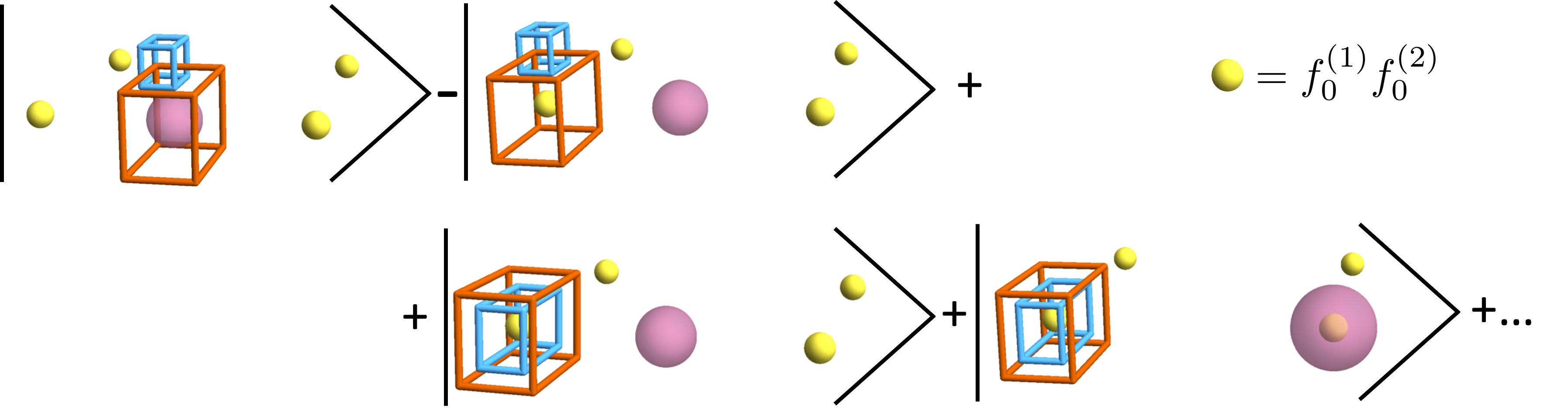}}\\
\subfloat[\label{fig:cageNetPhi}]{\includegraphics[width=1.6\columnwidth]{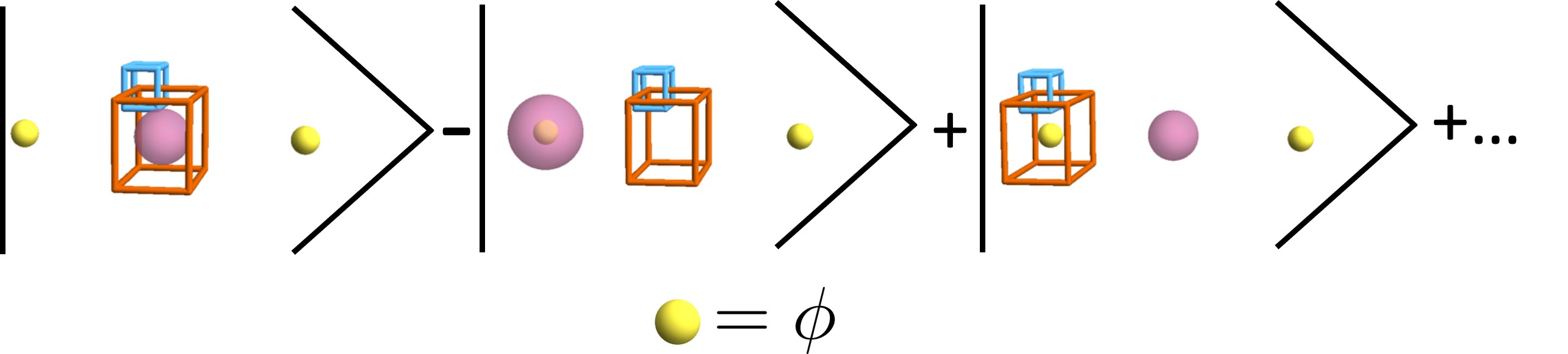}}
\caption{Sample terms in the cage-net wavefunction for Abelian quasiparticles in the gauged bilayer X-Cube model. (a) $e^{(1)}_xe^{(2)}_x$ excitations. (b) $f_0^{(1)}f_0^{(2)}$ excitations. (c) $\mathbb{Z}_2$ charge $\phi$.}
\end{figure*}

The bound states of excitations in both layers are Abelian and created in the same way as in the ungauged model; for example, a (rigid) string of $\sz^{(1)}\sz^{(2)}$ operators creates the one-dimensional particle $e_i^{(1)}e_i^{(2)}$ associated to the irrep $(a,c)\rightarrow (-1,1)$, which are, in the cage-net picture, the endpoints of a string of both colors, as in Fig.~\ref{fig:cageNetEE}.

A rectangular membrane of $\sx^{(1)}\sx^{(2)}$ operators creates four Abelian fractons $f_0^{(1)}f_0^{(2)}$. In the cage-net picture, this corresponds to having a minus sign in the ground state superposition whenever an odd number of cages surround the location of the fracton, as shown in Fig. \ref{fig:cageNetFF}.

Similarly, the $\mathbb{Z}_2$ charge $\phi$ is created by a string of $\tau_z$ operators acting on gauge qubits. Since this string commutes with $A_s$, $B_c$, $D_{\triangle}$, and $D_f$ and anticommutes with $C_e$, $\phi$ is easily checked to be fully mobile. This is not surprising since each $\tau$ spin appears in exactly two $C_e$ operators. In the wavefunction picture, this excitation corresponds to  minus signs in the ground state superposition whenever a branch membrane surrounds the excitation in question, as in Fig.~\ref{fig:cageNetPhi}.

\subsubsection{Non-Abelian fractons - cage-net picture}
\label{subsubsec:cageNetF0}

We begin with a cage-net picture of the wave function in the presence of the non-Abelian fracton $[f_0]$. We will find the
remarkable result that the degeneracy in the presence of many $[f_0]$ fractons depends on their relative positions.
We further use the wave function picture to understand how this degeneracy can be understood in terms of fusion
properties of the fractons. Subsequently we demonstrate certain fusion rules explicitly by studying the quasiparticle operator algebra. 

\begin{figure*}
\centering
\includegraphics[width=1.5\columnwidth]{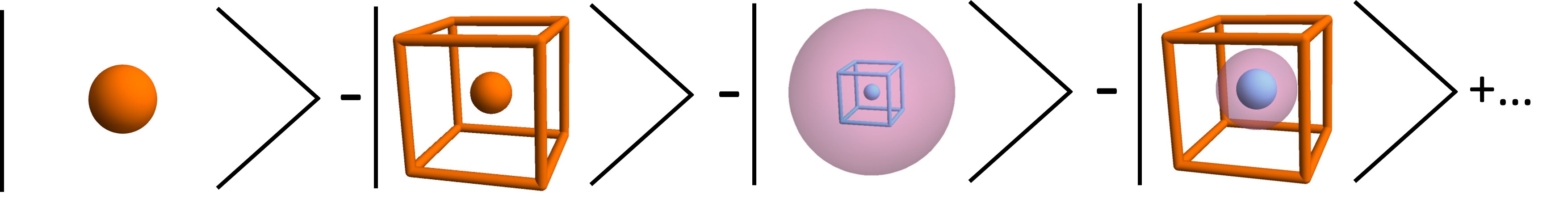}
\caption{Cage-net configurations near a single $[f_0]$ excitation illustrating the interaction of branch membranes and cages. The sphere color indicates whether layer-1 or layer-2 cages (in the absence of branch membranes) introduce minus signs into the wavefunction layer in that cage-net configuration. As before, orange (resp. blue) indicates layer 1 (resp. layer 2) cages and excitations and purple are branch membranes. The third and fourth terms show the interaction between branch membranes and cages surrounding $[f_0]$. All other excitations in the system are assumed to be very far from everything in these pictures.}
\label{fig:cageNetSingleFractonLocal}
\end{figure*}

The $[f_0]$ excitations are created in sets of four at the corners of rectangular membrane operators. The local cage-net configurations
near a single fracton are shown in Fig.~\ref{fig:cageNetSingleFractonLocal}. In particular, we start with a ``reference" configuration
where the fracton excitation is labeled with a definite layer index (color, in our pictures). When $[f_0]$ is surrounded by an odd
number of cages of its \textit{own color} (in the absence of any branch membranes), the term in the superposition switches sign,
as in the first two terms of Fig.~\ref{fig:cageNetSingleFractonLocal}. The $[f_0]$ excitation also switches colors
when surrounded by a branch membrane. Note that since branch membranes do not preserve the color of cages, the
configuration where an orange cage surrounds \textit{both} a blue fracton and a branch membrane has the \textit{same}
sign as the configuration where an orange cage surrounds an orange fracton, as in the
configurations in Fig.~\ref{fig:cageNetSingleFractonLocal}. This is because these two configurations are related by deforming the
branch membrane through the fracton in question, which switches the color of the fracton without introducing an additional minus sign.

Let us now consider the degeneracy associated with the excitations. It is important to realize that, just as in the ungauged model, there is a constraint at the level of the Hilbert space: in any reference configuration, the number of fractons of each color in \textit{every} plane perpendicular to a coordinate axis is even. This follows from the operator identity
\begin{align}
  \prod_{c \in \text{plane}}B_c^{(i)} = 1.
\end{align}
  That is, a cage of any fixed color which encloses an entire plane is exactly the same as no cage at all.

  Therefore, the number of ground states for a given spatial configuration of $[f_0]$ fractons is given by the number of ways of
  assigning layer labels to each fracton (modulo a global SWAP), subject to the above constraint. The requirement of satisfying the above constraints
  leads to a highly non-trivial position-dependence of the degeneracy, as we see explicitly in the following examples. 
  
  In the presence of four $[f_0]$ excitations at the corners of a rectangle, there is only one allowed reference configuration up to a global SWAP (in which all excitations have the same color), so there is only one state.

\begin{figure*}
\centering
	\subfloat[\label{fig:eightF0CageNet}]{\includegraphics[width=1.3\columnwidth]{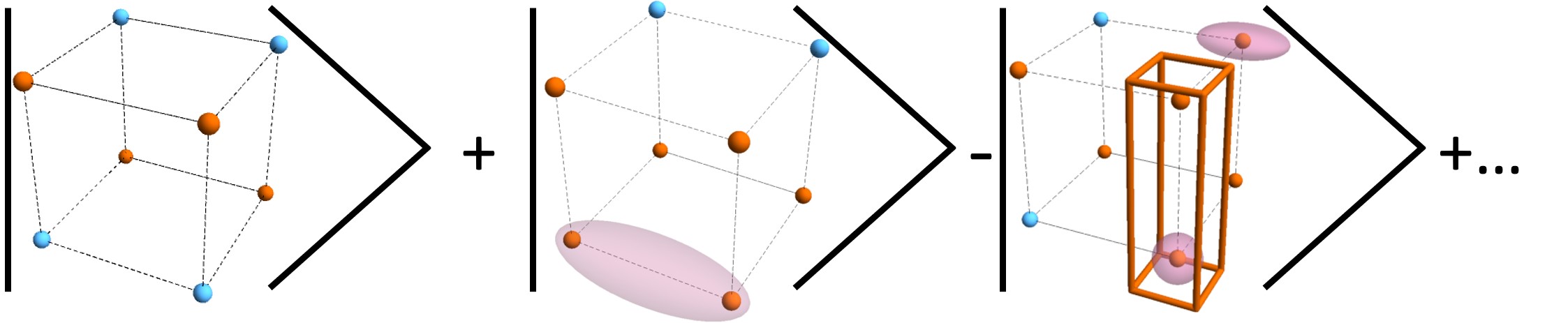}}\\
	\subfloat[\label{fig:layeredXCubeF0}]{\includegraphics[width=0.4\columnwidth]{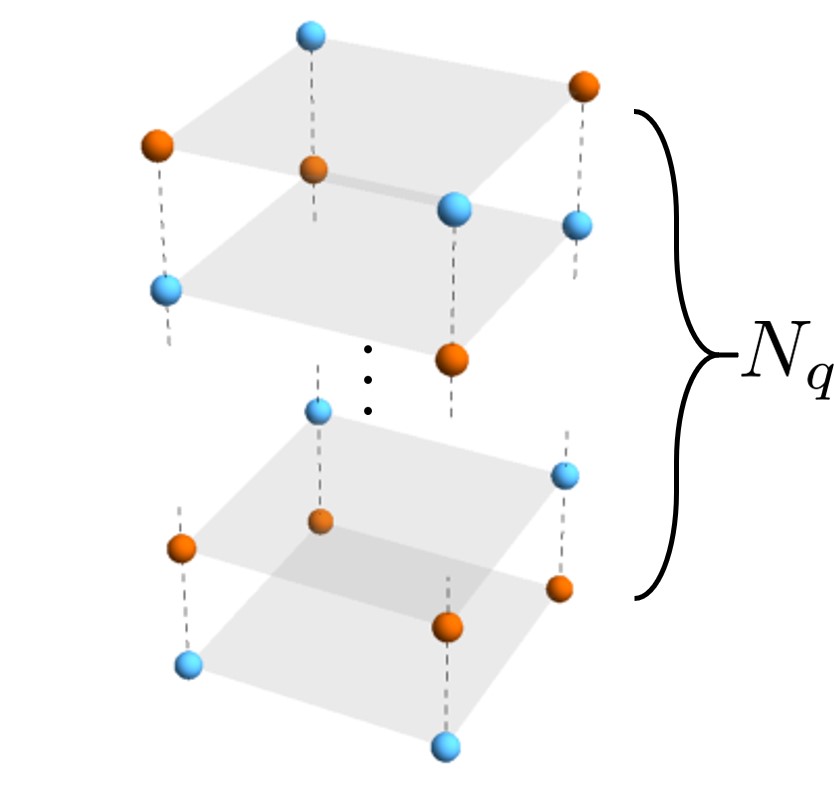}}
		\hspace{.1\columnwidth}
	\subfloat[\label{fig:eightF0StairstepGeometry}]{\includegraphics[width=0.4\columnwidth]{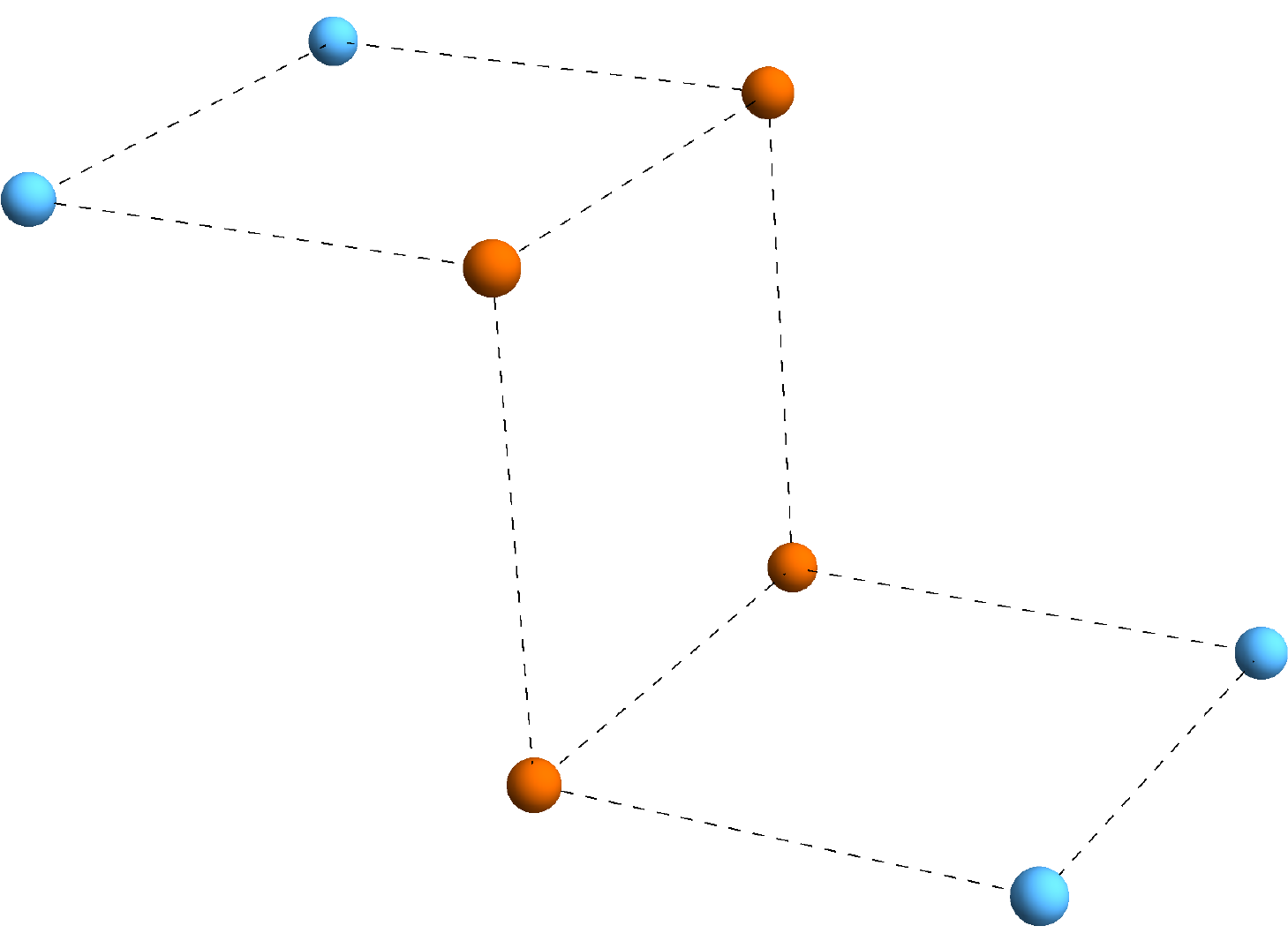}}
	\hspace{.1\columnwidth}
	\subfloat[\label{fig:eightF0shifted}]{\includegraphics[width=0.4\columnwidth]{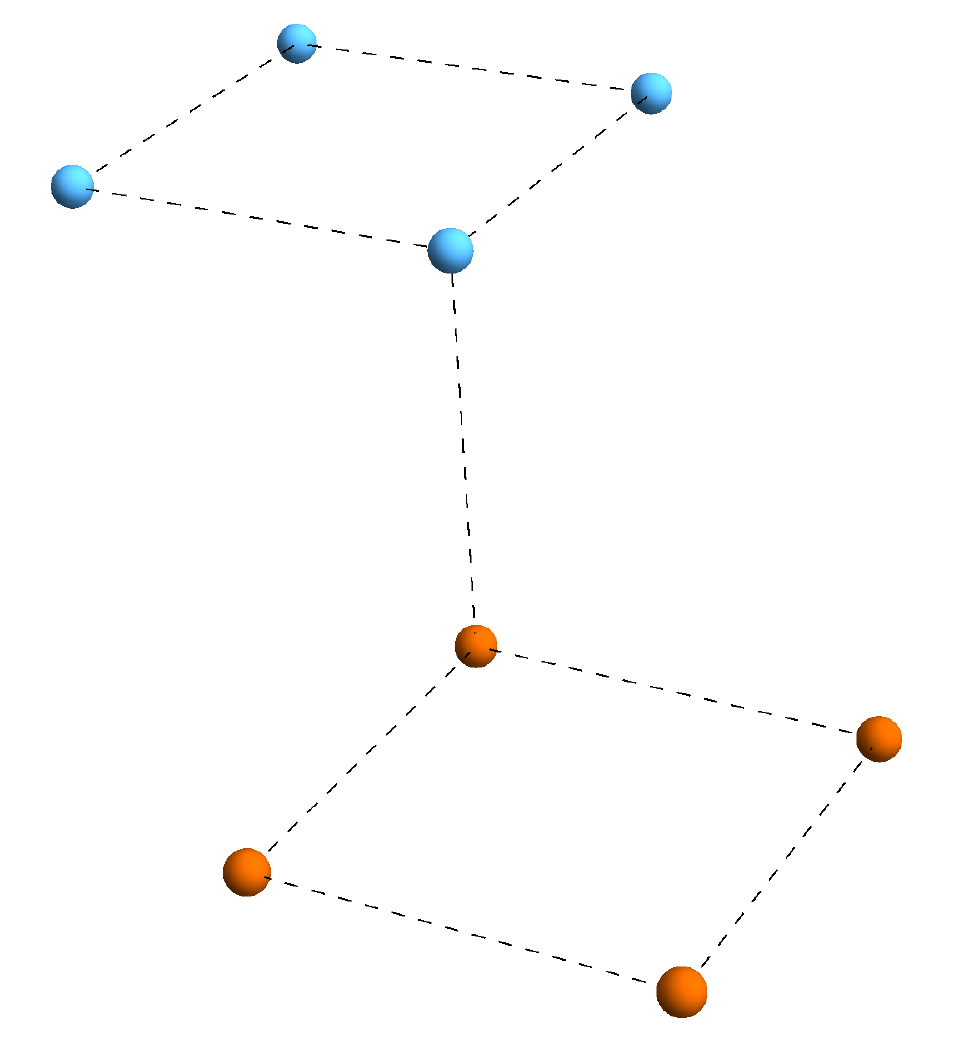}}
\hspace{.1\columnwidth}
	\subfloat[\label{fig:f0QuantumDimension}]{\includegraphics[width=0.4\columnwidth]{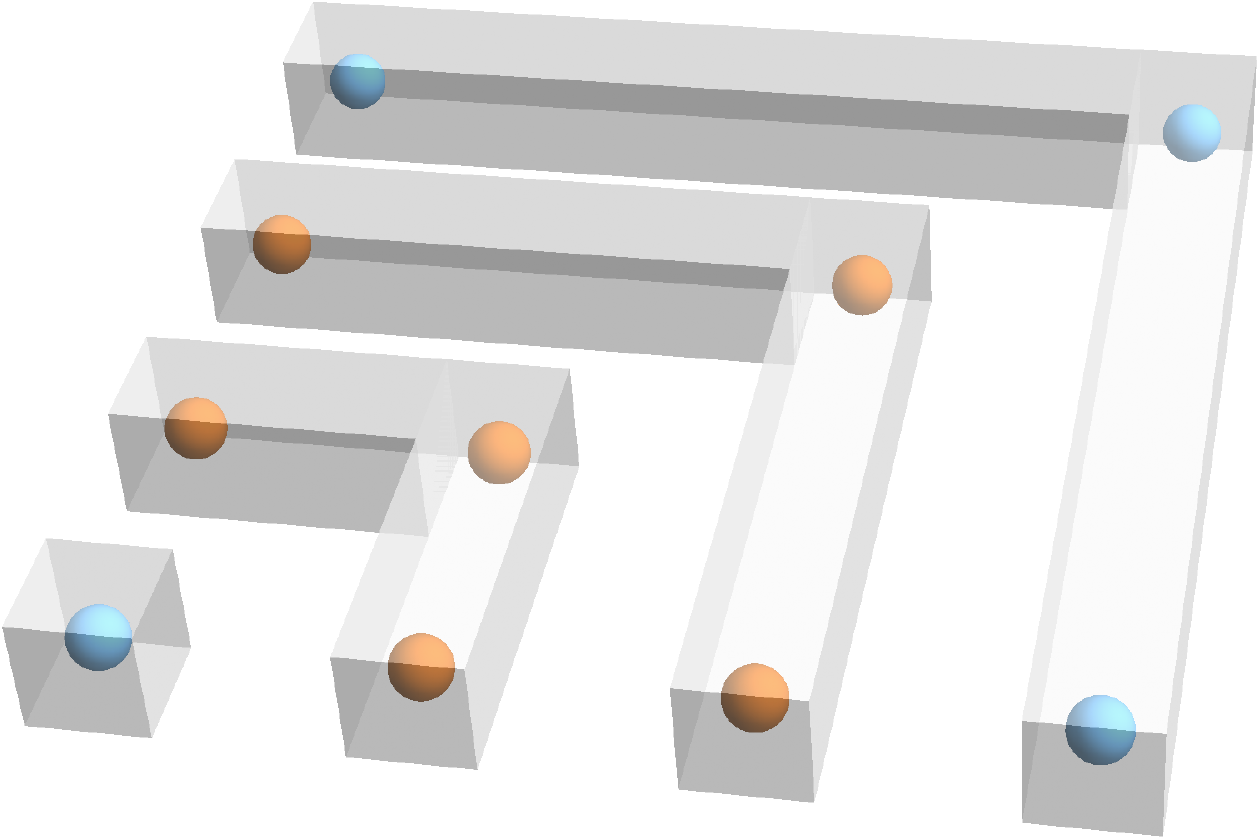}}
\caption{States associated to many $[f_0]$ excitations. Color coding is the same as Fig.~\ref{fig:cageNetSingleFractonLocal}, and dashed lines and grey rectangles are guides to the eye. (b) A reference configuration for $4N_q$ $[f_0]$ fractons created in $N_q$ ``layers"; there are $8^{N_q-1}$ degenerate states associated with these fractons in this geometry. (c,d) Example reference configurations for eight $[f_0]$ fractons in different geometries from (a). In (c) there are four degenerate states and in (d) there are only two. (e) Example reference configuration in the geometry used to calculate the quantum dimension of $[f_0]$.}
\label{fig:manyF0Geometry}
\end{figure*}

  Consider next eight $[f_0]$ excitations on the corners of a cube (or a rectangular prism). Then it is easy to check that, up to a global SWAP, there are eight possible reference configurations consistent with the aforementioned identity and which are inequivalent under proliferation of cage-nets. One example reference configuration is given in Fig.~\ref{fig:eightF0CageNet} along with a few terms in its cage-net superposition. Distinct configurations are locally indistinguishable because the layer index for an individual fracton is not a good quantum number, so there are eight degenerate states associated with these fractons. This is a general rule; the number of distinct reference configurations up to global SWAPs equals the degeneracy. Generalizing this, in the presence of $4N_q$ fractons created in $N_q$ quartets of four fractons in the geometry shown in Fig.~\ref{fig:layeredXCubeF0}, simple combinatorial arguments show that the number of inequivalent reference configurations and therefore the number of locally indistinguishable degenerate states is $8^{N_q-1}$.

When instead eight fractons are placed in the geometry shown in Fig.~\ref{fig:eightF0StairstepGeometry} , there are only \textit{four} allowed reference configurations, and in the geometry shown in Fig.~\ref{fig:eightF0shifted}, there are only \textit{two} allowed reference configurations. Hence the number of states associated with eight $[f_0]$ excitations depends strongly on their relative positions. This position-dependence of the degeneracy is consistent with the fact that in a fusion category description of fractons, their superselection sector of depends on their position\cite{PaiHermeleFusion}.

The locally indistinguishable, degenerate states arising from the $[f_0]$ fractons can be labeled in terms of fusion properties of the fractons, as we now describe. We note that a complete theory of fusion rules is complicated by the fact that fractons at different spatial positions are in distinct superselection
sectors, and a complete theory of fusion should properly take this into account. As such, here our discussion of the fusion properties is limited to a description of how to label the different states in terms of the eigenvalues of certain extended cage or membrane operators that can measure certain topological charges associated with a given region. We leave a more comprehensive theory of fusion rules in such non-Abelian fracton models as an open problem. 

Just as in the gauged bilayer toric code case, the layer label for each fracton is not definite, but the layer parity of any pair of fractons in a cage-net configuration depends only on the corresponding layer parity in the relevant reference configuration and the number of branch membranes surrounding those two fractons. Therefore, an operator which creates a cage of either color surrounding two excitations has a definite eigenvalue of $\pm 1$ in each of these eight reference configurations. We interpret these two possible eigenvalues as assigning to the given region an even or odd topological charge of the Abelian fracton $f_0^{(1)}f_0^{(2)}$.

\begin{figure}
\centering
	\subfloat[\label{fig:MSigma2Fractons}]{\includegraphics[width=.9\columnwidth]{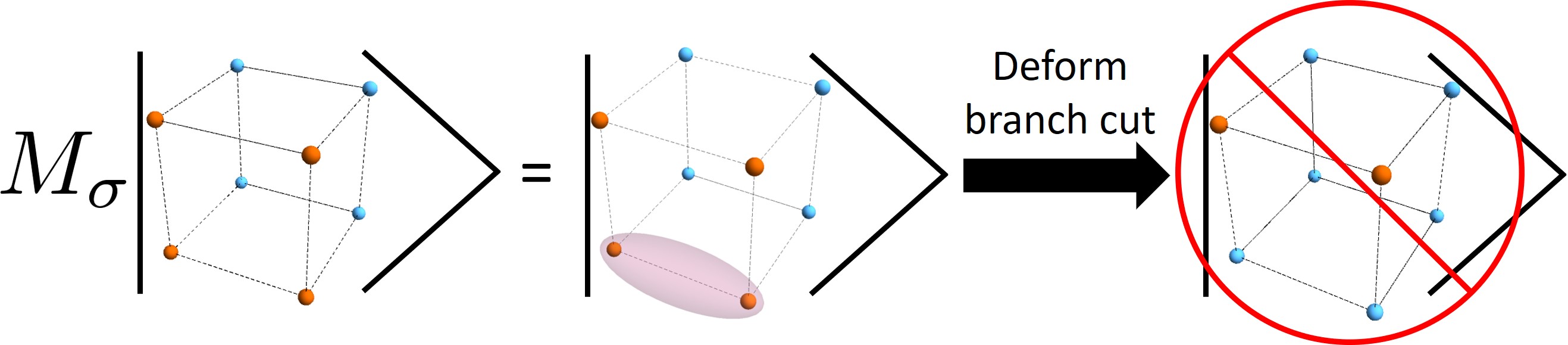}}\\
		\subfloat[\label{fig:MSigma4Fractons}]{\includegraphics[width=.9\columnwidth]{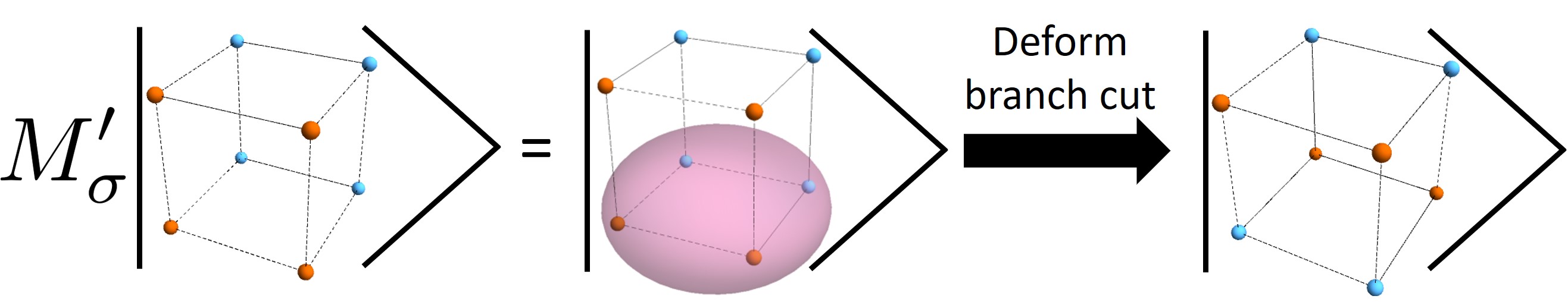}}\\
	\subfloat[\label{fig:eightF0DefiniteFusions}]{\includegraphics[width=0.9\columnwidth]{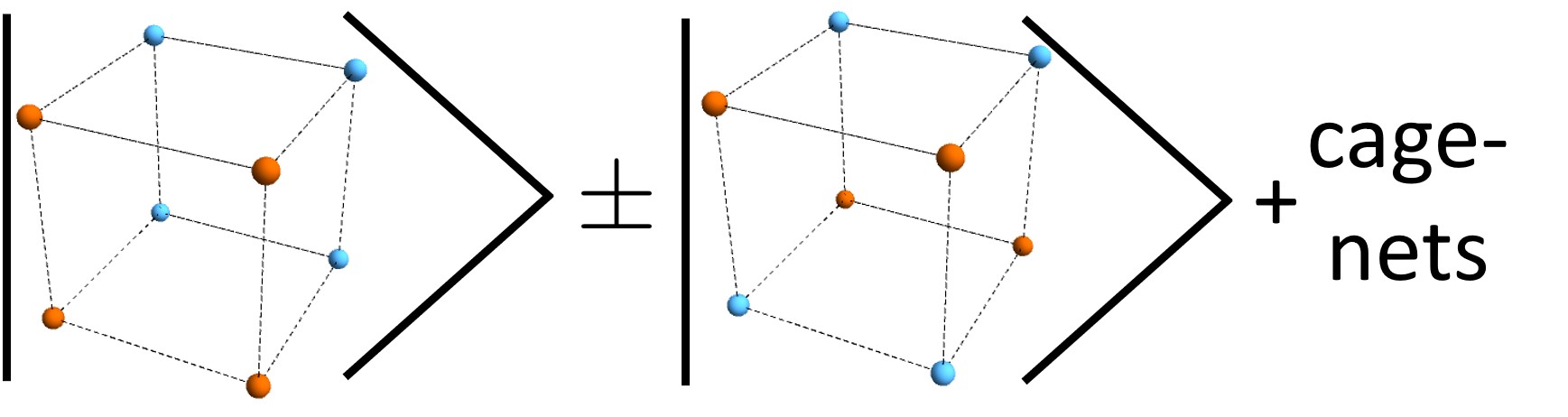}}
\caption{Color coding is the same as Fig.~\ref{fig:manyF0Geometry}. (a) Applying a membrane operator $M_{\sigma}$ which measures the $\mathbb{Z}_2$ charge contained by two fractons produces an ``illegal" state which violates the Hilbert space constraints. (b) Applying a membrane operator $M'_{\sigma}$ which measures the $\mathbb{Z}_2$ charge contained by four fractons produces a valid state. (c) States where the top (and bottom) four fractons fuse to two $f_0^{(1)}f_0^{(2)}$ fractons with (minus sign) and without (plus sign) a $\phi$. }
\end{figure}

One can also assign a $\mathbb{Z}_2$ charge $\phi$ to certain regions, which can be measured by the eigenvalue of a membrane
operator $M_{\sigma}$. However, not every region can be assigned such a definite topological charge. The desired operator $M_{\sigma}$
creates a $\mathbb{Z}_2$ flux string, moves it around the region in question, and attempts to re-annihilates the string. Analogous to the case
of $W_{\sigma}$ in the gauged bilayer toric code case, $M_{\sigma}$ does not change the color of fractons inside the region it is
associated with, since its support is far from those excitations; nevertheless, it leaves behind in the wave function a branch membrane.
Suppose first that $M_{\sigma}$ surrounds two $[f_0]$ excitations. After deforming back to the reference configuration, two fractons switch colors, which violates the constraint that all planes perpendicular to coordinate axes contain an even number of fractons of each color.
Therefore, such a process is actually not possible. This process is depicted in Fig.~\ref{fig:MSigma2Fractons}.  Stated differently,
$M_{\sigma}$ must take us out of the degenerate manifold of states to a different excited state rather than re-annihilating the string. A similar phenomenon
occurs in, for example, the (2+1)D Ising topological order, where a non-Abelian $\sigma$ loop cannot enclose a single non-Abelian
$\sigma$ particle, because of the difference in fusion channels between the initial and final state.

Nevertheless, there exist choices of four $[f_0]$ such that applying a membrane $M_{\sigma}'$ that surrounds them, as in Fig.~\ref{fig:MSigma4Fractons}, leads to a configuration which obeys the constraints. This means that a definite $\mathbb{Z}_2$ charge is only associated to certain choices of \textit{four} well-separated $[f_0]$ excitations, not two. 

Therefore, any pair of fractons can be assigned a definite $f_0^{(1)}f_0^{(2)}$ parity, but only certain sets of four fractons can have a definite $\phi$ parity. In particular, certain superpositions of the reference configurations, such as the one shown in Fig.~\ref{fig:eightF0DefiniteFusions}, are cage-net states with definite $f_0^{(1)}f_0^{(2)}$ (resp. $\phi$) parity for all pairs (resp. quartets) of fractons in the same $xy$-plane.  One can also check that this fusion information is sufficient to distinguish all of the eight degenerate states.

Let us now turn to a discussion of quantum dimension. In conventional topological order, the quantum dimension $d_a$ of a particle of type $a$ can be defined by considering the degeneracy in the presence of $N$ particles of type $a$, which scales as $d_a^N$. However since in the present context the degeneracy in the presence of $N$ fractons $[f_0]$ depends sensitively on their relative position, the definition of the quantum dimension of $[f_0]$ is more subtle.

In fact, because the superselection sector of fractons depends on their positions, the geometries considered above actually involve fusion of excitations in different superselection sectors. Thus, to define a quantum dimension for $[f_0]$, we wish to create a set of excitations which are all in the \it same \rm superselection sector. This can be done by considering a set of $[f_0]$ fractons in the geometry shown in Fig.~\ref{fig:f0QuantumDimension}. In this geometry, there is a membrane operator which turns the excitations in any one grey box into the excitations in any other grey box - in particular, each one can independently be turned into a copy of the grey box with a single $[f_0]$ inside. This means that each grey box should be thought of as living in the same superselection sector as the single $[f_0]$. In this geometry, for each reference configuration, all fractons in each grey box must be in the same layer in order to obey the global constraints. Simple counting shows that up to a global SWAP, there are $2^{N_b-2}$ degenerate states in this geometry, where $N_b$ is the number of grey boxes (provided $N_b$ is even, or no states are allowed at all). Therefore we find that the quantum dimension
\begin{align}
  d_{[f_0]} = 2 .
  \end{align}

\subsubsection{Non-Abelian fractons - explicit operators}

\begin{figure*}
\centering
\subfloat[\label{fig:membraneCubeComms}]{\includegraphics[width=1.5\columnwidth]{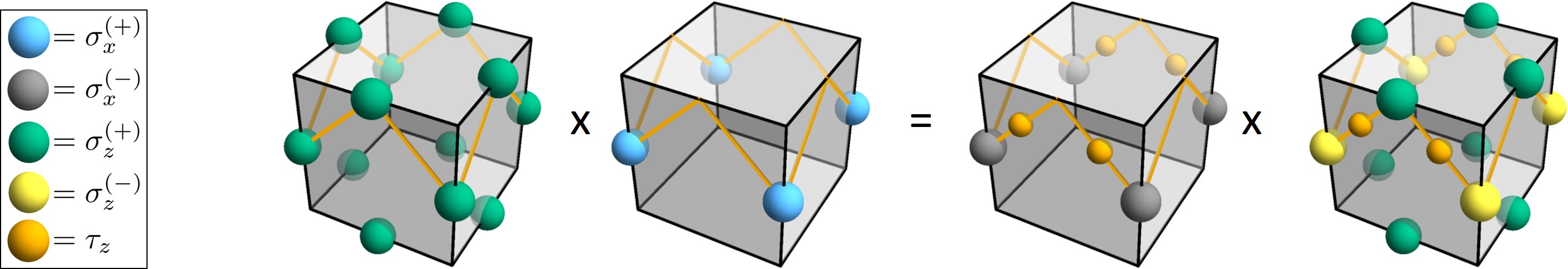}}\\
\subfloat[\label{fig:membraneAPlusB}]{\includegraphics[width=1.5\columnwidth]{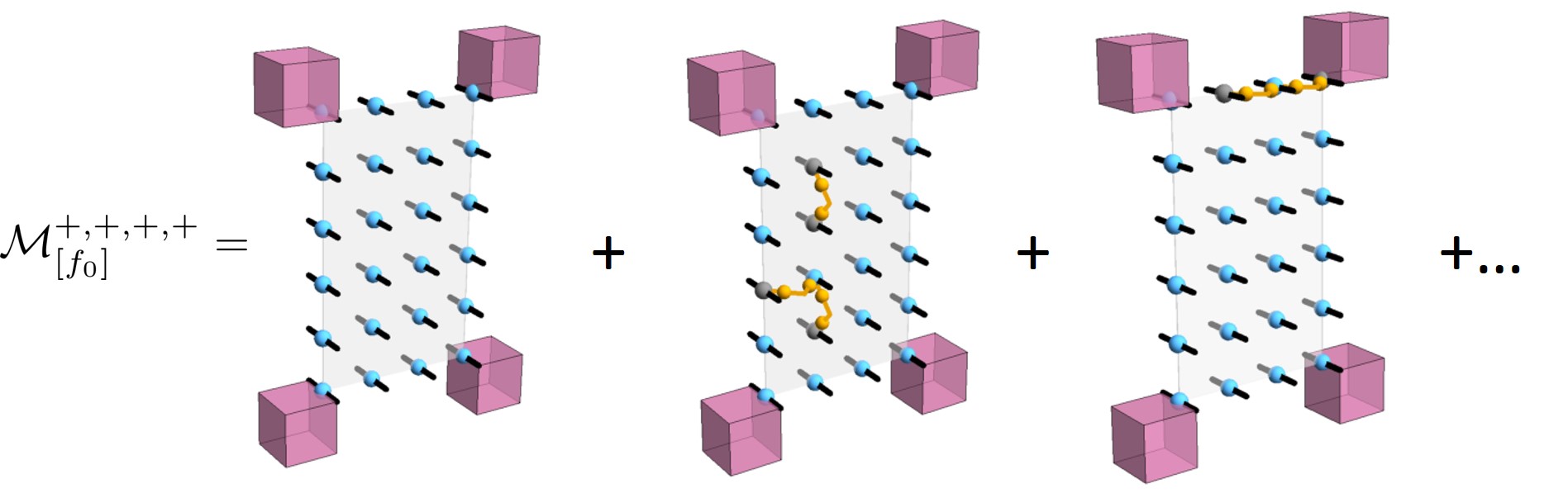}}\\
\subfloat[\label{fig:membraneAPlusB_ppmm}]{\includegraphics[width=1.5\columnwidth]{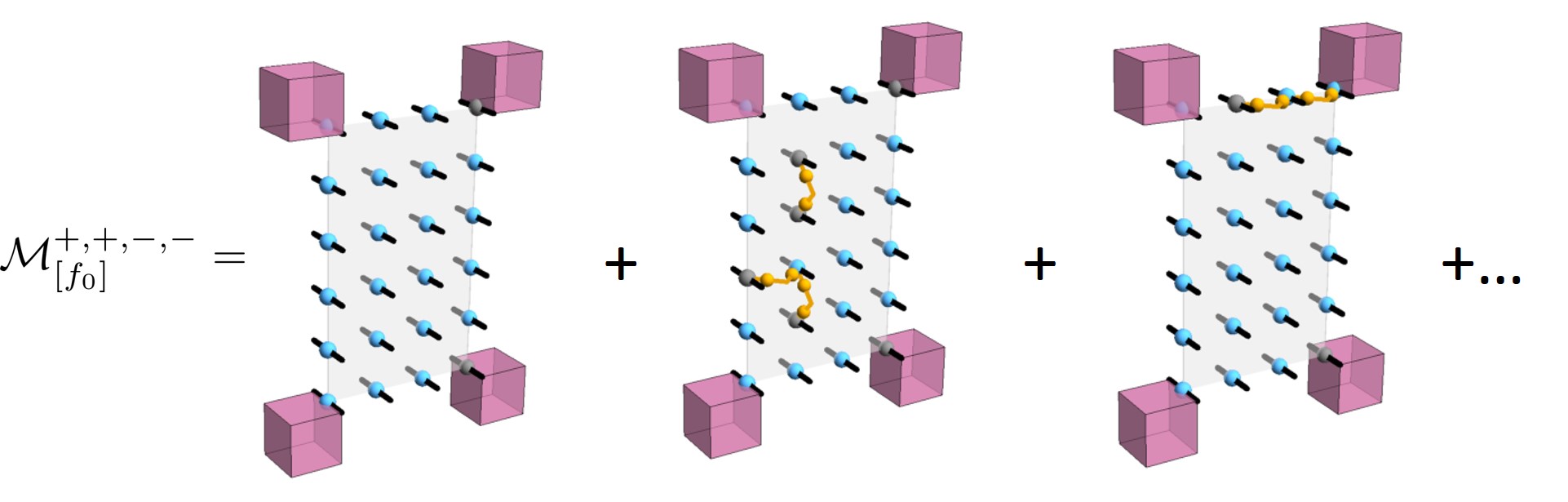}}\\
\caption{Constructing membrane operators for $[f_0]$ fractons. (a) Commutation relation between one term of $B_c$ (involving $\sz$ operators) and a small membrane of $\sx^{(+)}$ operators. Operator legend is in the bottom left. (b) and (c) Two choices of membrane operator that create $[f_0]$ fractons (purple cubes) in the gauged bilayer X-cube model. They differ only by the choice of operator (color) in the top right and bottom right corners of the membrane. All terms with an even number of grey spheres appear in each sum. The grey membrane is a guide to the eye representing the membrane-like support of the operator. The operator legend in (a) applies to all of the figures.}
\label{fig:membraneOperators}
\end{figure*}

With the cage-net understanding in hand, we now explicitly construct the operators that create the non-Abelian fractons $[f_0]$. Following the intuition from the gauged bilayer toric code, we expect that we should build a (gauge-invariant) membrane operator from $\sx^{(+)}$ and $\sx^{(-)}$ since these commute with $C_e$. However, the membrane should commute with the Hamiltonian everywhere except at its corners. Commuting $B_c$ past a membrane of $\sx^{(+)}$ operators will permute the various summands of $B_c$ at the cost of interchanging $\sx^{(+)}$ with $\sx^{(-)}$ and dressing with some $\tau_z$ operators, as in Fig.~\ref{fig:membraneCubeComms}. Therefore, superposing over all such interchanges will produce the desired membrane operator. Some examples have the form shown in Figs.~\ref{fig:membraneAPlusB} and \ref{fig:membraneAPlusB_ppmm}; these operators create four fractons $[f_0]$.

There are in fact sixteen different such operators; they can be obtained from the one shown in Fig.~\ref{fig:membraneAPlusB} by choosing any subset of the corners of the membrane and exchanging $\sx^{(+)}$ and $\sx^{(-)}$ in every term of the sum at the chosen corners. These operators differ from each other by the local action of $X^{(i)}$ at the corners. For future reference, we call these operators $\mathcal{M}_{[f_0]}^{\pm, \pm, \pm, \pm}$, where the signs refer to whether there is an $\sx^{(+)}$ or an $\sx^{(-)}$ at each corner in the term in the sum with no $\tau_z$ operators, as labeled in Fig.~\ref{fig:membraneOperators}.

As a suggestive calculation for the fusion channels, we explicitly compute products of the membrane operators. For the full operator algebra, see Appendix~\ref{app:bilayerXCubeExcitations}.

Let 
\begin{equation}
M = \sum_{\lbrace k,l,m,n=\pm | klmn=+\rbrace}\mathcal{M}_{[f_0]}^{k,l,m,n}
\end{equation} This is a particular choice of membrane operator which creates four fractons at its corners. Applying $M^2$ to the ground state creates two fractons at each corner of the membrane. We can fuse the two fractons at each corner into simple excitations; this amounts to re-expressing $M^2$ as a sum of operators creating simple excitations at each corner of the membrane. A tedious computation similar to the one done for the bilayer toric code shows that
\begin{widetext}
\begin{equation}
M^2 = \left(1 + \bigotimes_{\bv{r} \in \square} \sx^{(1)}\sx^{(2)}\right)\left(1+\sum_{i\neq j} \bigotimes_{\text{string}_{ij}} \tau_z + \bigotimes_{\text{string}_{ij},\text{string}_{kl}} \tau_z\right)
\end{equation}
\end{widetext}
where $i,j,k,l$ label distinct corners of the membrane and ``string$_{ij}$" means any string connecting the corners $i$ and $j$. In particular, we can identify this as a sum of the identity, the operator creating four $f_0^{(1)}f_0^{(2)}$ fractons, an operator creating any even number of $\mathbb{Z}_2$ charges $\phi$, and the bound states $\phi f_0^{(1)}f_0^{(2)}$. (Note that the string operators for $\phi$ can fluctuate between connecting any choice of pairs of $\phi$ excitations.) 

That is, there are four distinct outcomes for the fusion at each corner of the membrane: $1$, $f_0^{(1)}f_0^{(2)}$, $\phi$, and $\phi f_0^{(1)}f_0^{(2)}$, which can be chosen independently provided the total fusion is the identity. This naively seems to ``overcount" $\phi$ compared to the cage-net picture, where a definite parity of $\phi$ is only well-defined for a set of four fractons rather than a pair. However, in the present calculation, the fractons being fused are at the same location in space, so their fusion outcome can be distinguished by local operators (i.e. by arbitrarily small membrane / cage operators). It is not necessarily true that these locally distinguishable states all correspond to distinct degenerate states when the fractons are widely separated, especially given the geometry dependence of the non-Abelian degeneracy. However, when widely separated fractons are brought close together, the final fusion channels will of course be consistent with the fusion when the particles are well-separated.

\subsubsection{General operator structure}

We now exhibit the operators which create the full set of simple excitations in this model, deferring the full enumeration of their algebra to Appendix~\ref{app:bilayerXCubeExcitations}.

There are four types of operators which we need to consider: flexible string operators, which create $\mathbb{Z}_2$ charges $\phi$ (discussed previously), rigid string operators, which create excitations in the $e$ sector, rigid membrane operators, which create fractons, and flexible membrane operators, which create twist defect strings. We handle each one in turn, starting with string operators.

The structure of rigid string operators in the gauged bilayer X-cube model is the same as the string operators for pure charge excitations in the (2+1)D gauged bilayer toric code case (see Appendix~\ref{app:bilayerTCStrings}); the only difference is that the strings are rigid. In particular, the string operators are constructed from a set of local operators $\lbrace \mathcal{U}_{\pm},\mathcal{V}_{\pm},\mathcal{W}_{\pm} \rbrace$, although it so happens that in this sector the $\mathcal{V}_{\pm}$ are the identity. We will therefore ignore the $\mathcal{V}_{\pm}$ for the rest of the discussion.

As before, we start from a reference string as in Fig.~\ref{fig:generalString_reference} with only $\mathcal{U}_+$ and $\mathcal{W}_+$ appearing in the string except at its ends; either $\mathcal{U}_{\pm}$ may be chosen at the end, which correspond to local degrees of freedom of the quasiparticles, that is, this choice can be modified via the action of a local operator. Next, for every lattice site along the length of the string, we choose whether to flip $+ \leftrightarrow -$ for all links surrounding that site, as in Fig.~\ref{fig:generalString_charge}, obtaining a new operator. We then superpose over all such choices.

The quasiparticle types are: Abelian one-dimensional (1D) quasiparticles $e_i^{(1)}e_i^{(2)}$, where $i=x,y,z$ labels the direction of mobility, Abelian  1D quasiparticles $\phi e_i^{(1)}e_i^{(2)}$, and non-Abelian 1D quasiparticles $[e_i]$. A specification of their string operators is listed in Table~\ref{tab:XCubeStrings}.
\begin{table*}
\centering
\renewcommand{\arraystretch}{1.3}
\begin{tabular}{@{}lllllllllllllll}
\toprule[2pt]
\textbf{Label} && \textbf{Excitation type}&& $\mathcal{U}_+$ && $\mathcal{U}_-$ && $\mathcal{W}_+$ && $\mathcal{W}_-$ \\ \hline
$\phi$ && mobile boson && 1 && 1 && $\tau_z$ && $\tau_z$ \\
$e^{(1)}_{i}e^{(2)}_{i}$ && Abelian 1D particle && $\sx^{(1)}\sx^{(2)}$ && $\sx^{(1)}\sx^{(2)}$ && 1 && 1 \\
$\phi e^{(1)}_{i}e^{(2)}_{i}$ && Abelian 1D particle && $\sx^{(1)}\sx^{(2)}$ && $\sx^{(1)}\sx^{(2)}$ && $\tau_z$ && $\tau_z$ \\
$[e_i]$ && non-Abelian 1D particle && $\sz^{(+)}$ && $\sz^{(-)}$ && 1 && $\tau_z$ \\
\bottomrule[2pt]	
\end{tabular}
\caption{Specification of string operators which create $\mathbb{Z}_2$ charge and $e$-sector particles in the gauged bilayer X-cube model. The positions of the operators are exactly the same as in the (2+1)D gauged bilayer toric code case in Fig.~\ref{fig:generalString_reference}, and superpositions are formed similarly. The $\mathcal{V}_{\pm}$ are all 1 in this sector.}
\label{tab:XCubeStrings}
\end{table*}

We next turn to the rigid membrane operators, which create fractons, and describe them in a more systematic framework. The basic excitations in this sector are $f^{(1)}_{0}f^{(2)}_{0}$ and $\phi f^{(1)}_{0}f^{(2)}_{0}$, which are Abelian, and $[f_{0}]$, which is non-Abelian.

\begin{figure}
\includegraphics[width=0.6\columnwidth]{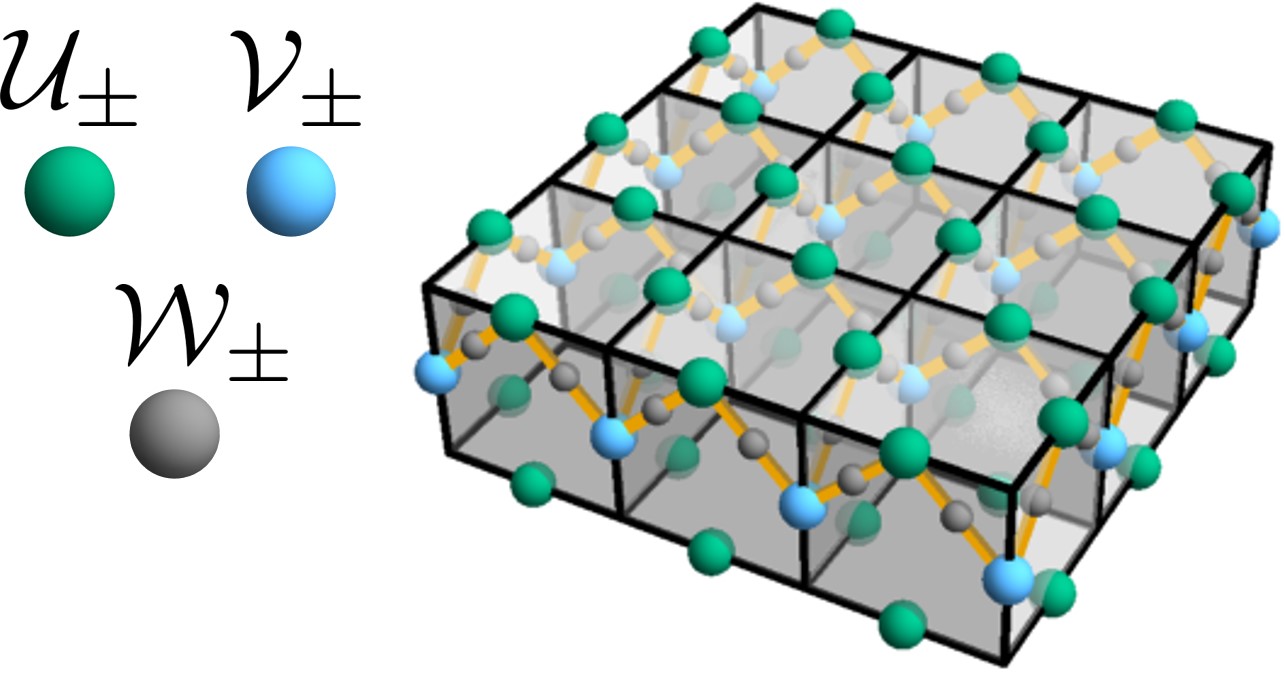}
\caption{Local operator locations for membrane operator in the gauged bilayer X-cube model. A superposition is formed from a reference configuration where every operator is chosen to be $+$ by choosing whether or not to flip $+ \leftrightarrow -$ on every spin touching a face (for an operator that creates fractons) or on an elementary cube (for an operator that creates a twist string).}
\label{fig:membraneOperatorLayout}
\end{figure}

We again take a reference operator and superpose flipped versions of that reference operator, as described in previous examples. The distinction is that this time, all of the local operators which border an elementary cube of the cubic lattice are affected by the flipping procedure.
We again define $\mathcal{U}_{\pm}$, $\mathcal{V}_{\pm}$, and $\mathcal{W}_{\pm}$ operators, which live now on the links in the plane of the membrane (both on top and on the bottom of the membrane), links puncturing the membrane, and $\tau$ on half of the links within the membrane, respectively, as in Fig.~\ref{fig:membraneOperatorLayout}. In the fracton sector, the $\mathcal{U}_{\pm}$ are always 1, so we ignore them in what follows, but we have defined the $\mathcal{U}_{\pm}$ for the purposes of discussing twist strings later.

The reference operators are defined to have $\mathcal{V}_+$ and $\mathcal{W}_+$ everywhere on the membrane, except we allow either $\mathcal{V}_+$ or $\mathcal{V}_-$ at the corners of the membrane. This leads to sixteen possible operators specified by four choices of $\mathcal{V}_{\pm}$ in the reference configuration; these sixteen operators will differ by the action of local operators at the corners of the membrane, that is, they specify purely local degrees of freedom of the fractons.

Now form more membrane operators by, for each face within the membrane, choosing whether or not to exchange $+ \leftrightarrow -$ on all the operators which touch that face.  Finally, superpose over all the membrane operators we have formed from a given reference configuration. 

The operator choices for the fractons and twist strings are given in Table~\ref{tab:XCubeMembranes}. Some example terms in the superposition for the excitation $[f_0]$ are given in Figs.~\ref{fig:membraneAPlusB} and \ref{fig:membraneAPlusB_ppmm}.
\begin{table*}
\centering
\renewcommand{\arraystretch}{1.3}
\begin{tabular}{@{}lllllllllllllll}
\toprule[2pt]
\textbf{Label} && \textbf{Excitation type}&& $\mathcal{U}_+$ && $\mathcal{U}_-$ && $\mathcal{V}_+$ && $\mathcal{V}_-$ && $\mathcal{W}_+$ && $\mathcal{W}_-$ \\ \hline
$f^{(1)}_{0}f^{(2)}_{0}$ && Abelian fracton && 1 && 1 && $\sx^{(1)}\sx^{(2)}$ && $\sx^{(1)}\sx^{(2)}$ && 1 && 1 \\
$[f_0]$ && non-Abelian fracton && 1 && 1 && $\sx^{(+)}$ && $\sx^{(-)}$ && 1 && $\tau_z$ \\
$\sigma$ && non-Abelian twist string && 1 && $\sz^{(1)}\sz^{(2)}$ && $1+\sx^{(1)}\sx^{(2)}$ && $1-\sx^{(1)}\sx^{(2)}$ && $\tau_x$ && $\tau_x$ \\	\bottomrule[2pt]	
\end{tabular}
\caption{Specification of membrane operators which create fractons and twist defects in the gauged bilayer X-cube model. The layout of the operators $\mathcal{U}$,$\mathcal{V}$, and $\mathcal{W}$ is shown in Fig.~\ref{fig:membraneOperatorLayout}. The excitation $\phi f_0^{(1)}f_0^{(2)}$ can be created using (flexible) string operators to create $\phi$ (in pairs) and fusing them onto $f_0^{(1)}f_0^{(2)}$ excitations created by the membrane operator given in the table.}
\label{tab:XCubeMembranes}
\end{table*}

Finally, the flexible membrane operators which create twist strings $\sigma$ are constructed similarly to the membranes which create fractons, but with some important differences. First, rather than only choosing between $\mathcal{V}_{\pm}$ at the corners of the membrane in the reference configuration, we may choose $\mathcal{V}_+$ or $\mathcal{V}_-$ \textit{anywhere} on the boundary of the membrane operator. This is reasonable because the string excitation it creates may generally have a local degree of freedom at each point on the string. Second, instead of interchanging $+ \leftrightarrow -$ on all operators touching a face within the membrane, we interchange $+ \leftrightarrow -$ on all operators touching \textit{cubes} within the membrane. Each cube affects eight $\mathcal{V}$s, four $\mathcal{U}$s, and eight $\mathcal{W}$s. Otherwise, the construction is the same. The appropriate operators for $\sigma$ are listed in Table~\ref{tab:XCubeMembranes}.

\subsection{String excitations and geometry-dependent degeneracy}
\label{subsec:stringDegen}

In this subsection, we show that a string excitation $\sigma_1$ is associated with a topological degeneracy that scales as $\sim 2^{2(\ell_x+\ell_y+\ell_z-1)}$ where $\ell_i$ is the linear extent of that excitation in the $i$ direction. This is qualitatively different from conventional topological order in (3+1)D, where non-Abelian string defects carry a degeneracy that is independent of their size and shape. We demonstrate this first by considering an extrinsic twist defect in the ungauged model. After gauging, the extrinsic twist defect becomes one of the string excitations. We will first compute the degeneracy in the ungauged case, with the degeneracy after gauging following straightforwardly.

For simplicity, we consider our system to be a large but finite-size cube. We choose boundary conditions where the one-dimensional ``$e$" particles (violations of star terms in $H$) in both layers of the original X-cube model are condensed at the boundary (``rough" boundary conditions). It is straightforward to check (in the wavefunction picture) that in the absence of excitations, the ground state is unique because there are no topologically nontrivial strings in the $m$ sector (associated to violations of the cube terms in $H$). Hence, any degeneracy will be associated with excitations. The exact degeneracy will depend on boundary conditions, as it does for conventional topological order, but the quantum dimension is independent of the boundary conditions.

\begin{figure}
\centering
\subfloat[\label{fig:braidAroundString}]{\includegraphics[width=0.4\columnwidth]{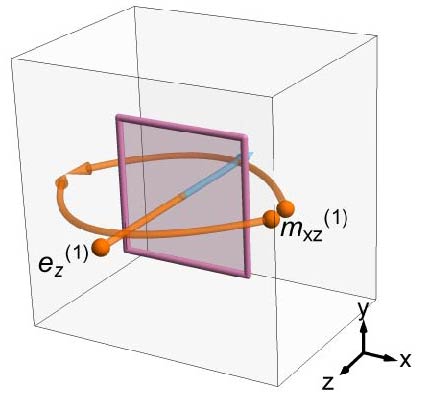}} \hspace{0.1\columnwidth}
\subfloat[\label{fig:independentLinesTwistDefect}]{\includegraphics[width=0.4\columnwidth]{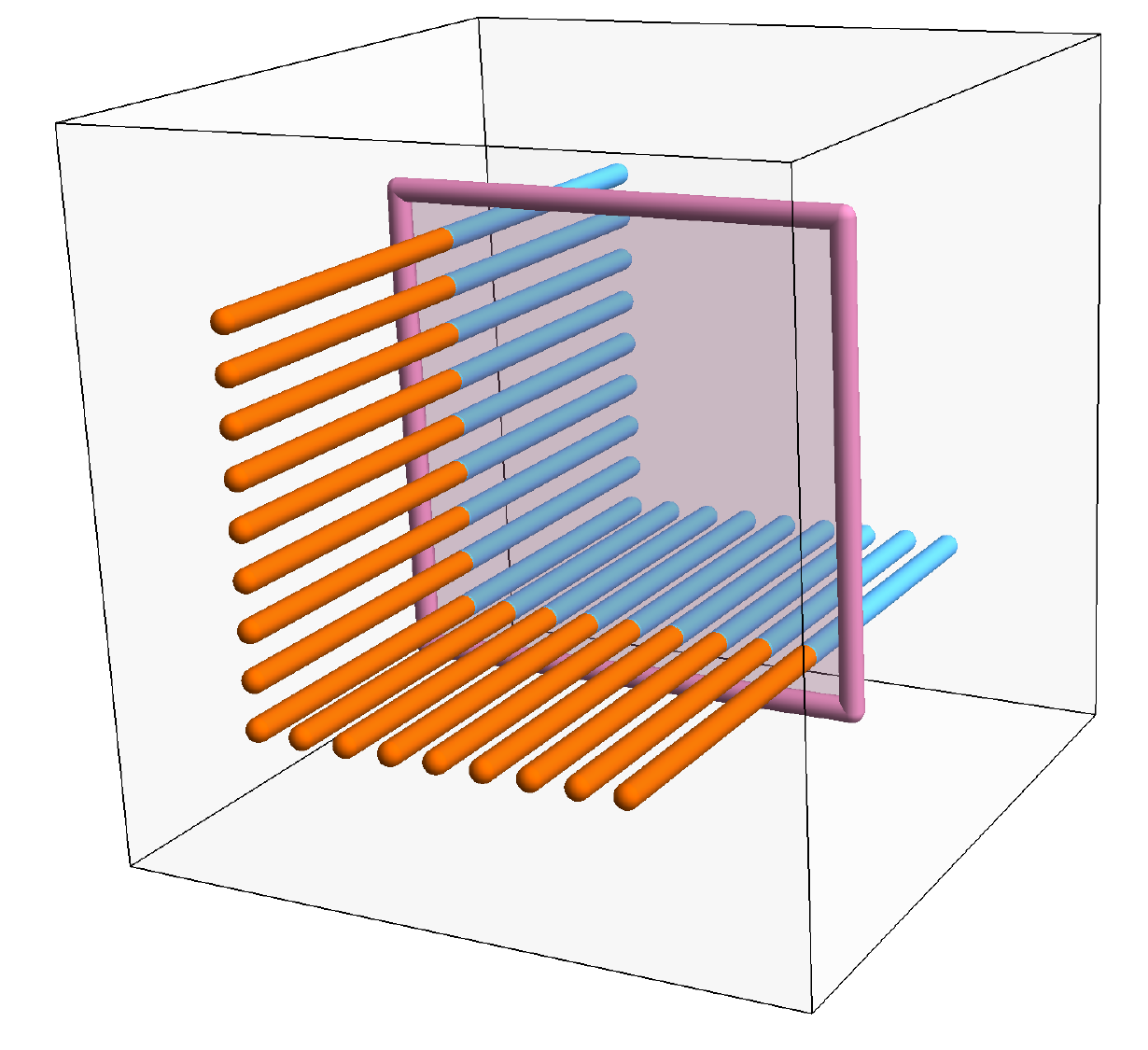}}\\
\subfloat[\label{fig:extraELines}]{\includegraphics[width=0.4\columnwidth]{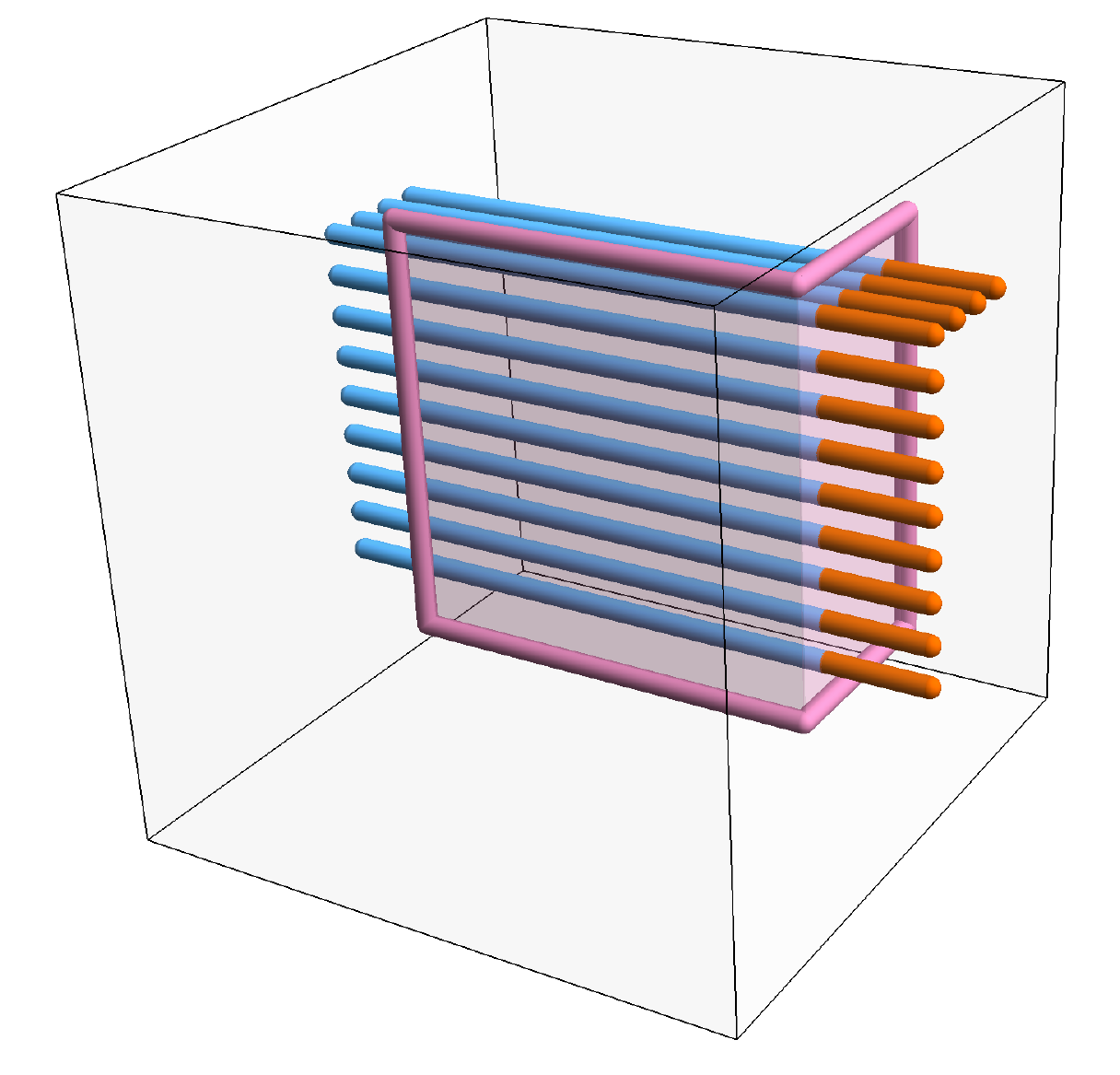}}
\caption{(a) Wilson lines in the presence of an extrinsic string defect (purple) in the bilayer X-Cube model with open boundary conditions where the $e$ excitations are condensed. The $e$ Wilson line consists of $\sz^{(i)}$ operators, where $i=1$ when the line is orange and $i=2$ when the line is blue, and the $m^{(1)}_{xz}$ Wilson line consists of $\sx^{(1)}$ operators. (b) Independent $e_z$ Wilson lines in the presence of a twist defect. The colors may be exchanged to find another set of independent Wilson lines. (c) Additional $e$ Wilson lines which appear when the string bends.}
\end{figure}

Consider first two decoupled copies of the X-cube model with an extrinsic $\ell_x \times \ell_y$ twist defect string inserted, as in Fig.~\ref{fig:braidAroundString}. We presently set notation. Call violations of the $A_s^{(i)}$ terms (see Eq. \eqref{eqn:ungaugedXCube}) as $e^{(i)}_{p}$, where $i=1,2$ labels a layer and $p=x,y,z$ labels the direction of motion of these one-dimensional particles. Bound states of excitations of two face-sharing $B_c^{(i)}$ terms are two-dimensional particles, which we call $m^{(i)}_{pq}$, where $pq$ means that this excitation is mobile in the $pq$-plane.

Note that $m_{pq}^{(i)}$ quasiparticles have closed Wilson loop operators (closed strings of $\sz^{(i)}$) associated to creating a pair of quasiparticles, moving them in a loop, and re-annihilating them. Because $m_{pq}^{(i)}$ is mobile only in the $pq$-plane, Wilson loops can be deformed in the $pq$-plane but not out of that plane. On the other hand, $e^{(i)}_p$ can form Wilson lines by creating a pair in the bulk, moving them to the boundary, and annihilating them. These Wilson lines are completely rigid because $e^{(i)}_p$ is only mobile in one dimension.

With the notation set, consider a closed Wilson loop for an $m^{(i)}_{xz}$ particle which surrounds the defect string, as shown in Fig.~\ref{fig:braidAroundString}. This Wilson loop \textit{cannot} be deformed to the identity because, due to the subdimensional nature of the excitations, the Wilson loop cannot be deformed out of the $xz$-plane. This is in contrast with conventional topological order, where, thanks to the full mobility of the particles, Wilson loops can be deformed arbitrarily. Such a Wilson loop anticommutes with any $e_z$ Wilson line which passes through the defect in the same $xz$-plane because the $e_z$ particle changes layers when it passes through the defect, as shown in Fig.~\ref{fig:braidAroundString}. There are therefore $2\ell_y$ nontrivial $m$ Wilson loops in the presence of the defect (the $2$ counts layers). The same holds for Wilson loops for $m^{(i)}_{yz}$ particles, so in total there are $2(\ell_x+\ell_y)$ nontrivial $m$ Wilson loops.

However, not all of these Wilson loops address independent states; we must count independent $e_z$ Wilson lines as well. The constraint discussed earlier also applies to Wilson lines which pass through the defect - the product of any four Wilson lines at the corners of a rectangle must equal the identity. Therefore, there are only $2(\ell_x+\ell_y-1)$ independent $e_z$ Wilson lines, as shown in Fig.~\ref{fig:independentLinesTwistDefect}, and each of them anticommutes with (at least) one $m$ Wilson line. The degeneracy in the presence of the defect is therefore $2^{2(\ell_x+\ell_y-1)}$.

We now wish to examine what happens when the loop bends in the $z$-direction as well, say in the $yz$-plane as in Fig.~\ref{fig:extraELines}. It is easy to see that $2(\ell_z + \ell_y-1)$ independent $e_x$ Wilson lines pass through the defect. Likewise, there are $2\ell_z$ additional $m_{xy}$ Wilson lines that surround the defect, which each anticommute with an $e_x$ Wilson line. However, there is no additional $m_{xz}$ Wilson loop which anticommutes with the other $e_x$ Wilson lines; those are already being ``used" as logical operators which anticommute with the $e_z$ Wilson lines. We therefore get an additional degeneracy of only $2^{2\ell_z}$, leading to a total degeneracy of $2^{2(\ell_x+\ell_y+\ell_z-1)}$.

\begin{figure}
\centering
\subfloat[\label{fig:twoStringLogicals1}]{\includegraphics[width=0.9\columnwidth]{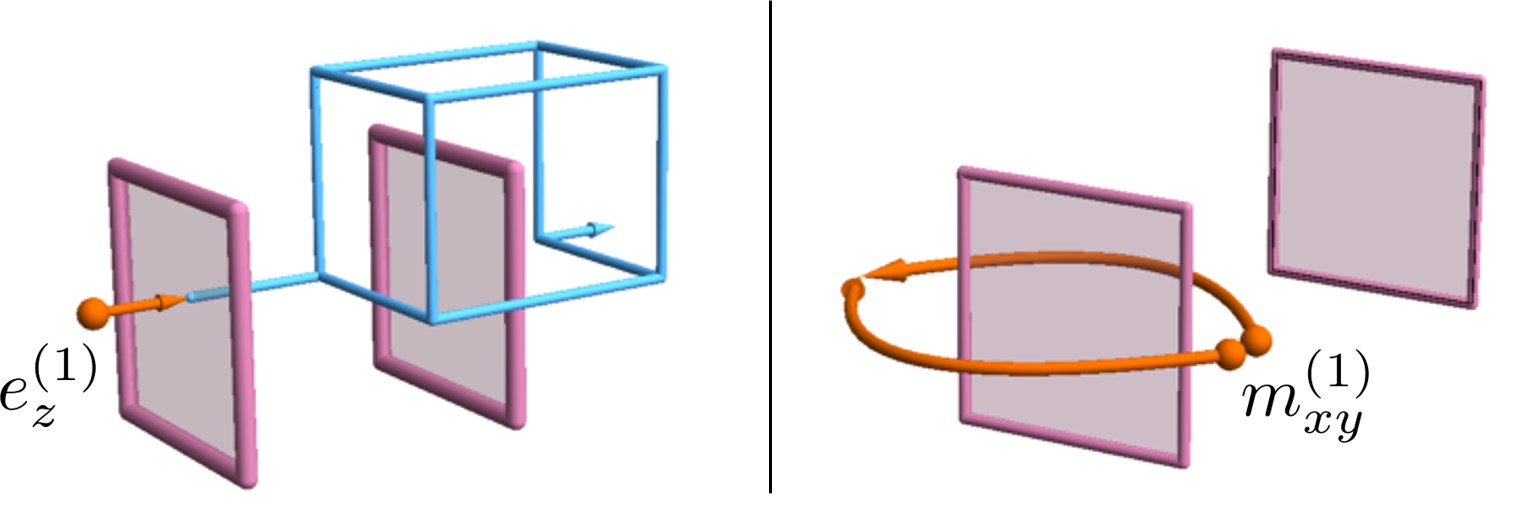}}\\
\subfloat[\label{fig:twoStringLogicals2}]{\includegraphics[width=0.9\columnwidth]{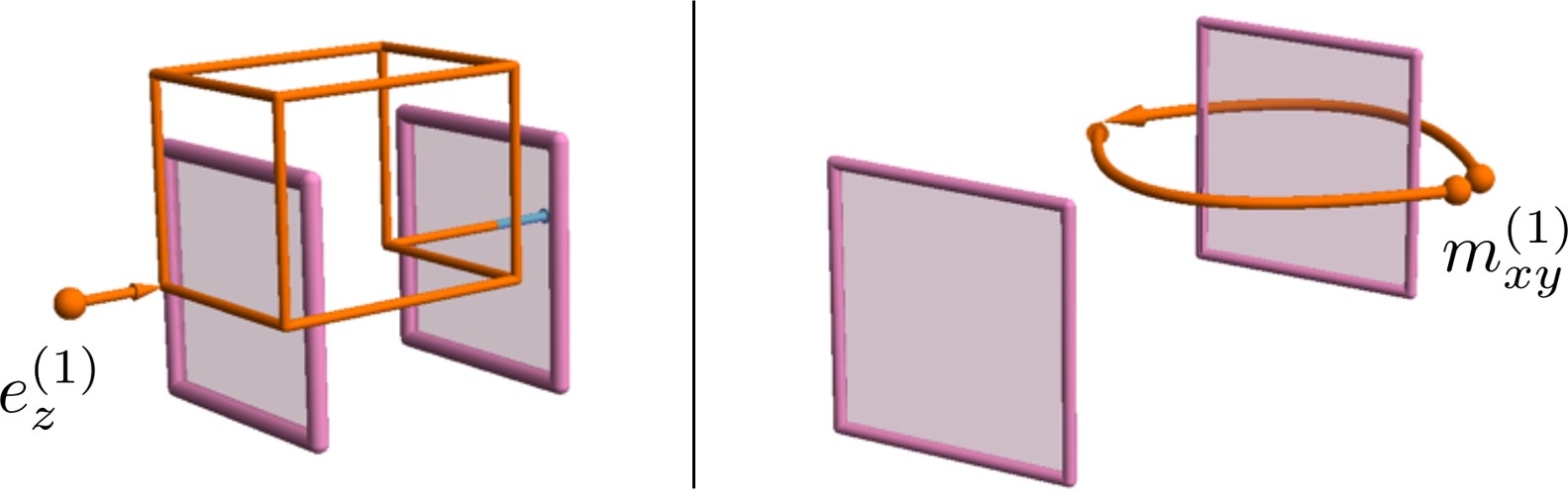}}\\
\subfloat[\label{fig:linkedString1}]{\includegraphics[width=0.9\columnwidth]{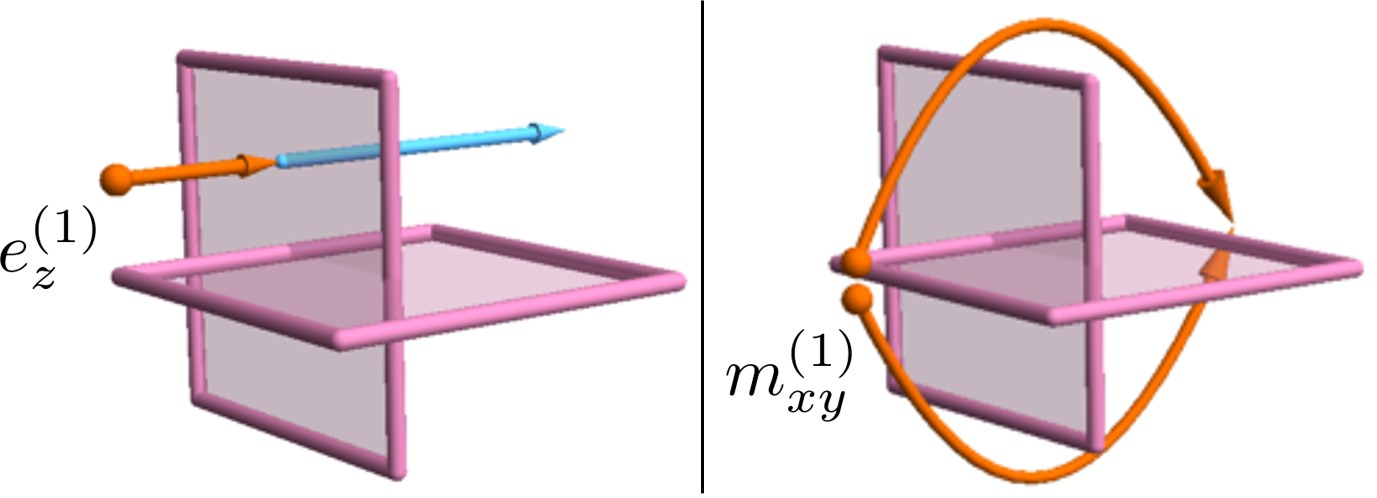}}\\
\subfloat[\label{fig:linkedString2}]{\includegraphics[width=0.9\columnwidth]{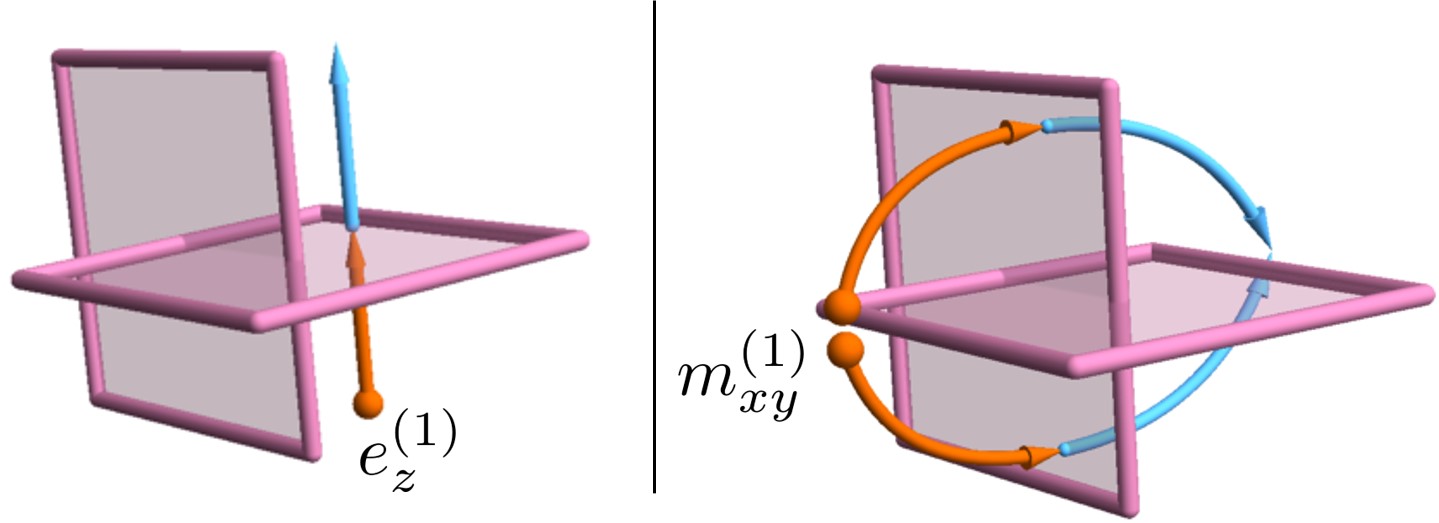}}
\caption{Wilson ``lines" in the presence of two twist defect strings (purple) when the strings are separated ((a) and (b)) and linked ((c) and (d)). The implication is that the $e_z^{(1)}$ particles are brought in from the boundary and annihilated on the opposite boundary, where we have chosen open boundary conditions with $e$ excitations condensed everywhere. (The details of the Wilson operators depend on the boundary conditions, although the number of independent Wilson operators does not.) Operators within a figure anticommute and operators in different figures (with the same defect configurations) commute.}
\end{figure}

Next, we briefly discuss what happens in the presence of multiple defect strings. If the strings are not linked, it is not too difficult to see that an independent set of $m$ Wilson lines for each string can be constructed, on the right side of Figs.~\ref{fig:twoStringLogicals1} and \ref{fig:twoStringLogicals2}. There is also an independent set of $e$ ``Wilson lines" for each string; in particular, by using the fact that $e_x$ can split into an $e_y$ and an $e_z$, the $e$ particles can ``go around" a string defect in the way shown on the left side of Figs.~\ref{fig:twoStringLogicals1} and \ref{fig:twoStringLogicals2}. Therefore, the degeneracies associated to each string are independent. 

If the strings are linked, the degeneracies are still independent; this time, the fact that the $e$ Wilson lines are distinct follows straightforwardly from the geometry, as in the left sides of Figs.~\ref{fig:linkedString1} and \ref{fig:linkedString2}, and the independent $m$ Wilson lines are obtained by wrapping around either one or both strings, as shown on the right sides of Figs.~\ref{fig:linkedString1} and \ref{fig:linkedString2}. Therefore, since each defect contributes independently to the degeneracy regardless of the relative configuration of the strings, the quantum dimension of the string defect is really $2^{2(\ell_x+\ell_y+\ell_z-1)}$ .

Finally, we promote our understanding of the ungauged model to the degeneracy in the gauged model. Before gauging, the number of degenerate states associated with a string excitation scales as $2^{2N} \equiv 2^{2(\ell_x+\ell_y+\ell_z-1)}$. Because the degeneracy is associated with the twist defect and not with the boundary conditions (although the boundary conditions can affect the total degeneracy), after gauging, all that happens is that we restrict to the states which are symmetric under a global SWAP. Following similar arguments used for the ground state degeneracy on the torus in Sec. \ref{subsec:GSDOutline}, there are $\sim (2^{2N-1}+2^{N-1}) \sim 2^{2N}$ states associated with the string. The quantum dimension is therefore $2^{2N}$ in the gauged model.

\subsection{Non-Abelian braiding analog for subdimensional particles}

Although a full theory of braiding for subdimensional particles has not yet been developed, we can
demonstrate that our model possesses an analog of non-Abelian braiding of subdimensional particles.

\begin{figure}
\centering
\subfloat[\label{fig:CDefinition}]{\includegraphics[width=0.9\columnwidth]{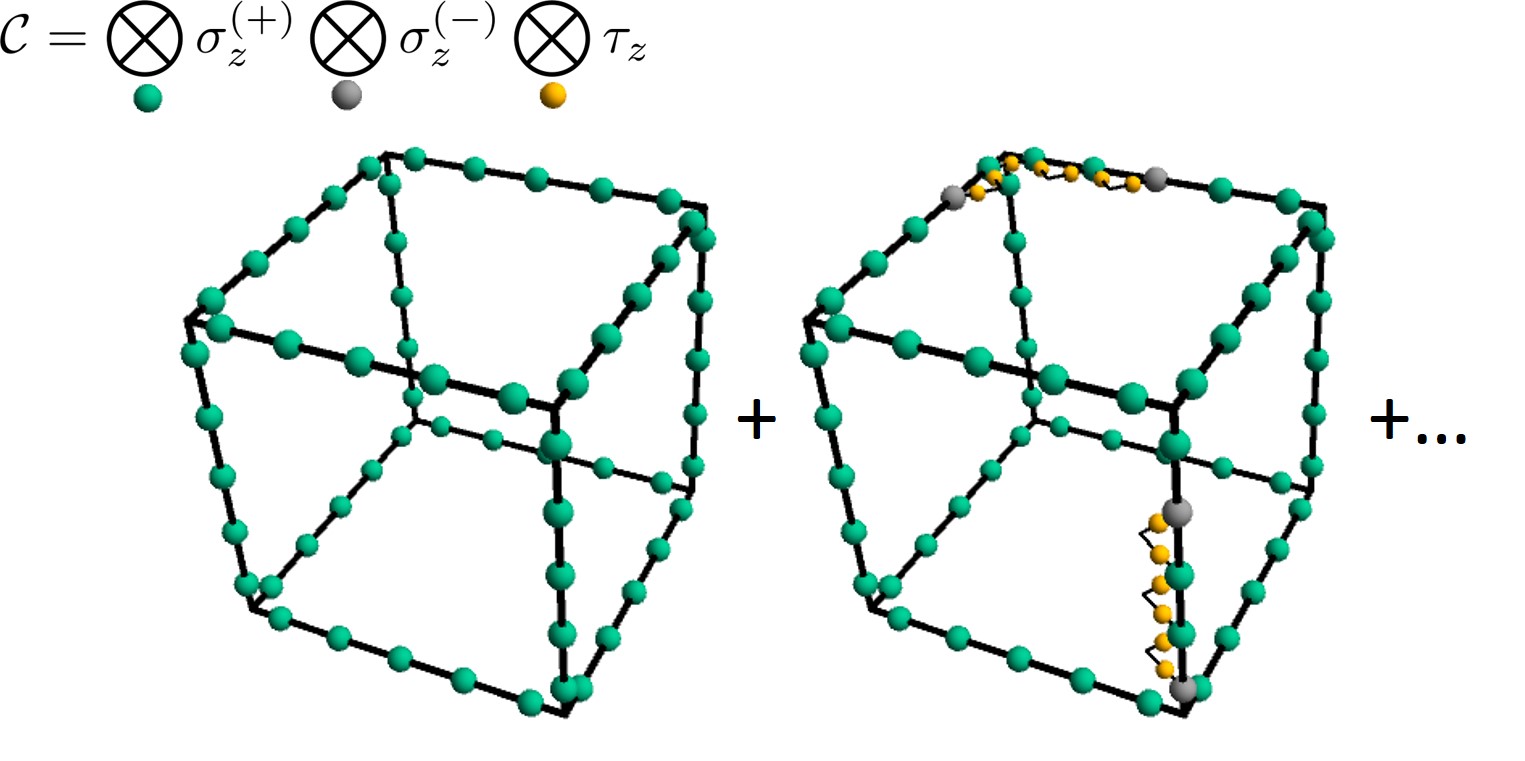}}\\
\subfloat[\label{fig:cageFractonBraid}]{\includegraphics[width=0.9\columnwidth]{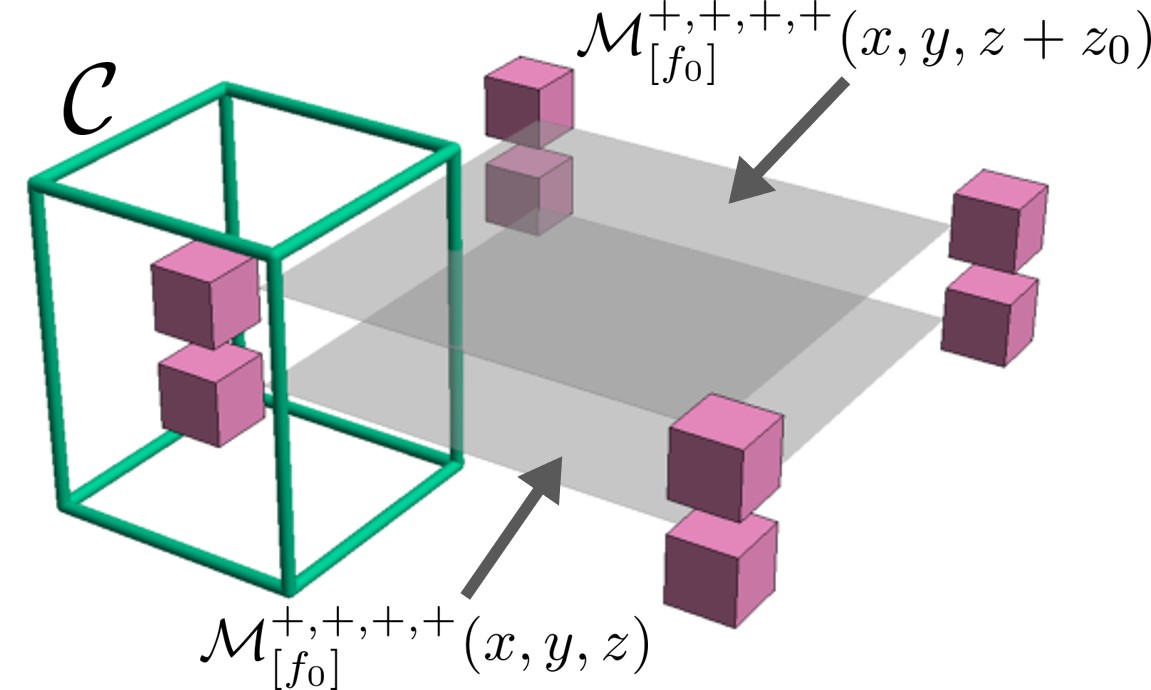}}
\caption{(a) Definition of the operator $\mathcal{C}$. (b) ``Braiding" of the cage $\mathcal{C}$ (green) around a pair of $[f_0]$ particles (purple). The fractons are all taken to be well-separated despite the fact the cubes representing the fractons are large in this figure.}
\end{figure}

Let $\ket{0}$ be a ground state, and consider the state $\ket{\psi} = \mathcal{M}_{[f_0]}^{+,+,+,+}(x,y,z)\mathcal{M}_{[f_0]}^{+,+,+,+}(x,y,z+z_0)\ket{0}$, where $\mathcal{M}_{[f_0]}(x,y,z)$ is a membrane in the $xy$-plane centered at $(x,y,z)$ and $z_0$ is large. Physically, this corresponds to creating eight $[f_0]$ excitations, one at each corner of a rectangular prism. One can check from the operator algebra (see Appendix \ref{app:bilayerXCubeExcitations}) that $\braket{\psi}{\psi} \neq 0$. Next, let $\mathcal{C}$ be a closed ``cage" operator for the $[e_i]$ particles; physically, we create from vacuum a triple of $[e_x]$, $[e_y]$, and $[e_z]$ at each corner of a wireframe cube, then move the particles and annihilate them pairwise. Explicitly, $\mathcal{C}$ is shown in Fig.~\ref{fig:CDefinition} and is simply an ``enlarged" version of $B_c$. One can check that $[\mathcal{C},H_{\text{gauged}}]=0$, and therefore $\mathcal{C}\ket{0} = \ket{0}$ (assuming that the linear size of $\mathcal{C}$ is small compared to the system size). One can also check explicitly that if $\mathcal{C}$ is positioned so that it surrounds exactly two fractons, as shown in Fig.~\ref{fig:cageFractonBraid}, then
\begin{equation}
\mathcal{C}(\mathcal{M}_{[f_0]}^{+,+,+,+})^2 = (\mathcal{M}_{[f_0]}^{+,+,+,-})^2\mathcal{C}
\end{equation}
Here the position indices on $\mathcal{M}_{[f_0]}$ are suppressed for legibility. It immediately follows that
\begin{equation}
\mathcal{C}\ket{\psi} = (\mathcal{M}_{[f_0]}^{+,+,+,-})^2\ket{0}
\end{equation}
so $\mathcal{C}\ket{\psi}$ and $\ket{\psi}$ have the same type of excitations at the same locations.  Hence (again using the operator algebra in Appendix \ref{app:bilayerXCubeExcitations})
\begin{align}
\bra{\psi}\mathcal{C}\ket{\psi} &=
\bra{0}(\mathcal{M}_{[f_0]}^{+,+,+,+})^2(\mathcal{M}_{[f_0]}^{+,+,+,-})^2\ket{0}\\
&=0
\end{align}
It is likewise straightforward to check that $\mathcal{C}\ket{\psi}$ has nonzero norm. Therefore, $\mathcal{C}$ performs a non-trivial unitary transformation on the set of degenerate states associated with the eight $[f_0]$ excitations created by $(\mathcal{M}_{[f_0]}^{+,+,+,+})^2$, such that $|\psi\rangle$ and $\mathcal{C} |\psi\rangle$ are orthogonal. This is precisely a consequence of an analog of non-Abelian braiding for subdimensional excitations.

\section{Gauged Bilayer Haah's Code}
\label{sec:Haah}

Our primary interest in gauging the layer-swap symmetry of Haah's code will be to show that there are non-Abelian fractal-type fractons in the model. (We call quasiparticles created at the corners of operators with fractal support, e.g. the quasiparticles in Haah's code, ``fractal-type". Our model does not fall cleanly into the ``type-I"/``type-II" dichotomy of Ref. \onlinecite{VijayFractons} because our model has both fractal-type fractons and a fully mobile quasiparticle, so we choose not to use that language.)  We will not attempt to discuss many more properties of the model in detail. This is either because they are extremely complicated (e.g. determining constraints between different fractal Wilson operators in order to compute the ground state degeneracy) or because they are not well-understood even for Abelian fractal-type fractons (e.g. the analog of particle-particle braiding processes).

\subsection{Construction of the model}

The model is constructed in an analogous way to the previous cases, although due to the Hilbert space of the ungauged model, the gauge qubits are a bit simpler than the X-Cube case.

\begin{figure}
\centering
\includegraphics[width=0.9\columnwidth]{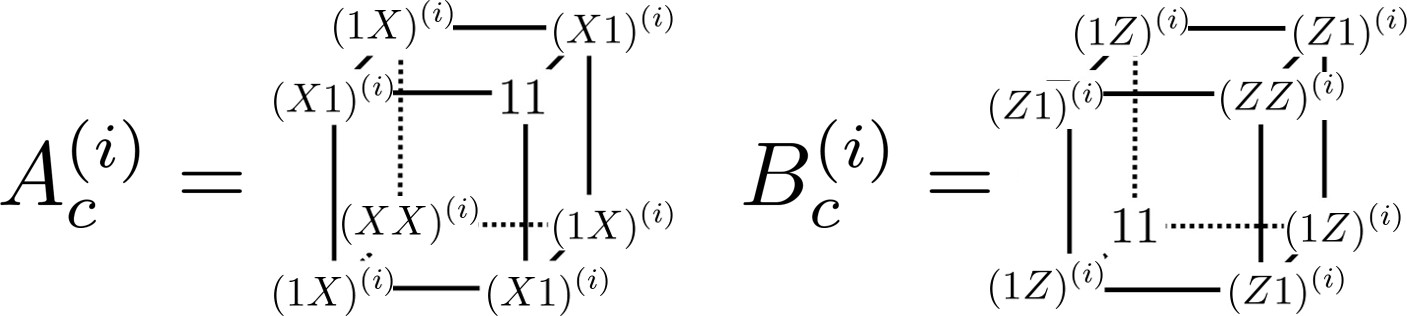}
\caption{Operators appearing in the ungauged bilayer Haah's code model. The Hilbert space consists of two qubits per layer for a total of four qubits per site. The Pauli operators are written as $X$ and $Z$ for legibility.}
\label{fig:HaahTerms}
\end{figure}

The Hilbert space of the ungauged model consists of four qubits per site, two per layer of Haah's code. The model has the Hamiltonian
\begin{equation}
H_{\text{ungauged}} = -\sum_{i,c} (A_c^{(i)}+B_c^{(i)})
\end{equation}
where $i$ labels a layer and $A_c^{(i)}$ and $B_c^{(i)}$ are eight-qubit operators shown in Fig.~\ref{fig:HaahTerms}. The notation is that, for example, $(XX)^{(i)}$ means a Pauli $X$ operator on both spins in the $i$th layer.

\begin{figure*}
\centering
\subfloat[\label{fig:HaahDecomp}]{\includegraphics[width=1.7\columnwidth]{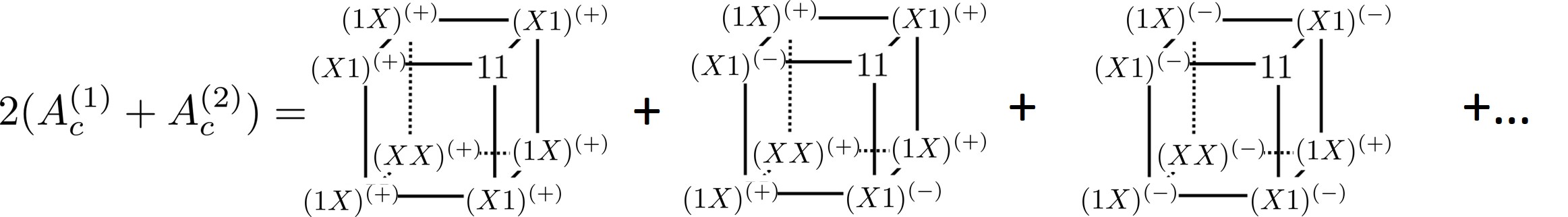}}\\
\subfloat[\label{fig:gaugedHaahHilbertSpace}]{\includegraphics[width=0.4\columnwidth]{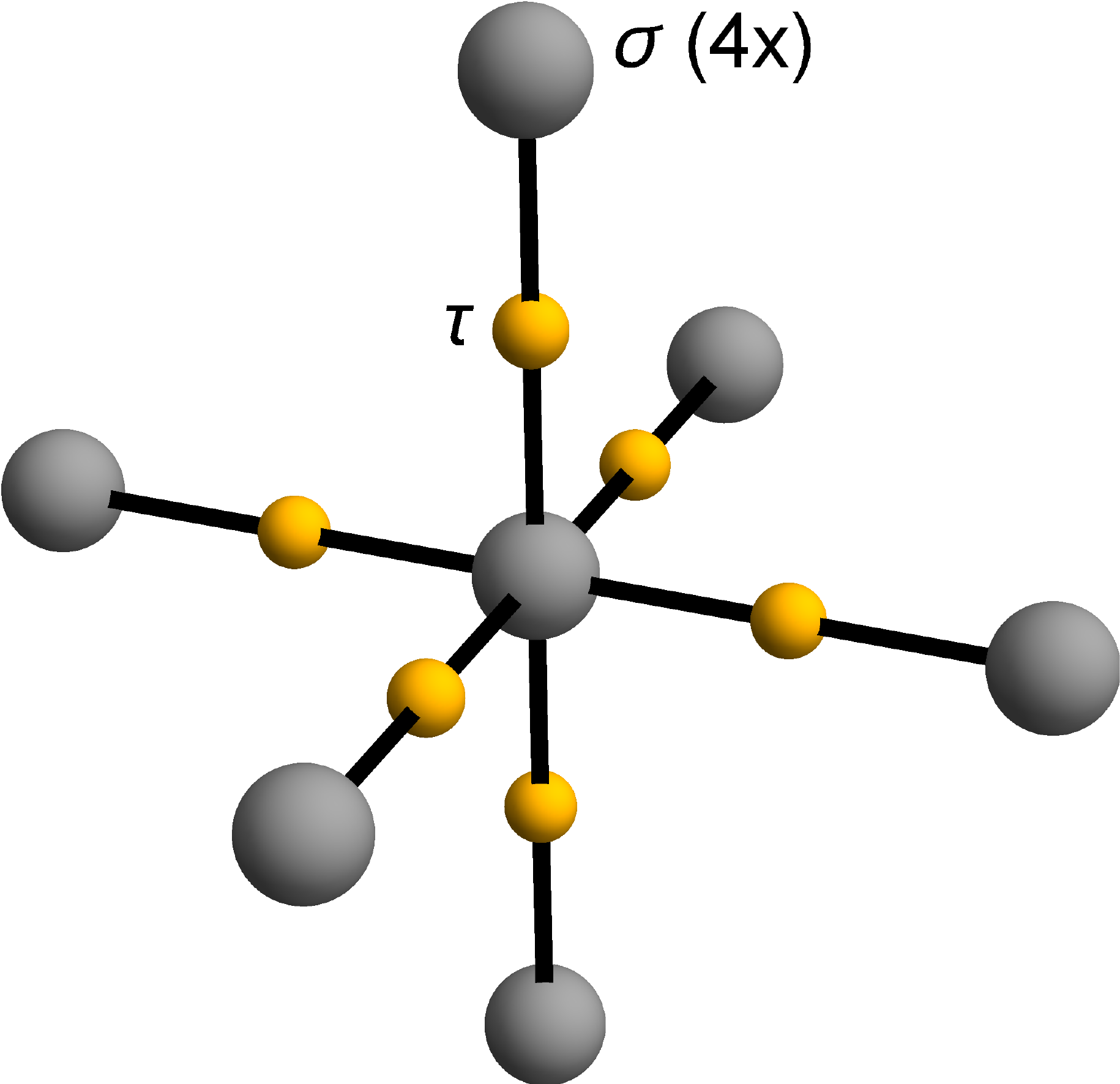}}
\hspace{0.3\columnwidth}
\subfloat[\label{fig:HaahGenerator}]{\includegraphics[width=0.4\columnwidth]{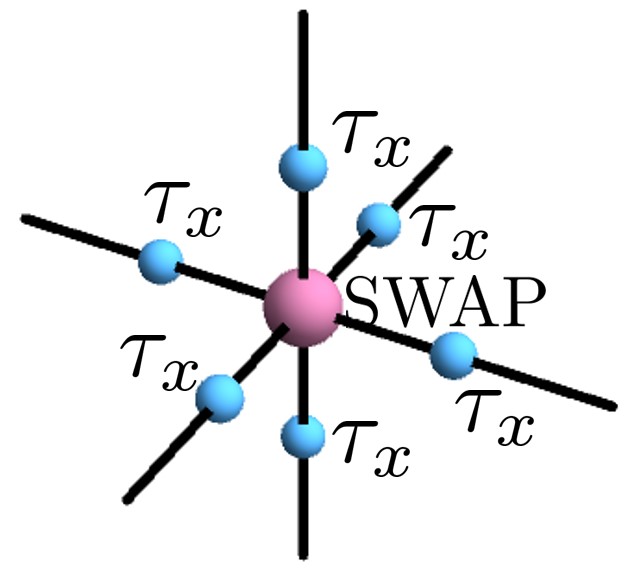}}
\\
\subfloat[\label{fig:gaugedHaahTerms}]{\includegraphics[width=1.7\columnwidth]{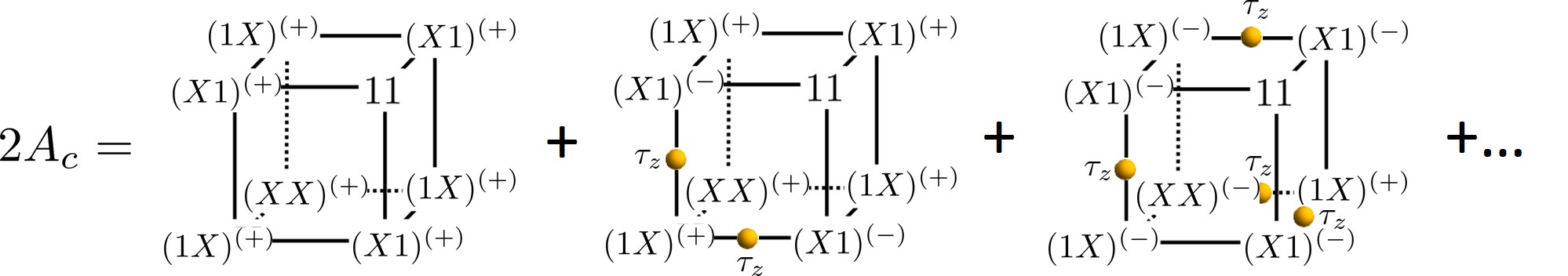}}
\caption{Gauging procedure for bilayer Haah's code. (a) Decomposition of the ungauged Hamiltonian terms $A_c^{(i)}$ into terms even and odd under local SWAPs (of the sort in Eq. \eqref{eqn:HaahSWAP}). The sum is over all $64$ terms with an even number of $(-)$ operators; three are shown. (b) Hilbert space arrangement for the gauged model. There are four ``matter" spins (dark grey) per site of the lattice and one ``gauge" spin (orange) per link of the lattice. (c) Generator $C_s$ of local gauge transformations. (d) Gauged cube term $A_c$. The only difference from (a) is the addition of $\tau_z$ operators (orange) acting on gauge spins.}
\end{figure*}

We pick the particular SWAP operator which exchanges \textit{both} qubits in layer 1 with both qubits in layer 2, that is, on a single site 
\begin{equation}
\SWAP\ket{\psi_1 \psi_2}_1 \otimes \ket{\phi_1 \phi_2} = \ket{\phi_1 \phi_2}_1 \otimes \ket{\psi_1 \psi_2}_2
\label{eqn:HaahSWAP}
\end{equation}
where the outer subscript labels a layer and the inner subscript labels a spin within a layer. Obviously a global SWAP of this type is a global symmetry of the ungauged model. As before, we can decompose the Hamiltonian into a sum of terms which are even or odd under onsite SWAPs; $64$ terms appear, a few examples of which are shown in Fig.~\ref{fig:HaahDecomp}. Each is invariant under a global SWAP. We have defined the $(\pm)$ superscripts as before; for any operator $\mathcal{O}^{(i)}$ within a layer,
\begin{equation}
\mathcal{O}^{(\pm)} = \frac{1}{\sqrt{2}}\left(\mathcal{O}^{(1)} \pm \mathcal{O}^{(2)}\right)
\end{equation}

To gauge the symmetry, we again add gauge qubits $\tau$, this time one per link of the cubic lattice. We then demand that all terms in the Hamiltonian commute with the local symmetry generator 
\begin{equation}
C_s = \SWAP_s \bigotimes_{\text{star}} \tau^x
\end{equation}
where $s$ labels a site. This operator is shown pictorially in Fig.~\ref{fig:HaahGenerator}. This is, as before, achieved by adding $\tau^z$ operators on paths which connect disjoint pairs of SWAP-odd operators. Some examples are shown in Fig.~\ref{fig:gaugedHaahTerms}. This is always possible because the ungauged Hamiltonian is invariant under a global SWAP; this means that the number of operators in a term which are odd under onsite SWAP is always even.

Finally, we define the usual $\mathbb{Z}_2$ flux operators which are associated with plaquettes $p$ of the lattice
\begin{equation}
D_p = \bigotimes_{\bv{r} \in p} \tau_{z,\bv{r}}  
\end{equation}
With these definitions, the gauged model has the Hamiltonian
\begin{equation}
H_{\text{gauged}}= -\sum_{c}(A_c+B_c) -\sum_s C_s - \sum_p D_p
\end{equation}
where $A_c$ and $B_c$ are the gauged versions of the ungauged cube operators, each of which contains $64$ different terms of the sorts shown in Fig.~\ref{fig:gaugedHaahTerms}. 

\subsection{Excitations and non-Abelian fractal-type fractons}

\begin{table*}
\centering
\renewcommand{\arraystretch}{1.3}
\begin{tabular}{@{}lllllll}
\toprule[2pt]
\textbf{Label} && \textbf{Excitation}&& \textbf{Mobility} &&\textbf{$d_a$}\\ \hline
$\phi$ && Point $\mathbb{Z}_2$ charge && 3D && 1 \\
$a^{(1)}a^{(2)}$ && Abelian violation of $A_c$ constraint && fractal-type fracton && 1 \\
$b^{(1)}b^{(2)}$ && Abelian violation of $B_c$ constraint && fractal-type fracton && 1 \\
$[a]$ && non-Abelian violation of $A_c$ constraint && fractal-type fracton && 2? \\
$[b]$ && non-Abelian violation of $B_c$ constraint && fractal-type fracton && 2? \\
$\sigma$ && Flux (twist) string && Flexible string excitation&& Unknown
\\	\bottomrule[2pt]	
\end{tabular}
\caption{Nontrivial simple excitations in the gauged bilayer Haah's code model, their mobility, and their quantum dimensions $d_a$. }
\label{tab:HaahExcitations}
\end{table*}

Each layer of the ungauged $\mathbb{Z}_2$ Haah's code model has three fracton excitations corresponding to $A_c^{(i)}=-1$, which we shall call $a^{(i)}$, $B_c^{(i)}=-1$, which we shall call $b^{(i)}$, and the ``dyonic" bound state $(ab)^{(i)}$. Upon gauging, the point-like particles are organized into symmetry orbits $a^{(1)}a^{(2)}$, $b^{(1)}b^{(2)}$, $[a]$, $[b]$, and their dyonic bound states. The first two are Abelian fractal-type fractons and the last two are non-Abelian fractal-type fractons. There is also be a fully mobile $\mathbb{Z}_2$ point charge $\phi$ and a twist string excitation $\sigma$. The properties of these particles are tabulated in Table~\ref{tab:HaahExcitations}. 

\begin{figure}
\centering
\subfloat[\label{fig:smallTetrahedra}]{\includegraphics[width=0.4\columnwidth]{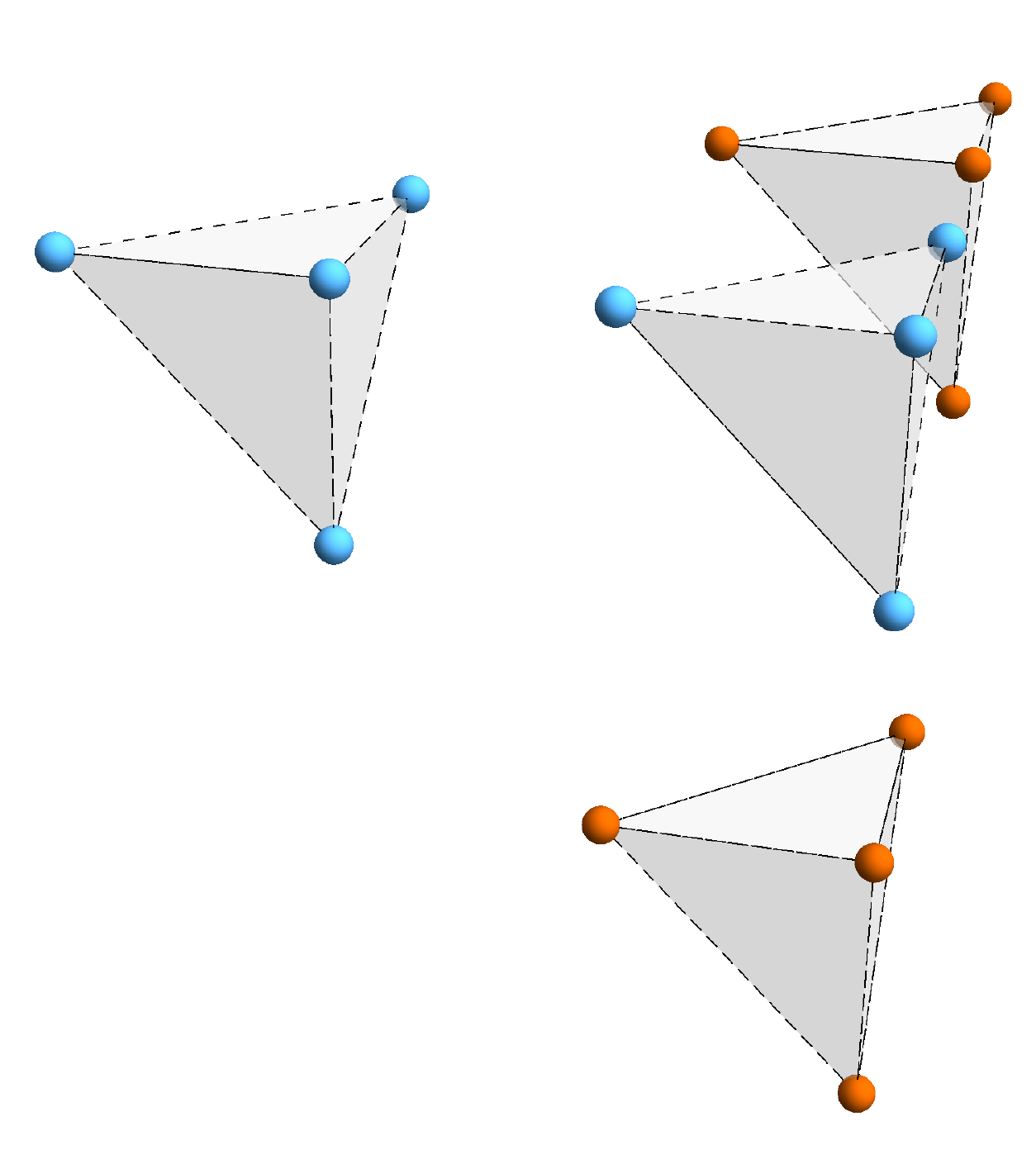}}\hspace{0.1\columnwidth}
\subfloat[\label{fig:largeTetrahedra}]{\includegraphics[width=0.4\columnwidth]{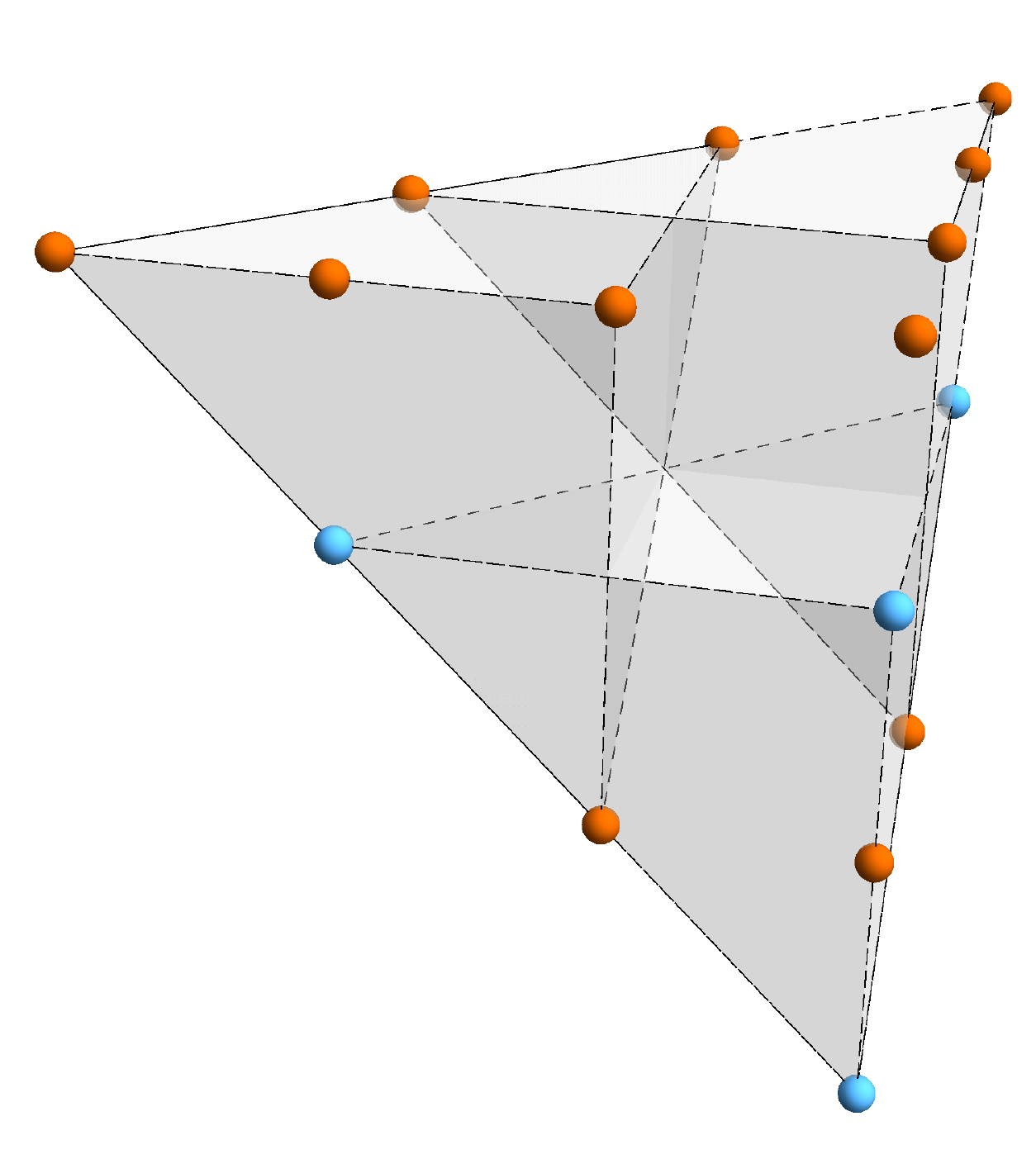}}\\
\subfloat[\label{fig:aQuantumDimension}]{\includegraphics[width=0.4\columnwidth]{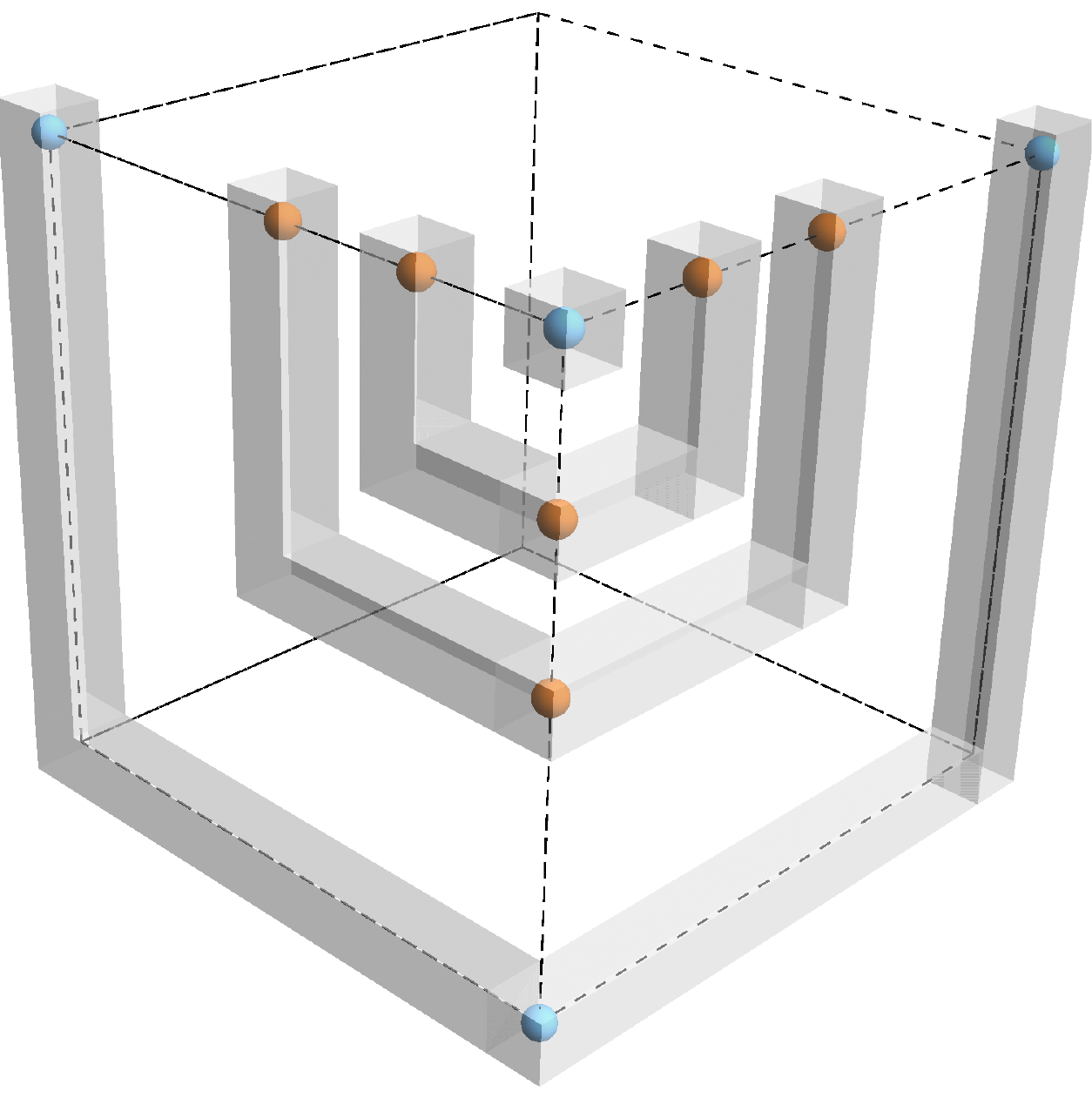}}
\caption{Reference cage-net-like configurations for $[a]$ fractons (spheres) in the gauged bilayer Haah's code. In these configurations which lack branch membranes, fractons can be assigned definite layers (orange and blue for layers 1 and 2 respectively). Grey tetrahedra have fractons at their corners and are guides to the eye (one not shown for legibility in (b)). Configuration (b) is only possible when the four small tetrahedra of fractons are arranged at the corners of a larger tetrahedron. (c) Geometry used to compute the quantum dimension of $[a]$. Fractons within each grey prism fuse to the same superselection sector.}
\end{figure}

We are not aware of a cage-net construction for Haah's code, so we are mostly forced to work entirely with the explicit operators. However, before doing so, we can demonstrate that the $[a]$ and $[b]$ excitations do carry nontrivial (presumably configuration-dependent) topological degeneracy, although we only know how to lower-bound this degeneracy. The reason is that although we do not have a cage-net picture for Haah's code, the $\mathbb{Z}_2$ gauge sector of the theory can still be represented by a membrane-net, so we can still consider ``reference" configurations where the membranes are all absent. In such configurations, the excitations have a definite layer index. In these configurations, the Hilbert space-level constraints descend from those of the ungauged model. This means that in these ``reference" configurations, four fractons of a fixed color at the corners of (Sierpinski) tetrahedra of side length $2^n$ for integer $n$ is definitely a valid configuration, just as they are in the ungauged model. If $N_t$ such tetrahedra of fractons are present, then there are at least $2^{N_t-1}$ reference configurations allowed up to global SWAPs wherein all fractons on a given tetrahedron have the same color. An example reference configuration is shown in Fig.~\ref{fig:smallTetrahedra}. This is sufficient to show that the fractons are non-Abelian.

Because it is not simple to understand the full set of Hilbert space constraints, in an arbitrary geometry it is not clear if there are additional configurations which are allowed. However, some geometries definitely admit additional configurations, in particular when four tetrahedra are placed at the corners of a larger tetrahedron. In this geometry there are multiple ways to group the fractons into tetrahedra, leading to additional reference configurations such as that in Fig.~\ref{fig:largeTetrahedra}. In this geometry, the two ways of grouping sixteen fractons into four tetrahedra leads to a lower bound on the degeneracy of $2^4-1=15$ states, compared to the $8$ states for four tetrahedra at generic positions. This strongly suggests, although we cannot prove rigorously, that the degeneracy associated with the non-Abelian fractons depends strongly on the geometry in which they are created. 

The quantum dimension of the non-Abelian fractons can be lower-bounded in a similar manner. Since fractons can be created on the corners of tetrahedra, in the geometry shown in Fig.~\ref{fig:aQuantumDimension}, all of the grey boxes contain excitations in the same superselection sector as the box with a single $[a]$ fracton in it. We expect that the only valid reference configurations are where all fractons in a given grey box are of the same color. Hence the degeneracy in this configuration is lower bounded by $2^{N_b-2}$, where $N_b$ is the number of grey boxes, and the quantum dimension of the $[a]$ and $[b]$ fractons is lower bounded by 2. We expect that indeed $d_{[a]} = d_{[b]} =2$, but we need a full understanding of the constraints in order to prove this.

\begin{figure*}
\centering
\subfloat[\label{fig:HaahAbelian}]{\includegraphics[width=\columnwidth]{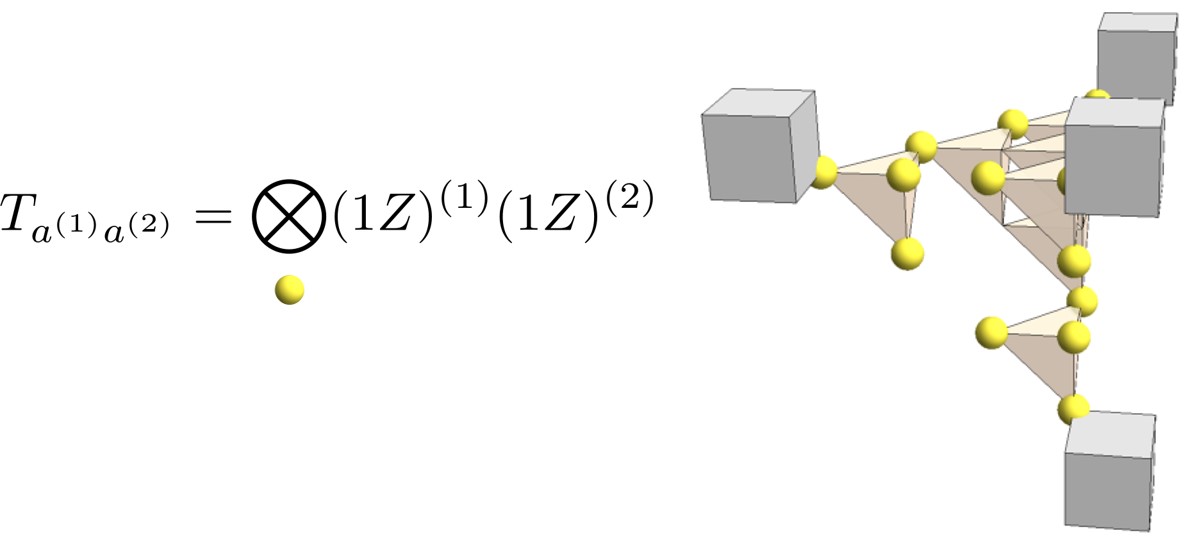}} \hspace{0.2\columnwidth}\\
\subfloat[\label{fig:HaahCommRelations}]{\includegraphics[width=1.5\columnwidth]{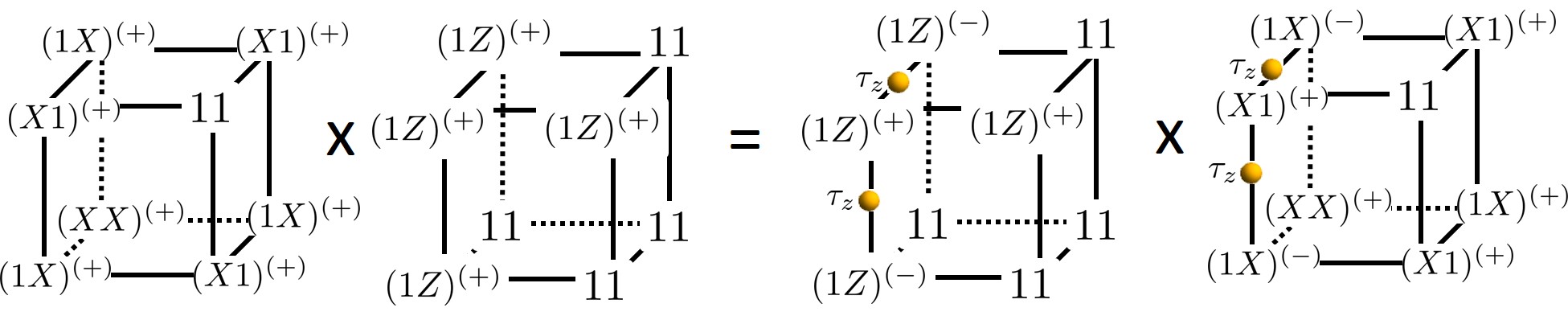}}\\
\subfloat[\label{fig:HaahNonAb}]{\includegraphics[width=1.5\columnwidth]{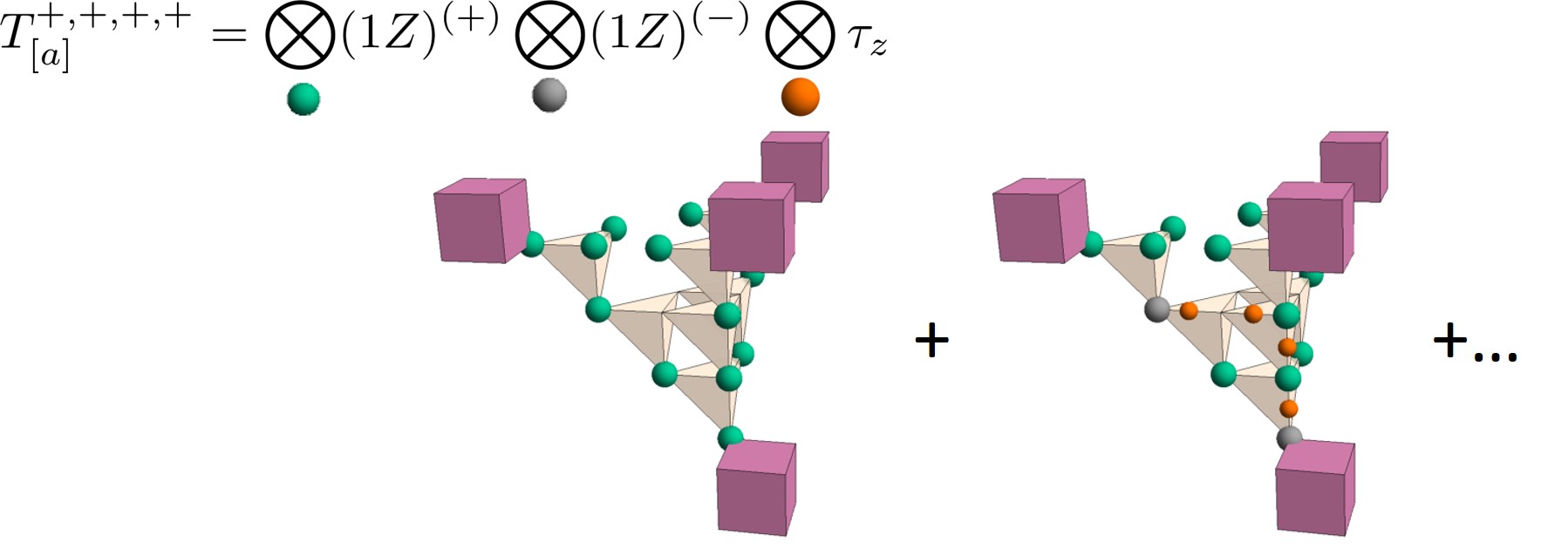}}\\
\subfloat[\label{fig:HaahZ1NonAb}]{\includegraphics[width=1.3\columnwidth]{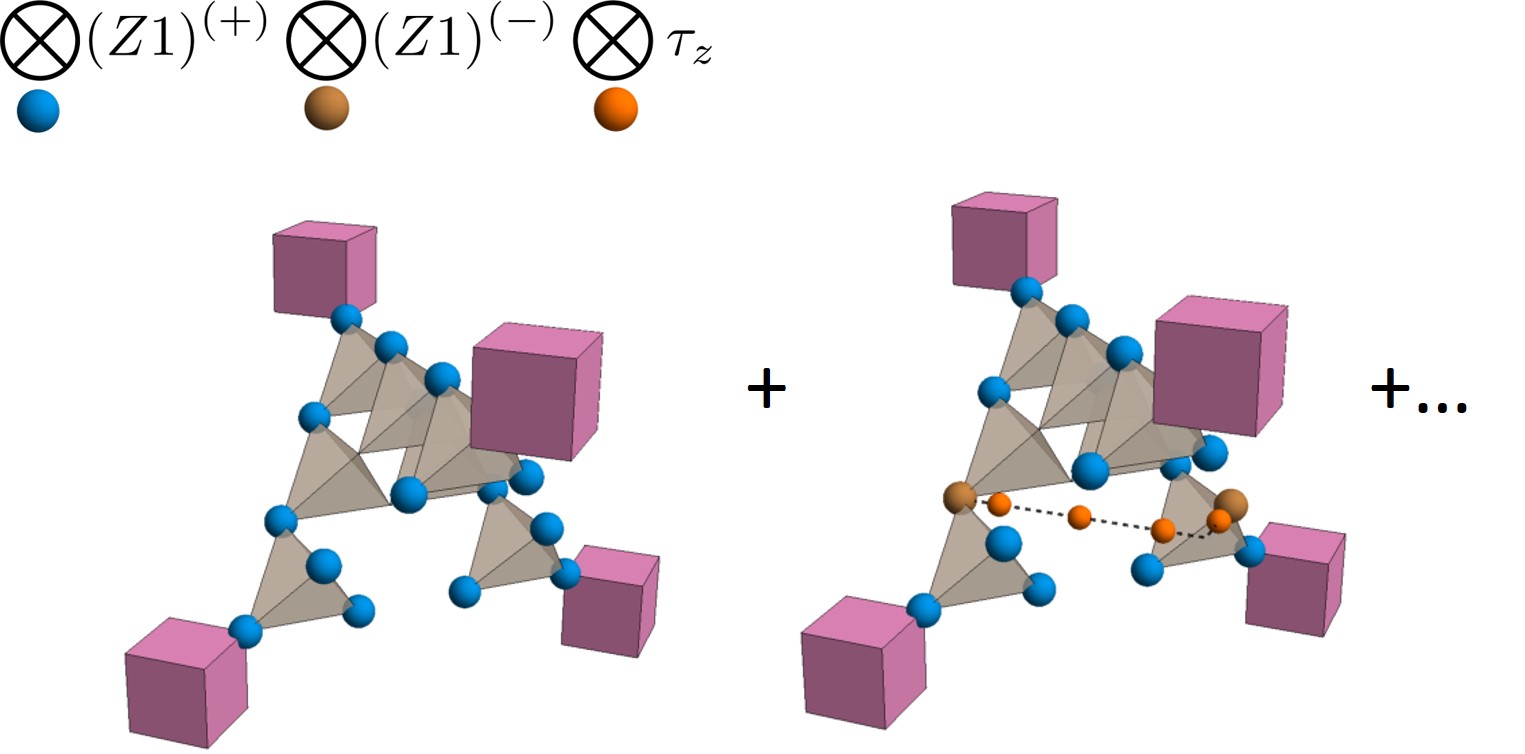}}
\caption{(a) Operator which creates Abelian fractons (grey) in Haah's code. (b) Commutation relations between one term in $A_c$ and a small piece of a Sierpinski prism operator. (c) and (d) Operators which create non-Abelian $[a]$ fractons (purple) in Haah's code in different geometries. The tetrahedra are guides to the fractal structure; the edges of the tetrahedra are not necessarily links of the cubic lattice. Hence in (d) the dashed line consists of several edges of the actual lattice. In (a), (c), and (d), these operators do not commute with the $A_c$ terms in the Hamiltonian at the location of the excitations.}
\end{figure*}

We now explicitly construct the operators which create the excitations. As usual, $\phi$ is created by a string of $\tau^z$ operators. The Abelian fractons, since they are manifestly SWAP-symmetric, are created in the same way as in the ungauged models; for example, $a^{(1)}a^{(2)}$ is created by acting with $(1Z)^{(1)}(1Z)^{(2)}$ on a Sierpinski tetrahedron as shown in Fig.~\ref{fig:HaahAbelian}. We call this operator $T_{a^{(1)}a^{(2)}}$.

One of our main results is the existence of the $[a]$ and $[b]$ fractons, that is, this model contains non-Abelian fractons which correspond to the symmetry orbit of a fracton in only one of the original layers. The operator creating $[a]$ is constructed in an analogous way to the non-Abelian excitations in the previous models. We begin with a ``reference" operator consisting of a Sierpinski tetrahedron of $(1Z)^{(+)}$ operators, as in the first term of Fig.~\ref{fig:HaahNonAb}. To ensure that there are only excitations at the corners of the strings, we observe what happens when commuting terms of $A_c$ past this operator; as usual, the terms of $A_c$ are permuted at the cost of converting $(1Z)^{(+)} \leftrightarrow (1Z)^{(-)}$ and dressing with some $\tau_z$ operators, as shown in Fig.~\ref{fig:HaahCommRelations}. Superposing all the fractal-shaped operators resulting from such interchanges produces the desired fractal operator for $[a]$, shown in Fig.~\ref{fig:HaahNonAb}. 

One can check explicitly, at least for small operators, that each term in the superposition is obtained by choosing any even number of $(1Z)^{(+)}$ operators to turn into $(1Z)^{(-)}$ operators, then connecting disjoint pairs of $(1Z)^{(-)}$ operators by $\tau_z$ strings. One can also check that every such term appears with equal weight. By the self-similarity of the Sierpinski tetrahedron, this should also be the case for arbitrarily large operators, although we do not have a formal proof.

As in the previous cases, additional fractal operators can be obtained by choosing at each corner of the tetrahedron whether or not to interchange the $(1Z)^{(+)}$ and $(1Z)^{(-)}$ operators. By inspection, these different operators are locally distinguishable, for example by the action of $(1X)^{(-)}$. There are $2^4$ such choices.

Analogously to the ungauged model, there are also operators with fractal support built from a $(Z1)^{(+)}$ reference operator which creates four $[a]$ fractons in a slightly different geometry; an example is shown in Fig.~\ref{fig:HaahZ1NonAb}.

The calculation of the fusion rules is done in exactly the same way as for the membrane operators for X-Cube. In particular, label the sixteen different operators $\mathcal{T}_{[a]}^{\pm, \pm, \pm, \pm}$ where the $\pm$ refers to which of $(1Z)^{(\pm)}$ appears at each corner of the Sierpinski tetrahedron in the reference configuration; $\mathcal{T}_{[a]}^{+,+,+,+}$ is shown in Fig.~\ref{fig:HaahNonAb}.

Define $T=\mathcal{T}_{[a]}^{+,+,+,+}+\mathcal{T}_{[a]}^{-,-,-,-}$. As in the X-Cube case, $T^2$ creates two of the non-Abelian $[a]$ excitations at each corner of the Sierpinski tetrahedron on which $T$ has support. This operator can be re-expressed as a sum of operators which create simple excitations which correspond to fusion outcomes of the pairs of fractons. A tedious but straightforward calculation shows that
\begin{equation}
T^2 = (1+T_{a^{(1)}a^{(2)}})\left(1+\bigotimes_{\bv{r} \in \text{two strings}}\tau_{z,\bv{r}}\right)
\end{equation}
where the ``two strings" of $\tau_z$ operators connect disjoint pairs of corners of the tetrahedron. This means that the fractons either all fuse to the identity, to a $\mathbb{Z}_2$ charge $\phi$, to an Abelian fracton $a^{(1)}a^{(2)}$, or to $\phi a^{(1)}a^{(2)}$. As in the gauged bilayer X-cube model, since these topological charges can be determined by local measurements, the number of fusion outcomes when the fractons are at the same point in space does not map directly onto degeneracy when the fractons are separated in space, but it is suggestive.

The full operator algebra for the $\mathcal{T}_{[a]}$ operators (and likewise for the $\mathcal{T}_{[b]}$ operators) turns out to be identical to the algebra given in Appendix~\ref{app:bilayerXCubeExcitations} for the $\mathcal{M}_{[f_0]}$ operators in the gauged bilayer X-cube model.

Finally, the flux string excitations $\sigma$ are created by membrane operators which are constructed in a manner which is spiritually similar to the analogous operator in the X-cube model, but a systematic construction of the operator is sufficiently complicated that we leave discussion of these operators to Appendix~\ref{app:HaahStringOp}.

\section{Discussion}
\label{sec:Discussion}

We have exhibited a general method for obtaining new classes of fracton models: start from a fracton model with a global symmetry which permutes nontrivial excitations, and then gauge the symmetry. This procedure produces a model with non-Abelian versions of the excitations in the ungauged model by identifying each excitation with its symmetry orbit. Notably, the mobility of the excitations is not affected. This allows us to generate models with, for example, non-Abelian fractal-type fractons. This comes with a side effect which places these models outside the simplest framework for fractons: not only do our models generically have at least one fully mobile particle (the pure symmetry charge), but they also necessarily come with a non-Abelian string-like excitation which braids nontrivially with some of the subdimensional particles. As such, these models in some sense lie between conventional fracton models with only point-like, subdimensional excitations and TQFTs.

A number of extensions of our work follow naturally. We have focused on gauging $\mathbb{Z}_2$ SWAP symmetries, but the generalization to, say, $\mathbb{Z}_n$ subgroups of the permutation group should be straightforward. Another simple generalization of our results would be to gauge the $\mathbb{Z}_2$ global symmetry in the bosonic checkerboard model which exchanges the electric and magnetic sectors (generated by a Hadamard transformation on every spin); we expect this to lead to a fractonic version of Ising $\times$ Ising$^{\ast}$ topological order. One could also gauge the ``charge conjugation" symmetry of the $\mathbb{Z}_3$ X-cube model or its generalizations.  Relatedly, a single layer of Haah's code has a symmetry consisting of a Hadamard transformation on every spin combined with spatial inversion; although gauging a symmetry which involves inversion is not obviously meaningful, one could imagine that some other type-II fracton model would have an internal Hadamard-type transformation which could be gauged to produce a Haah-like Ising$\times$Ising$^{\ast}$ model. Our procedure can be further generalized to gauge any finite subgroup $G$ of the permutation group on $n$ copies of a fracton model, with a further choice of a Dijkgraaf-Witten twist $\mathcal{H}^4(G, U(1))$; it would be interesting to develop an understanding of the properties of the models for more general $G$. 

Our non-Abelian Haah's code model is in itself of considerable further interest. Many properties of the model apart are challenging to compute, even numerically (since the Hilbert space has 7 qubits per site and is not a Pauli Hamiltonian, even small system size calculations will be computationally intensive). Its ability to serve as a topological quantum memory or qubit would also be worthy of investigation, particularly since the interplay between mobile quasiparticles, immobile fractons, and flexible non-Abelian strings in the model is quite complicated. We generically expect that exchanging a (fully mobile) $\phi$ particle between certain sets of fractons can split the degeneracy associated with non-Abelian fractons, so the full degenerate manifold of states does not inherit the same robustness at finite temperature as, say, the ground state manifold of the Abelian model on a torus. However, fully splitting the degenerate manifold of states associated to non-Abelian fractons should require exchanging Abelian fractons among non-Abelian fractons, so we expect that some portion of the degeneracy would possess a memory lifetime at finite temperature that diverges with the distance between the non-Abelian fractons. 

Finally, it would also be interesting to connect our models to several constructions from the literature. For example, the non-Abelian twisted fracton models in Ref. \onlinecite{SongTwisted} do not contain the unusual string excitations that we have found, but one could imagine coupling them to conventional $3+1$D topological order in a nontrivial way so that the string excitations braid with fractons, perhaps producing a model similar to our gauged bilayer X-cube model. It could also be enlightening to see if, in the spirit of Refs.~\onlinecite{MaLayer,VijayLayer}, coupling layers of 2D $\mathbb{D}_4$ quantum double models in some way could also produce the gauged bilayer X-cube model.

\begin{acknowledgements}
  
We thank Michael Hermele and Thomas Iadecola for helpful discussions.  This work is supported by NSF CAREER (DMR- 1753240), Alfred P. Sloan Research Fellowship, and JQI- PFC-UMD.
  
\end{acknowledgements}

\appendix

\section{Alternate gauging procedure for the $\mathbb{D}_4$ quantum double model}
\label{app:alternateGauge}

In this appendix, we explain an alternate way to gauge the SWAP symmetry of the bilayer $2+1$D toric code model (strictly speaking, the $\mathbb{Z}_2 \times \mathbb{Z}_2$ quantum double model) which explicitly produces the $\mathbb{D}_4$ quantum double model.

For later convenience, we choose to start with the $\mathbb{Z}_2 \times \mathbb{Z}_2$ quantum double model rather than decoupled toric codes. The Hamiltonian is
\begin{widetext}
\begin{equation}
H = -\sum_s\left(\sum_{i=1,2}A_s^{(i)}+A_s^{(1)}A_s^{(2)}\right) - \sum_p \left(\sum_{i=1,2}B_p^{(i)} +B_p^{(1)}B_p^{(2)}\right)
\end{equation}
\end{widetext}
where $A_s^{(i)}=\bigotimes_+ \sx^{(i)}$ and $B_p^{(i)}=\bigotimes_{\square}\sz^{(i)}$ as in Fig.~\ref{fig:toricCodeH}. The only difference from the decoupled bilayer toric code model Eq.~\eqref{eqn:ungaugedBilayerTC} are the $A_s^{(1)}A_s^{(2)}$ and $B_p^{(1)}B_p^{(2)}$ terms, which simply makes the energy of single charges (resp. fluxes) equal to the energy of a bound state of the charges (resp. fluxes) in the two layers. We emphasize that this does not change the eigenstates; it only changes the energies of some of the excited states.

\begin{figure}
\centering
\subfloat[\label{fig:alternateTCGaugeGenerator}]{\includegraphics[width=0.4\columnwidth]{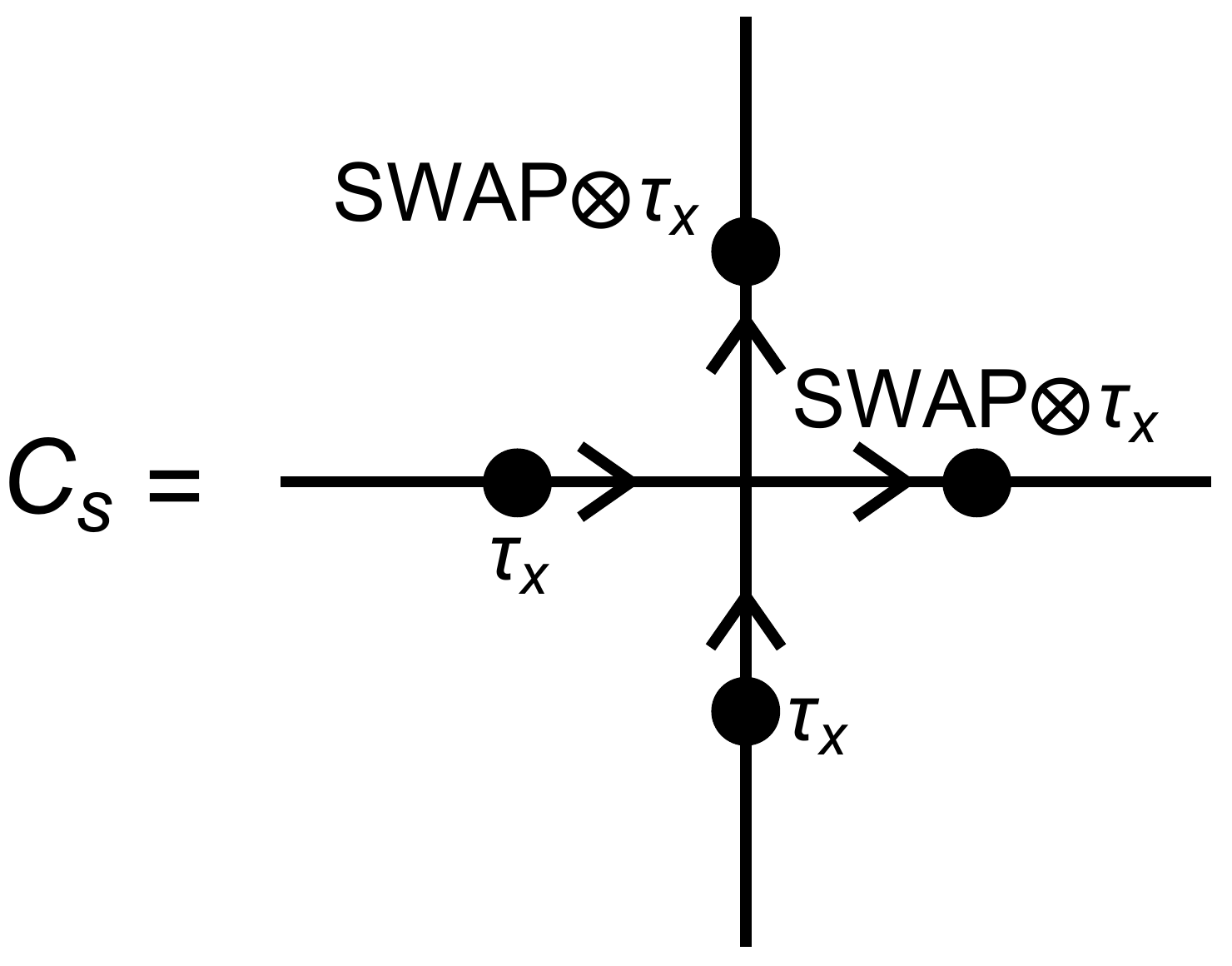}}\\
\subfloat[\label{fig:alternateGaugeAs}]{\includegraphics[width=0.9\columnwidth]{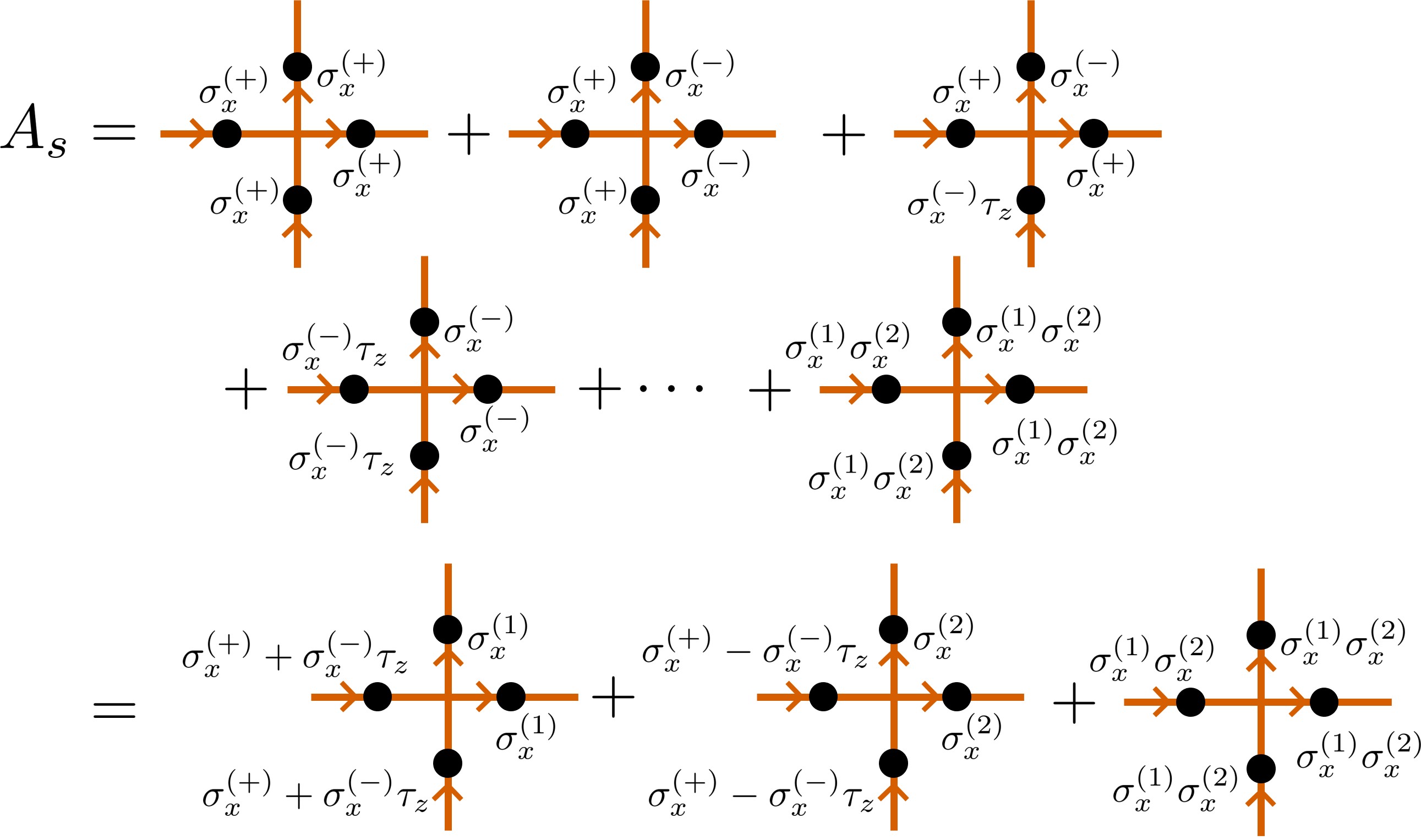}}\\
\subfloat[\label{fig:alternateGaugeBp}]{\includegraphics[width=0.9\columnwidth]{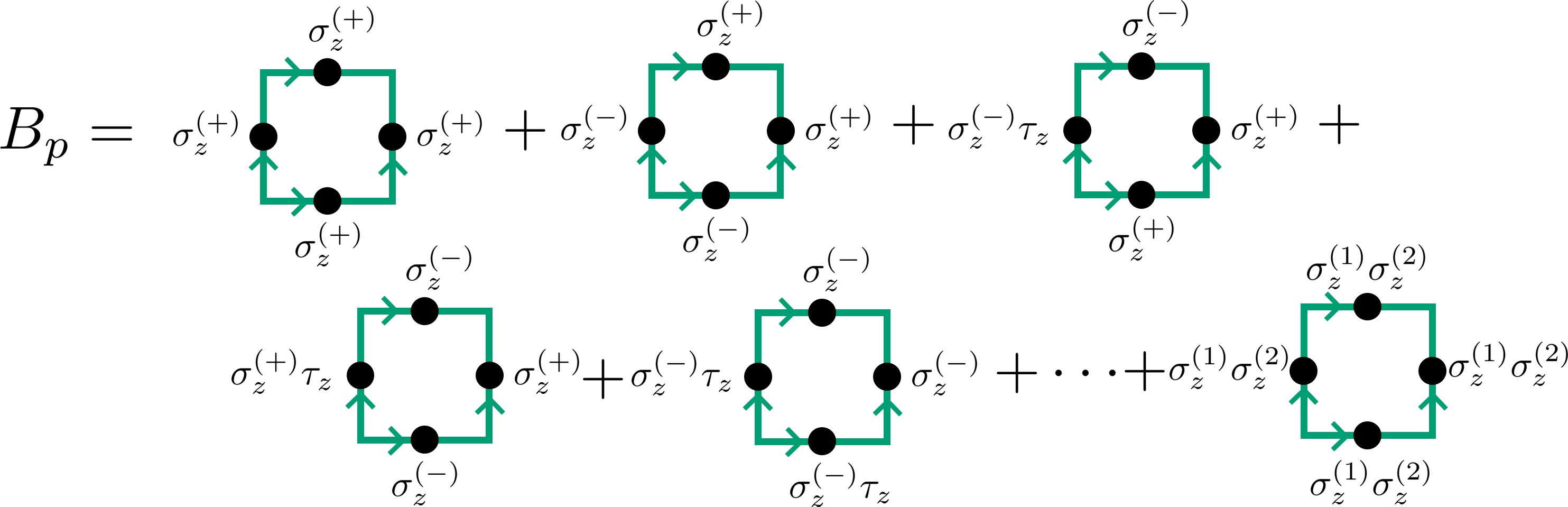}}
\caption{Operator definitions in the alternate gauging procedure for the bilayer toric code. (a) Generator $C_s$ of the gauge symmetry and orientation convention for links. (b) Star operator $A_s$. (c) Plaquette operator $B_p$.}
\label{fig:alternateGaugeOps}
\end{figure}

In this gauging procedure, we make a different choice for the generator of local symmetries - the local SWAP operator acts on \textit{two} links, one oriented in each lattice direction, instead of just one. Accordingly, we define the Hilbert space of the gauged model to have three spins per link of the square lattice, and define the generator of the local gauge symmetry $C_s$, where $s$ is a site of the lattice, via
\begin{equation}
C_s = \bigotimes_{\textsf{L}} \SWAP \bigotimes_{+}\tau_x
\end{equation}
as shown in Fig.~\ref{fig:alternateTCGaugeGenerator}. We have included an orientation on the links for later convenience.

We again reexpress $A_s^{(1)}+A_s^{(2)}+A_s^{(1)}A_s^{(2)}$ and $B_p^{(1)}+B_p^{(2)}+B_p^{(1)}B_p^{(2)}$ in terms of the $\sigma^{(\pm)}$ operators, just as we did in Eq.~\eqref{eqn:plaquetteRewrite}. The difference from Sec.~\ref{sec:ToricCode} is in the way that the matter spins are dressed by the gauge spins $\tau$. The resulting gauged operators $A_s$ and $B_p$ are shown in Figs.~\ref{fig:alternateGaugeAs} and \ref{fig:alternateGaugeBp}, respectively. The simplified form for $A_s$ on the bottom line of Fig.~\ref{fig:alternateGaugeAs} can be verified with straightforward algebra. The Hamiltonian simplifies to
\begin{equation}
H_{\text{gauged}}= -\sum_s A_s- 2\sum_s C_s - \sum_p B_p - \sum_p D_p
\label{eqn:simpleGauged}
\end{equation}
where $D_p = \bigotimes_{\square} \tau_z$ is the toric code flux operator on a plaquette. Since $A_s,B_p$ and $D_p$ commute with $C_s$ and with $D_p$, the factor of 2, which we have added for convenience, only changes the energy of $\mathbb{Z}_2$ charge excitations without changing any topological properties of the model. 

\begin{figure}
\centering
\includegraphics[width=0.9\columnwidth]{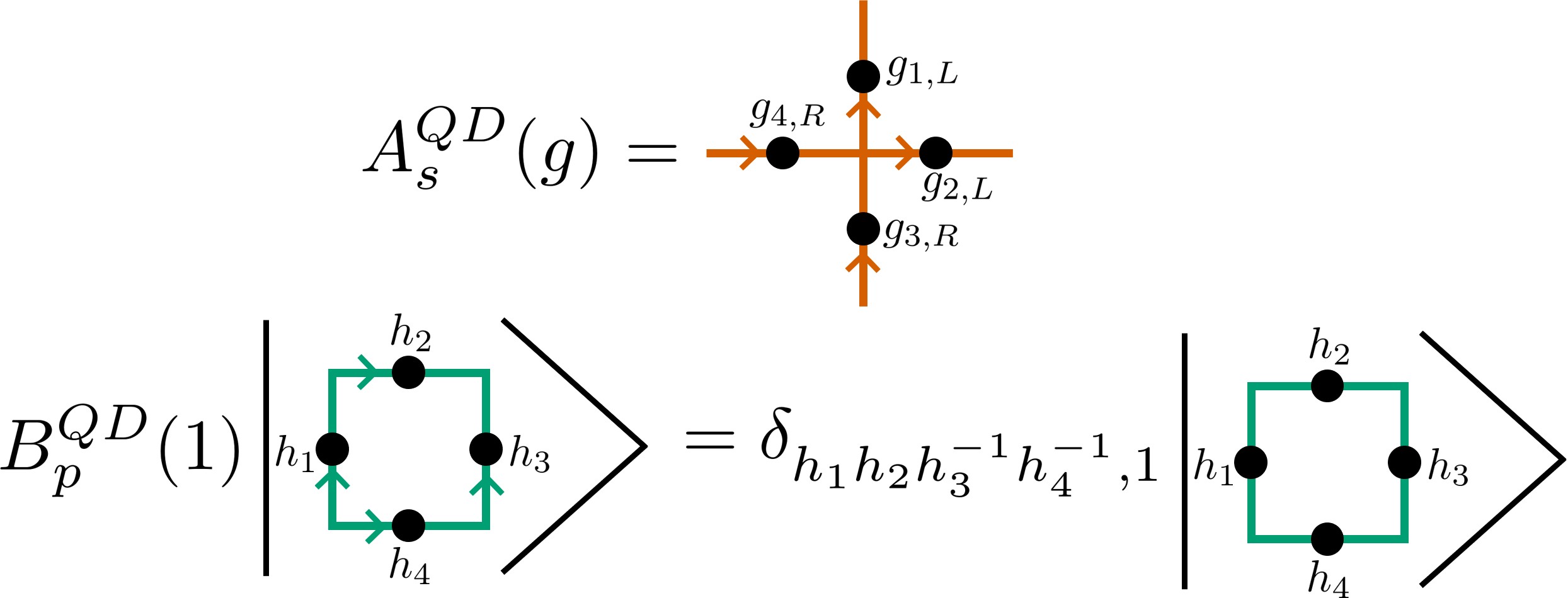}
\caption{Operator definitions for the $\mathbb{D}_4$ quantum double Hamiltonian Eq.~\eqref{eqn:trueD4QD}.}
\label{fig:QDHamiltonian}
\end{figure}

We will now build the correspondence with the $\mathbb{D}_4$ quantum double model, which we briefly review now. The $\mathbb{D}_4$ quantum double model is defined on a Hilbert space where group elements live on (oriented) links of the square lattice. The states obey $\braket{g}{h} = \delta_{g,h}$, and there are operators $g_L$ and $g_R$ ($L^g_{\pm}$ in Kitaev's notation) which implement left and right group multiplication respectively; these operators obey $g_L \ket{h} = \ket{gh}$ and $g_R\ket{h} = \ket{hg}$, which implies 
\begin{align}
g_Lh_L &= (gh)_L \label{eqn:gLConsistency}\\
g_Rh_R &=(hg)_R \label{eqn:gRConsistency}
\end{align} for all $g,h \in \mathbb{D}_4$. 
We also need the operators $T^g_{\pm}$ which have the action $T^g_+\ket{h} = \delta_{g,h}\ket{h}$ and $T^g_-\ket{h}=\delta_{g,h^{-1}}\ket{h}$. The model is defined using star operators $A_s(g)$ and plaquette operators $B_p(g)$ via
\begin{equation}
H_{\text{QD}} = -\sum_s \sum_g A^{QD}_s(g) - \sum_p B^{QD}_p(1)
\label{eqn:trueD4QD}
\end{equation}
where 
\begin{align}
A^{QD}_s(g) &= g_{1,L}g_{2,L}g_{3,R}g_{4,R}\\
B^{QD}_p(1)\ket{h_1,h_2,h_3,h_4}&=\delta_{h_1h_2h_3^{-1}h_4^{-1},1}\ket{h_1,h_2,h_3,h_4}
\end{align}
where the numerical subscripts label different links and the L/R indices/choice of $h^{-1}$ follow the link orientations as shown in Fig.~\ref{fig:QDHamiltonian}. We have included a ``QD" superscript to indicate that these are the star and plaquette operators in the quantum double as opposed to the terms we have obtained via gauging the bilayer toric code. Our goal is to map the three-spin-per-link Hilbert space of the gauged bilayer toric code model, for which we represent states in the basis of $\ket{\sigma^{(1)}_z \sigma^{(2)}_z \tau_z}$, to the Hilbert space labeled by elements of $\mathbb{D}_4$ so that the terms in Eq.~\eqref{eqn:simpleGauged} directly map to $A^{QD}_s(g)$ and $B^{QD}_p(1)$.

To construct this mapping, we use the group presentation $\mathbb{D}_4 = \langle a,b,c | a^2=b^1=c^2=1, cac=b\rangle$; the distinct group elements can be labeled $\lbrace 1,a,b,c,ab,ac,bc,abc\rbrace$. We define the mapping so that if $a$ (resp. $b,c$) is present in this choice of labeling of the group elements, then $\sz^{(1)}$ (resp. $\sz^{(2)}$, $\tau_z$) is down; otherwise $\sz^{(1)}$ (resp. $\sz^{(2)}$, $\tau_z$) is up. For example, $\ket{\downarrow \uparrow\downarrow} \rightarrow \ket{ac}$. Using this mapping of states, the operators $g_{L,R}$ can simply be read off in the $\sigma,\tau$ basis from their desired action on the states. Consistency with the group multiplication rules, i.e. Eqs.~\eqref{eqn:gLConsistency} and \eqref{eqn:gRConsistency} can be checked straightforwardly. The explicit mapping is given in Table~\ref{tab:operatorMapping}. 

\begin{table*}
\centering
\renewcommand{\arraystretch}{1.3}
\begin{tabular}{@{}lllllll}
\toprule[2pt]
\textbf{Group element} $g \in \mathbb{D}_4$ &\phantom{ab}& $\ket{g}$ &\phantom{a}& $g_L$ &\phantom{a}& $g_R$  \\ \hline
$1$ && $\ket{\ua \ua \ua}$ && 1 && 1\\
$a$ && $\ket{\da \ua \ua}$ && $\sx^{(1)}$ && $\sx^{(+)} + \tau^z \sx^{(-)}$ \\
$b$ && $\ket{\ua \da \ua}$ && $\sx^{(2)}$ && $\sx^{(+)} - \tau^z \sx^{(-)}$ \\
$c$ && $\ket{\ua \ua \da}$ && $\tau_x\SWAP$ && $\tau_x$ \\
$ab$ && $\ket{\da \da \ua}$ && $\sx^{(1)}\sx^{(2)}$ && $\sx^{(1)}\sx^{(2)}$\\
$ac=cb$ && $\ket{\da \ua \da}$ && $\sx^{(1)}\tau_x\SWAP$ && $\left(\sx^{(+)} -\tau^z \sx^{(-)}\right)\tau_x$ \\
$bc=ca$ && $\ket{\ua \da \da}$ &&  $\sx^{(2)}\tau_x\SWAP$ && $\left(\sx^{(+)} + \tau^z \sx^{(-)}\right)\tau_x$ \\
$abc$ && $\ket{\da \da \da}$ && $\sx^{(1)}\sx^{(2)}\tau_x\SWAP$ && $\sx^{(1)}\sx^{(2)} \tau_x$
 \\	\bottomrule[2pt]	
\end{tabular}
\caption{Identification of the three-spin Hilbert space $\ket{\sigma^{(1)}_z \sigma^{(2)}_z \tau_z}$ with the Hilbert space labeled by elements of $\mathbb{D}_4$.}
\label{tab:operatorMapping}
\end{table*}

The consistency with the group algebra can be checked. Given these operator identifications, we immediately find that the star terms in the Hamiltonian Eq.~\eqref{eqn:simpleGauged} (shown in Figs.~\ref{fig:alternateTCGaugeGenerator} and \ref{fig:alternateGaugeAs}) can be rewritten 
\begin{equation}
\tilde{A}_s \equiv A_s+2C_s = -\sum_{s} \sum_{g=a,b,c,ab} (1+\delta_{g,c})A^{QD}_s(g)
\end{equation}
so that the Hamiltonian in the $\mathbb{D}_4$ basis becomes
\begin{equation}
H_{\text{gauged}}=-\sum_s \tilde{A}_s - \sum_p (B_p+D_p)
\label{eqn:D4QDgauge}
\end{equation}

From the standard algebra of the $A^{QD}(g)$ and $B^{QD}(h)$ operators, one can check that $\tilde{A}_s=\sum_{g=a,b,c,ab}(1+\delta_{g,c})A^{QD}_s(g)$ and $B_p$
commute with each other, so Eq.~\eqref{eqn:D4QDgauge} is indeed a commuting projector model. However, it is not quite the $\mathbb{D}_4$ quantum double model Eq.~\eqref{eqn:trueD4QD}. To understand the difference, note that adding a term such as $\tilde{A}_s^2$ or $\tilde{A}_s^3$ to the Hamiltonian does not change the eigenstates; it simply changes the energy of some of the excitations. In fact, up to an overall constant, one can check by brute force that
\begin{equation}
\sum_{g \in \mathbb{D}_4} A_s(g) = \tilde{A}_s+\frac{19}{52}\tilde{A}_s^2-\frac{1}{26}\tilde{A}_s^3
\end{equation}
It can also be explicitly checked that eigenstates corresponding to the largest eigenvalue of $\tilde{A}_s$ also correspond to the largest eigenvalue of $A_s$. Hence, replacing $\tilde{A}_s$ by $A_s$ in the gauged Hamiltonian Eq.~\eqref{eqn:D4QDgauge} only changes the energies of some excitations without changing the eigenstates or ground state manifold. As for the plaquette terms, we note that $B_p$ and $D_p$ commute with each other and with $\tilde{A}_s$, and a very tedious computation shows that
\begin{equation}
B_p(1) = 1+\tilde{B}_p+D_p+D_p\tilde{B}_p
\end{equation}
where the left-hand side is the $\mathbb{D}_4$ quantum double flux operator. Therefore, in Eq.~\eqref{eqn:D4QDgauge}, replacing $\tilde{B}_p+D_p$ by $B_p(1)$ only changes the energies of some excitations without changing the eigenstates or ground state manifold.

Hence, the only difference between Eq.~\eqref{eqn:D4QDgauge} and the $\mathbb{D}_4$ quantum double model Eq.~\eqref{eqn:trueD4QD} is changes to the quasiparticle energies; the ground state manifold and the eigenstates of the models are otherwise the same.

Tracing through the argument, one can check that choosing different links to contain the SWAP operators in the local symmetry generators of the gauged bilayer toric code will lead to quantum double models with different choices of link orientations.

\section{String operator algebra in the gauged bilayer toric code}
\label{app:bilayerTCStrings}

In this appendix, we exhibit the algebra of string operators in the gauged bilayer toric code.

As discussed in the main text, products of strings for anyons $a$ and $b$ produce (sums of) string operators of the fusion products $a \times b$. One benefit of our parametrization of the string operators is that the product of a string operator for an Abelian anyon with any other string operator is obtained simply by multiplying the corresponding $\mathcal{U}$, $\mathcal{V}$,  and $\mathcal{W}$ operators. This allows the algebra of the Abelian anyons to be read off straightforwardly. For example, it is immediately clear that the Abelian anyons all square to the identity, that any two Abelian (pure) charges fuse to the third, and that bound states of the Abelian flux $m^{(1)}m^{(2)}$ with Abelian charges are distinct anyon types (Abelian dyons). Likewise, we can read off that, for example, $\phi e^{(1)}e^{(2)} \times \sigma_m = \sigma_m$ (although the role of all the 1 operators and all the 2 operators are interchanged, which can be reversed with the action of local operators at the ends of the string).

The fusion of non-Abelian anyons is more delicate, and the precise operators which result depend on the particular choices of local degrees of freedom at the ends of the strings. We presently enumerate the string algebra, obtained from direct computation. Denote the string operators by $\mathcal{S}_a^{k,l}$ where $a$ is an anyon type and $k,l=\pm$ correspond to choosing $\mathcal{U}_k$ and $\mathcal{U}_l$ on the left and right (respectively) ends of the string. Note that $\mathcal{S}_{[e]}^{k,l}=\mathcal{O}_{k,l}$ in the notation in the main text. Multiplication is implied if $kl$ appears outside of superscripts.

\begin{widetext}
\begin{align}
(\mathcal{S}_{\phi})^2&=(\mathcal{S}_{e^{(1)}e^{(2)}})^2=(\mathcal{S}_{m^{(1)}m^{(2)}})^2 = 1\\
\mathcal{S}_{\phi}\mathcal{S}_{e^{(1)}e^{(2)}} &= \mathcal{S}_{\phi e^{(1)}e^{(2)}}\\
\mathcal{S}_{\phi}\mathcal{S}_{[e]}^{k,l}&=\mathcal{S}_{[e]}^{-k,-l}\\
\mathcal{S}_{e^{(1)}e^{(2)}}\mathcal{S}_{[e]}^{k,l} &= kl \mathcal{S}_{[e]}^{k,l}\\
\mathcal{S}_{[e]}^{k,l}\mathcal{S}_{[e]}^{k',l'} &= \begin{cases}
0 & klk'l'=-\\
1 + kl \mathcal{S}_{e^{(1)}e^{(2)}} & klk'l'=+,kk'=+\\
\mathcal{S}_{\phi}+kl\mathcal{S}_{\phi e^{(1)}e^{(2)}} & klk'l'=+,kk'=+
\end{cases}\\
\mathcal{S}_{\phi}\mathcal{S}_{m^{(1)}m^{(2)}} &= \mathcal{S}_{\phi m^{(1)}m^{(2)}}\\
\mathcal{S}_{\phi}\mathcal{S}_{[m]}^{k,l}&=\mathcal{S}_{[m]}^{-k,-l}\\
\mathcal{S}_{[m]}^{k,l}\mathcal{S}_{[m]}^{k',l'} &= \begin{cases}
0 & klk'l'=-\\
1 + kl \mathcal{S}_{m^{(1)}m^{(2)}} & klk'l'=+,kk'=+\\
\mathcal{S}_{\phi}+kl\mathcal{S}_{\phi m^{(1)}e^{(2)}} & klk'l'=+,kk'=-
\end{cases}\\
\mathcal{S}_{e^{(1)}e^{(2)}}\mathcal{S}_{\sigma_1}^{k,l} &= \mathcal{S}_{\sigma_1}^{-k,-l}\\
\mathcal{S}_{m^{(1)}m^{(2)}}\mathcal{S}_{\sigma_1}^{k,l} &= kl \mathcal{S}_{\sigma_1}^{-k,-l}\\
\mathcal{S}_{\sigma_1}^{k,l}\mathcal{S}_{\sigma_1}^{k',l'} &= \begin{cases}
0 & klk'l'=-\\
1+kl \mathcal{S}_{m^{(1)}m^{(2)}} & klk'l'=+,kk'=+\\
\mathcal{S}_{e^{(1)}e^{(2)}m^{(1)}m^{(2)}} + kl \mathcal{S}_{f^{(1)}f^{(2)}} & klk'l'=+,kk'=-
\end{cases}
\end{align}
\end{widetext}
where $f^{(i)}=e^{(i)}m^{(i)}$ is a toric code fermion. This operator algebra reproduces the $\mathbb{D}_4$ fusion rules, such as
\begin{align}
e^{(1)}e^{(2)} \times [e] &= \phi \times [e] = [e]\\
m^{(1)}m^{(2)} \times [m] &= \phi \times [m] = [m]\\
[e] \times [e] &= 1+\phi+e^{(1)}e^{(2)} + \phi e^{(1)}e^{(2)}\\
[m] \times [m] &= 1+\phi+m^{(1)}m^{(2)} + \phi m^{(1)}m^{(2)}\\
e^{(1)}e^{(2)} \times [\sigma_1] &= m^{(1)}m^{(2)} \times [\sigma_1] = \sigma_1\\
\sigma_1 \times \sigma_1 &= 1+e^{(1)}e^{(2)}+m^{(1)}m^{(2)}+f^{(1)}f^{(2)}
\end{align}

\section{Review of the X-cube model}
\label{app:XCubeReview}

In this appendix, we briefly review the properties of the $\mathbb{Z}_2$ X-cube model\cite{CastelnovoFirstXCubePaper,VijayGaugedSubsystem} which are relevant for our discussion and establish notation for some of its excitations.

The Hilbert space of the X-cube model consists of spin-1/2s on the links of the cubic lattice. The Hamiltonian is
\begin{equation}
H = -\sum_s \sum_{i=x,y,z}A_{s,i} - \sum_c B_c
\end{equation}
where $A_{s,i} = \otimes_+ \sx$ is a product $\sx$ on the four spins that touch site $s$ within the plane perpendicular to the $i$ direction and $B_c = \otimes_{\text{\mancube}}\text{ }\sz$ is the product of all $\sz$ operators on an elementary cube, as shown in Fig.~\ref{fig:XCubeTerms}. The terms in the Hamiltonian are commuting projectors, so the ground state subspace consists of states with $A_{s,i}=B_c=+1$. On an $L_x\times L_y \times L_z$ 3-torus, the ground state degeneracy is $2^{2(L_x+L_y+L_z)-3}$.

\begin{figure}
\centering
\subfloat[\label{fig:singleLayerXCubeF0}]{\includegraphics[width=0.4\columnwidth]{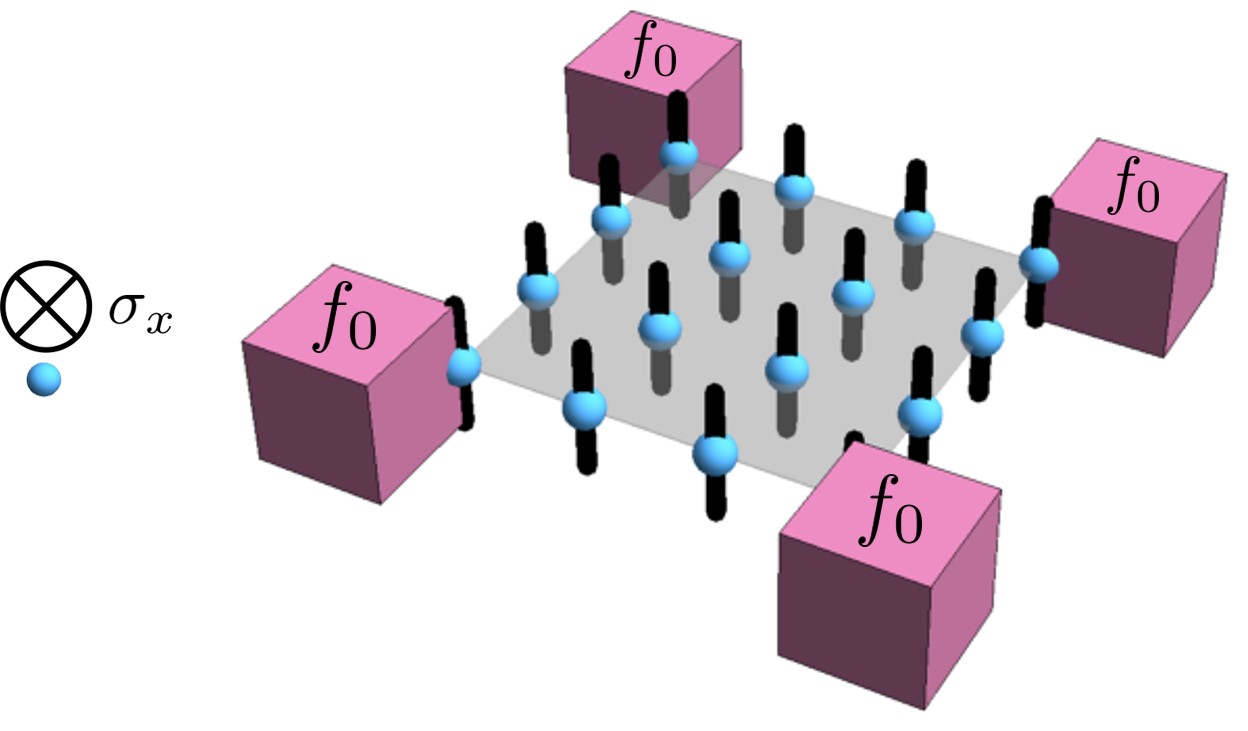}}\\
\subfloat[\label{fig:singleLayerXCubeMxy}]{\includegraphics[width=0.4\columnwidth]{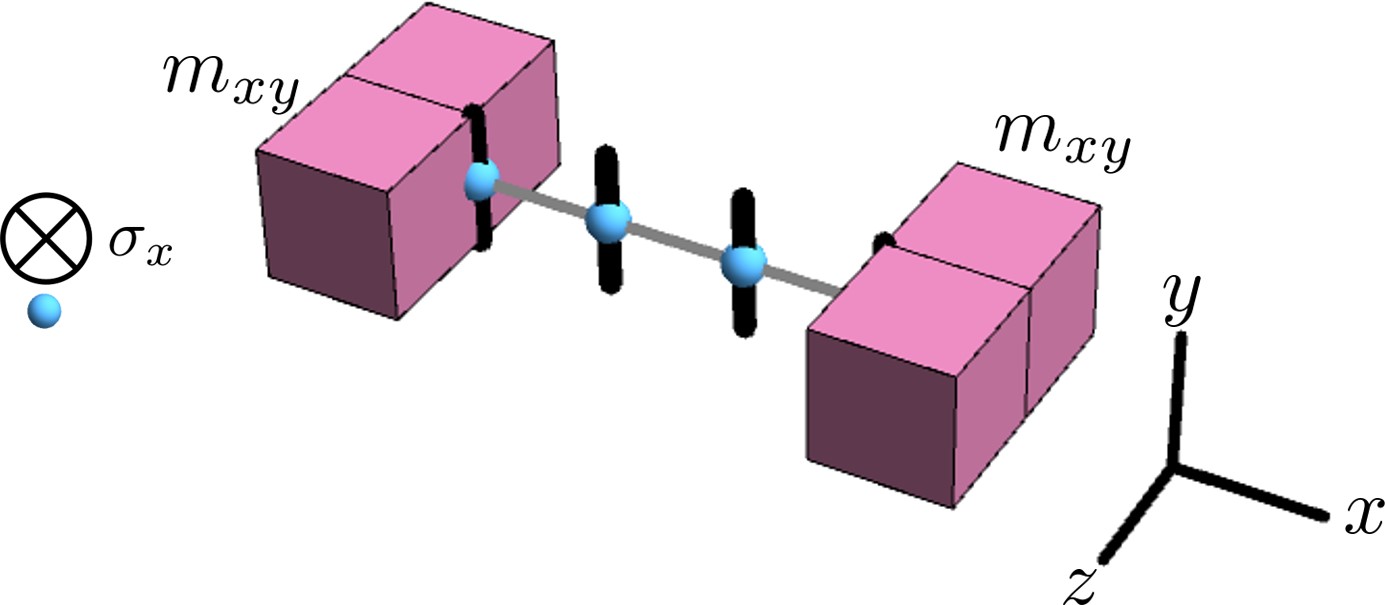}}\\
\subfloat[\label{fig:singleLayerXCubeEx}]{\includegraphics[width=0.4\columnwidth]{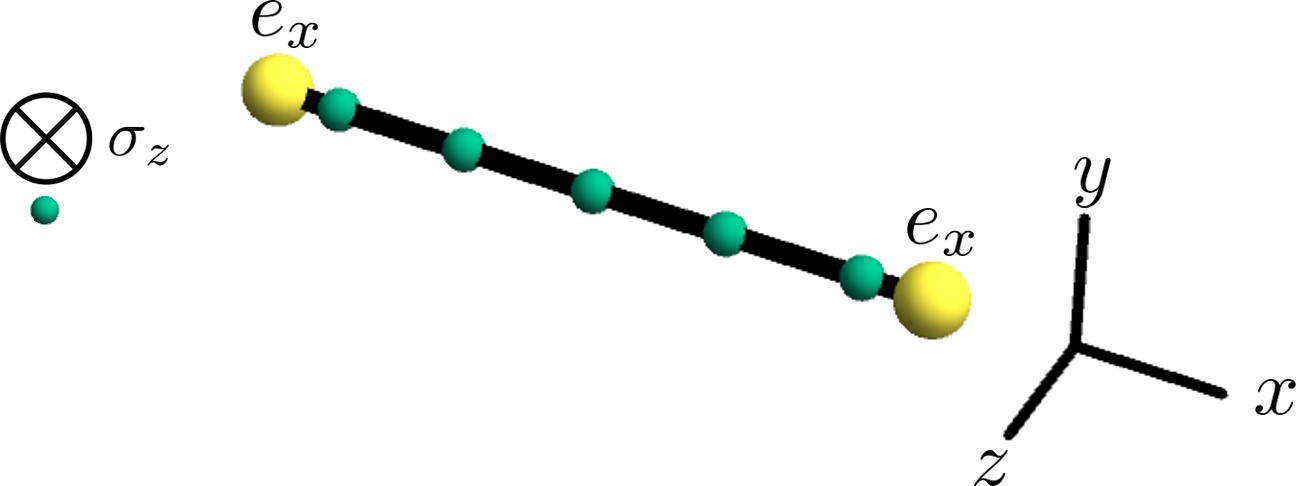}}
\caption{Excitations and operators that create them in the X-cube model. (a) Four fractons $f_0$ (purple cubes), which are cubes where $B_c=-1$, created at the end of a membrane (light grey) of $\sx$ operators (blue spheres). (b) Two-dimensional bound state $m_{xy}$ of two fractons, created at the end of a string of $\sx$ operators. (c) One-dimensional quasiparticles $e_x$ (yellow spheres) where $A_y=A_z=-1$, created at the end of a string of $\sz$ operators (green spheres).}
\end{figure}

A single cube with $B_c=-1$ is an (immobile) fracton excitation, which we call $f_0$. These excitations are created in sets of four by acting on a ground state with a rigid rectangular membrane of $\sx$ operators; the fractons are created at the corners of the membrane, as shown in Fig.~\ref{fig:singleLayerXCubeF0}. If one of the membrane's sides is reduced to zero size, or equivalently we apply a string of $\sx$ operators instead as in Fig.~\ref{fig:singleLayerXCubeMxy}, a two-fracton bound state is formed with $B_c=-1$ on two face-sharing cubes. This excitation, which we call $m_{ij}$, is mobile in the plane of the shared face. Here $i,j=x,y,z$ labels the directions in which the excitation is mobile.

A bound state of $A_{s,i}=A_{s,j}=-1$ for $i\neq j$ is a one-dimensional excitation which we call $e_k$ (here $k\neq i,j$). (Note that as $A_{s,x}A_{s,y}A_{s,z}=+1$ as an operator identity, so violations of the $A_{s,i}=+1$ ground state constraint must occur in pairs on every site.) These excitations are created by rigid strings of $\sz$ operators, as in Fig.~\ref{fig:singleLayerXCubeEx}, and mobile in the direction of the extent of the string.

\section{Algebra of quasiparticle creation operators in the gauged bilayer X-cube model}

\label{app:bilayerXCubeExcitations}

In this appendix, we enumerate the algebra of operators which create simple excitations in the gauged bilayer X-Cube model. The operators have been defined in the main text.

We label the strings operators $\mathcal{S}_a^{k,l}$ where $a$ is a quasiparticle type and $k,l=\pm$ specify choices of $\mathcal{U}_k$ and $\mathcal{U}_l$ appear on the left and right (respectively) ends of the string operators in the reference configuration. The string operator algebra is (with multiplication implied if $kl$ appears outside the superscripts)
\begin{align}
(\mathcal{S}_{e_i^{(1)}e_i^{(2)}})^2 &= (\mathcal{S}_{\phi})^2 =1\\
\mathcal{S}_{\phi} \mathcal{S}_{e_i^{(1)}e_i^{(2)}}& = \mathcal{S}_{\phi e_i^{(1)}e_i^{(2)}}\\
\mathcal{S}_{e_i^{(1)}e_i^{(2)}} \mathcal{S}_{[e_i]}^{k,l} &= kl \mathcal{S}_{[e_i]}^{k,l}\\
\mathcal{S}_{\phi} \mathcal{S}_{[e_i]}^{+,+} &= \mathcal{S}_{[e_i]}^{-,-}\\
\mathcal{S}_{\phi} \mathcal{S}_{[e_i]}^{+,-} &= \mathcal{S}_{[e_i]}^{-,+}\\
\mathcal{S}_{[e_i]}^{k,l}\mathcal{S}_{[e_i]}^{k',l'} &= \begin{cases}
0 & klk'l'=-\\
1 + kl\mathcal{S}_{e_i^{(1)}e_i^{(2)}} & klk'l'=+, kk'=+\\
\mathcal{S}_{\phi} + kl \mathcal{S}_{\phi e_i^{(1)}e_i^{(2)}} & klk'l'=-, kk'=-
\end{cases}
\end{align}

We label the rigid membrane operators $\mathcal{M}_{a}^{k,l,m,n}$ where $a$ is an excitation type and $k,l,m,n =\pm$ specify choices of $\mathcal{V}_{1,2}$ at each corner of the membrane. Here $+$ corresponds to $\mathcal{V}_1$ and $-$ corresponds to $\mathcal{V}_2$. If $klmn$ appears outside an operator superscript, multiplication is implied. 
\begin{widetext} 
\begin{align}
(\mathcal{M}_{f^{(1)}_{0}f^{(2)}_{0}})^2 &= 1\\
\mathcal{M}_{f^{(1)}_{0}f^{(2)}_{0}} \mathcal{M}_{[f_0]}^{k,l,m,n} &= klmn \mathcal{M}_{[f_0]}^{k,l,m,n}\\
\mathcal{M}_{[f_0]}^{k,l,m,n} \mathcal{M}_{[f_0]}^{k',l',m',n'} &= \begin{cases}
0 & klmnk'l'm'n'=-\\
(\mathcal{S}_{\phi})^{\alpha}(1+klmn \mathcal{M}_{f^{(1)}_{0}f^{(2)}_{0}}) & klmnk'l'm'n' = +
\end{cases}
\end{align}
\end{widetext}
The $\alpha$ superscript in the last line means that if $kk'=-$ (resp. $ll'=-,mm'=-,nn'=-)$, then an $\mathcal{S}_{\phi}$ string operator terminates at that corner of the membrane. Depending on $k,l,m,n,k',l',m',n'$, this means that there are zero, one, or two string operators present in the product.

As discussed in the main text, for a flexible membrane operator of linear size $\ell$, there are $O(\ell)$ choices of local degrees of freedom (one for every point on the string) and $\mathcal{O}(2^{\ell})$ degenerate states associated with the fusion of two such strings. As such, we will not attempt to systematically enumerate the very large algebra of these string operators explicitly and simply remark by inspection that fusion of $\sigma$ strings will generally cause the strings to ``shatter" into a collection of Abelian excitations and $\mathbb{Z}_2$ charges along the length of the strings.

\section{Ground state degeneracy of the gauged bilayer X-Cube model}
\label{app:GSD}

In this appendix, we perform the detailed calculation of the ground state degeneracy of the gauged bilayer X-Cube model outlined in Sec. \ref{subsec:GSDOutline}. We work in the cage-net wavefunction picture with the ``color" of a string denoting its layer, and separate the calculation into sectors which depend on how many twist membranes are present in the ground-state wavefunction. 

As discussed in the main text, we do the calculation in each sector by counting ``reference configurations" for the ground states of the ungauged model where no topologically trivial membranes or strings are present, then determine additional constraints which result from the boundary conditions and gauging. A basis for the reference configurations is labeled by specifying the presence or absence of $2(L_x+L_y+L_z)-3$ strings of each color; $L_x+L_y-1$ of them live in the $x=0$ and $y=0$ planes and extend in the $z$ direction, and the same with $x \rightarrow y \rightarrow z \rightarrow x$.

The result for the untwisted sector was given in Eq. \eqref{eqn:untwistedGSD}, which we replicate here for completeness:
\begin{equation}
\text{GSD}_{\text{untwisted}}=2^N + \frac{1}{2}\left(4^N-2^N\right) = 2^{2N-1}+2^{N-1}
\end{equation}
where $N$ is given in Eq.~\eqref{eqn:XCubeN} and is equal to the number of independent (rigid) Wilson loops in a single copy of the X-cube model.

For conciseness, given a type (i.e. orientation and position) of Wilson loop, if both layers' strings are present in a cage-net configuration, we refer to the string as ``bichromatic." Accordingly, if exactly one layer's string is present, we refer to the string as ``monochromatic."

\subsection{Single-twist sector}

Without loss of generality, assume the twist defect spans the $x-$ and $y-$direction handle of the torus.

Every $z$-oriented string must pass through the twist defect, while no other string does. If a string in the wavefunction passes through a twist defect, it changes color; once such a string goes all the way around the torus, it must pass around again to return to its original color and become a closed string. Therefore, any string type which intersects the twist defect passes around the torus twice (see Fig.~\ref{fig:3TorusDoubleWrap}). The presence or absence of such a string is obviously symmetric under a global SWAP, so there are therefore $2^{L_x+L_y-1}$ possible configurations for the $z$-oriented strings. 

\begin{figure}
\centering
\subfloat[\label{fig:oneTwistBichromaticConstraint}]{\includegraphics[width=0.9\columnwidth]{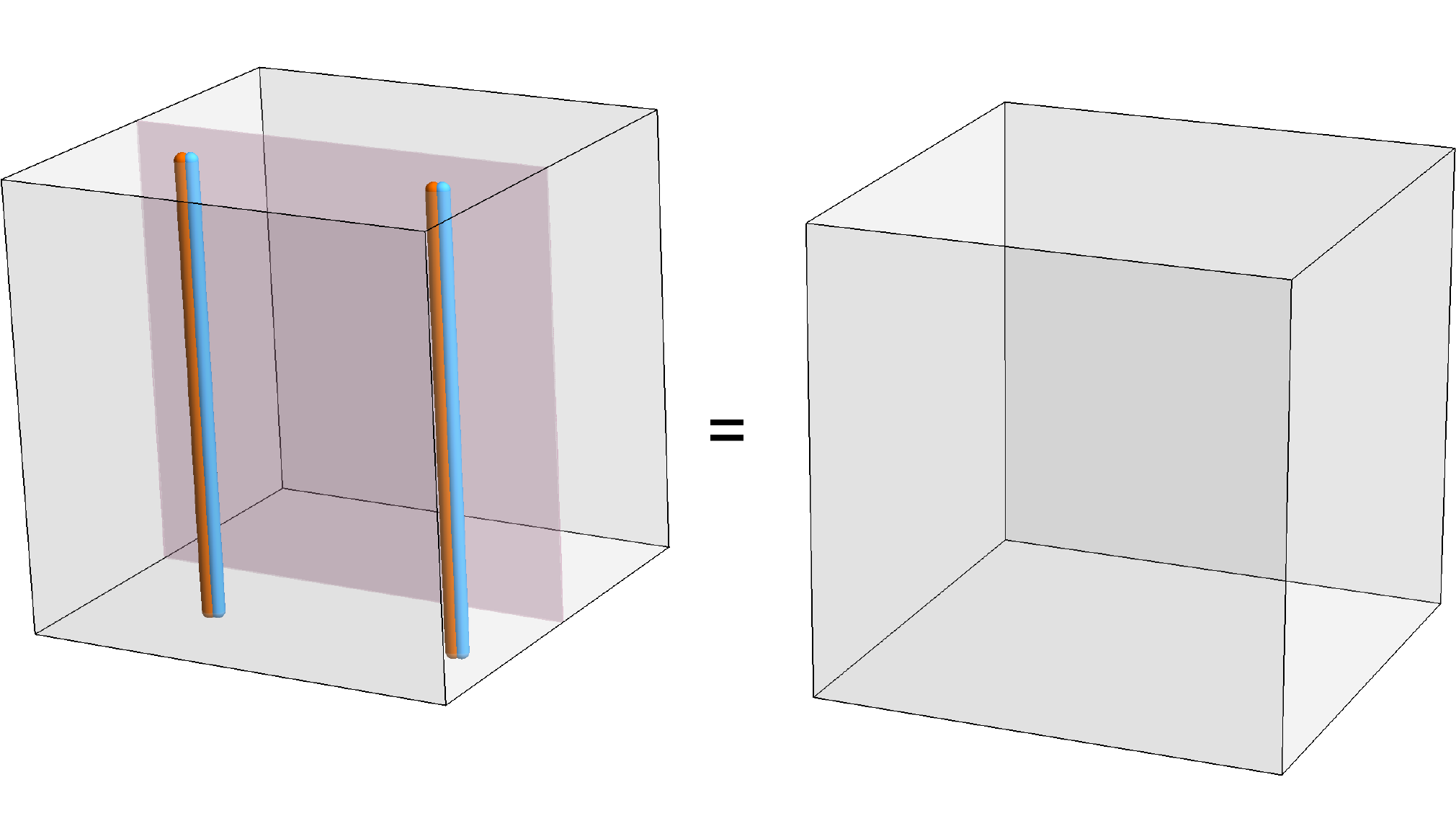}}\\
\subfloat[\label{fig:oneTwistThreeStringConstraint}]{\includegraphics[width=0.9\columnwidth]{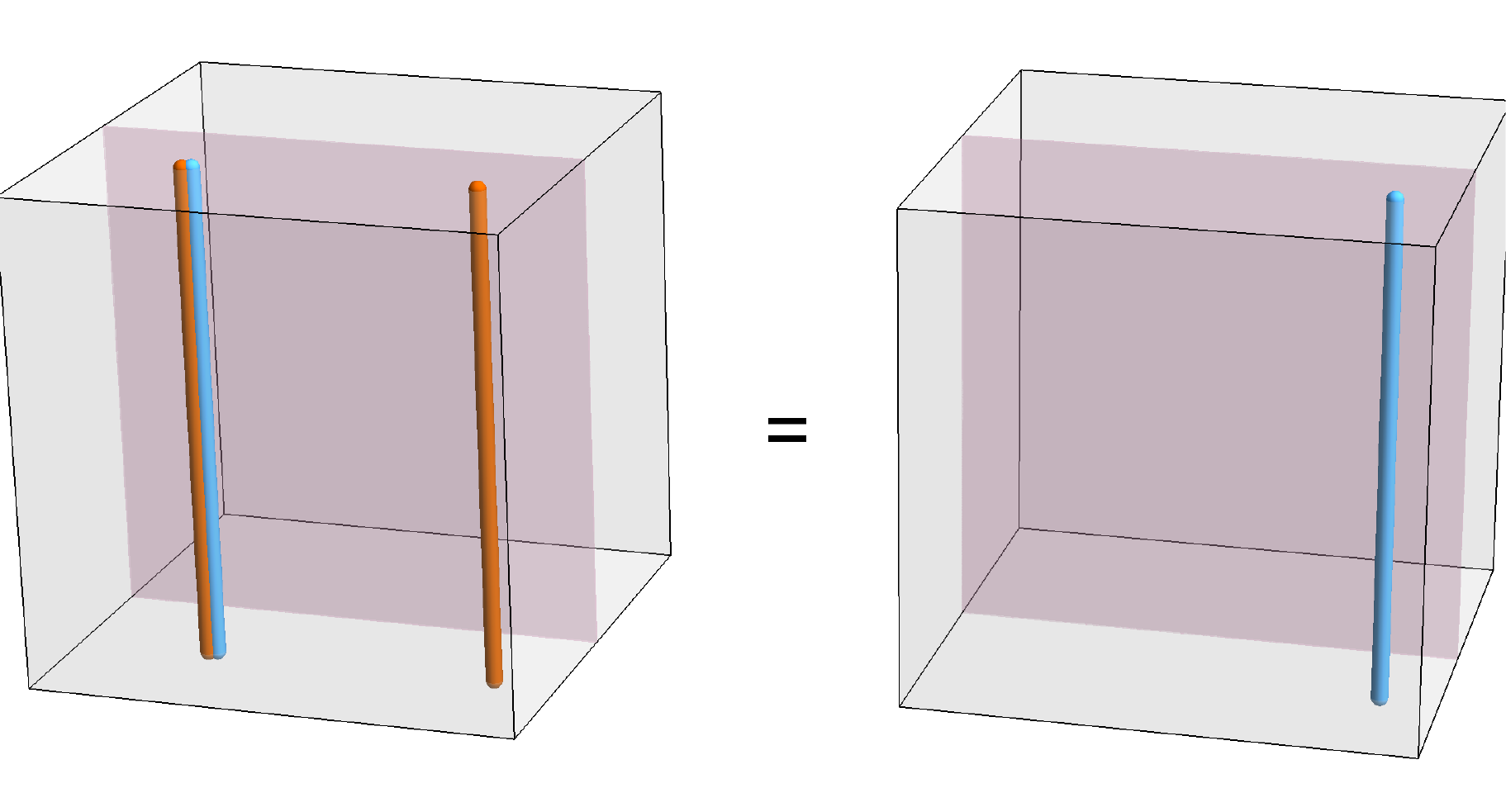}}
\caption{Constraints in the single-twist sector arising from the constraint in Fig.~\ref{fig:oneTwistPrismConstraint} and process in Fig.~\ref{fig:pairTranslation}. (a) Two bichromatic strings can mutually annihilate (equivalently, a single bichromatic string can move in the $z=0$ plane). (b) A bichromatic string can annihilate on a monochromatic string at the cost of flipping the color of the monochromatic string.}
\end{figure}

In the presence of the defect, additional constraints appear for the $x$- and $y$-oriented strings. Recall that, in the absence of twisted boundary conditions, the product of four strings of the same color on the edges of a rectangular prism is the identity (see Fig.~\ref{fig:prismConstraint}). However, in the presence of twisted boundary conditions, this is modified when strings live on opposite sides of the branch cut, becoming the condition in Fig.~\ref{fig:oneTwistPrismConstraint}. This allows us to move (for example) two $y-$oriented strings of the same color at the same $z$ coordinate (say $z=0$), all the way around the $z$-handle of the torus at the cost of changing their color, as in Fig.~\ref{fig:pairTranslation}. Notably, this leads to a constraint on bichromatic strings given in  Fig.~\ref{fig:oneTwistBichromaticConstraint}.

Therefore, many reference configurations in the ungauged theory become equivalent. Note that in our basis, the aforementioned constraints apply to all of the strings in the $z=0$ plane, but there is no such constraint on the strings at $x=0,z>0$.

Given a reference configuration of the $x$- and $y$-oriented Wilson loops in the ungauged model, we can algorithmically implement these constraints for each orientation of $z=0$ string types as follows. (In what follows, we refer only to strings of a given orientation at $z=0$.)

\textit{Step 1:}  Destroy pairs of bichromatic strings using the constraint in Fig.~\ref{fig:oneTwistBichromaticConstraint}. We are left with at most one bichromatic string, which can be moved freely using the same constraint.

\textit{Step 2:} If there are no monochromatic strings, stop. Otherwise, if a $z=0$ string is bichromatic, it can be annihilated at the cost of swapping the layer of one of the monochromatic strings, as in Fig.~\ref{fig:oneTwistThreeStringConstraint}.

\textit{Step 3:} Since pairs of monochromatic strings of the same color can be moved around the $z$ handle of the torus and change colors, the remaining states are labeled by the \textit{parity} of the number of strings of each color.

We now count the remaining string configurations, sorting by whether or not they are invariant under a global SWAP. For the $z=0$ strings which are $x$-oriented, a configuration is invariant under a global SWAP if:
\begin{itemize}
\item It has no strings present
\item It has one bichromatic string
\item The number of strings in layer 1 has the same parity as the number of strings in layer 2
\end{itemize}
There are 
\begin{equation}
1 + 1 + 2\sum_{m \text {even},>0}^{L_y} {L_y \choose m}
\end{equation}
such configurations (the factor of 2 is for both layers having an even vs. odd number of strings). The configuration transforms under a global SWAP if the number of strings in each layer has opposite parity; there are
\begin{equation}
2\sum_{m \text{ odd},>0}^{L_y} {L_y \choose m}
\end{equation}
such configurations (the factor of 2 is for layer 1 having an even vs. odd number of strings). The same argument holds with $x \leftrightarrow y$.

For the $2L_z-2$ strings at $z \neq 0$ (counting both the $x$ and $y$ orientations), there are $2^{2L_z-2}$ configurations which are invariant under a global SWAP (each string is absent or bichromatic) and 
\begin{equation}
\sum_{m=1}^{2L_z-2} {2L_z-2 \choose m} 2^m \times 2^{2L_z-2-m}
\end{equation}
 configurations which are not (choose a color for $m$ of the string types, and the other $2L_z-2-m$ string types either have no string or are bichromatic).
 
We can therefore count the total GSD in this sector by adding two quantities: the number of configurations for which all string types are invariant under a global SWAP, and half the number of configurations for which some string types are not invariant under a global SWAP. An example term is
\begin{widetext}
\begin{equation}
\frac{1}{2}\times 2^{L_x+L_y-1}\times \left(2\sum_{m' \text{ odd},>0}^{L_y} {L_y \choose m'}\right)  \times \left(2 + 2\sum_{m \text{ even},>0}^{L_x} {L_x \choose m}\right)  \times 2^{2L_z-2}
\end{equation}
\end{widetext}
The $1/2$ in front indicates this term transforms nontrivially under a global SWAP. The $2^{L_x+L_y-1}$ factor is for the $z$-oriented strings, which we have chosen to be invariant under SWAP.  The next factor is the number of $z=0$ $x$-oriented strings which transform under a global SWAP (which is why the overall configuration transforms under a global SWAP). The next factor is the number of the $z=0$ $x$-oriented strings which are invariant under a global SWAP. The final factor means we have chosen the $z \neq 0$ strings which are $x-$ and $y-$oriented to be invariant under SWAP. One must add all such terms for all choices of SWAP-invariant/non-invariant configurations for each set of strings.

The sum is messy but can be evaluated; it simplifies dramatically to $2^{N-1}(1+2^{2L_z})$. After summing over the other orientations of the twist defects, we obtain
\begin{equation}
\text{GSD}_{1-twist} = 2^{N-1}\left(3+\sum_{i=x,y,z}2^{2L_i}\right)
\end{equation}

\subsection{Double-twist sector}

For concreteness, we take the case where the twist defects span the $xy$ and $xz$ planes. Then the $y-$ and $z-$oriented strings can each either be absent or wrap the torus twice; there are $2^{2L_x+L_y+L_z-2}$ such configurations for those strings, all of which are invariant under a global SWAP.

We now count the configurations for the $x$-oriented strings, of which there are $L_y+L_z-1$. This time, pairs of same-colored strings at $z=0$ ($y=0$) can be moved around the $y$ ($z$) handle of the torus and swap layers. This time \textit{any} pair of bichromatic strings can annihilate since any bichromatic string can be moved to the $y=z=0$ location. Likewise, if any monochromatic strings are present, a bichromatic string can fuse with it at the cost of changing the color of the monochromatic string.

After annihilating as many bichromatic strings as possible, it will be convenient to count states depending on the state of the $y=z=0$ string.

\textit{Bichromatic string present at $y=z=0$:} By the process for moving or annihilating bichromatic strings, this reduces to the case where no string is present.

\textit{No string present at $y=z=0$:} There are two states in this sector which are invariant under a global SWAP: no strings at all, or a single bichromatic string.

In any other case, there are only monochromatic strings. Any pair of same-colored strings at $y=0$ or any pair of same-colored strings at $z=0$ can move around the torus and change colors, so only the positions of the monochromatic strings and the parity of the number of strings in each layer is well-defined.

Let there be $m$ monochromatic strings at $z=0$ and no strings at $y=0$. There are ${L_y - 1 \choose m}$ choices for the positions of these strings. If $m$ is even, there are two states, each invariant under a global SWAP, determined by the parity of the number of strings in layer 1, which equals the parity of the number of strings in layer 2. If $m$ is odd, then the two parity choices transform into each other by a global SWAP, so there is only one state. The same argument holds with $y \leftrightarrow z$.

Finally, let there be $m>0$ and $m'>0$ monochromatic strings at $z=0$ and $y=0$ respectively, with fixed positions (there are ${L_y-1 \choose m}{L_z-1 \choose m'}$ choices for the positions). Then if $m$ and $m'$ are both even, there are four states which are each invariant under a global SWAP, labeled by separately choosing the parity of the number of layer-1 strings at $y=0$ and at $z=0$. Otherwise, the global SWAP acts nontrivially and so we only obtain 2 states.

Hence, if there is no string at $y=z=0$, we obtain
\begin{widetext}
\begin{equation}
2+\sum_{m=1}^{L_y-1}{L_y-1\choose m}\left(\frac{3+(-1)^m}{2}\right) + \sum_{m=1}^{L_z-1}{L_z-1\choose m}\left(\frac{3+(-1)^m}{2}\right) + \sum_{m=1}^{L_y-1}\sum_{m'=1}^{L_z-1}{L_y-1\choose m}{L_z-1\choose m'} \times \begin{cases} 4 & m,m' \text{ even}\\
2 & \text{ else}
\end{cases}
\label{eqn:noCornerString}
\end{equation}
\end{widetext}
states.

\textit{Monochromatic string present at $y=z=0$:}

If there are no other strings present, there is obviously one state (the superposition of both string colors at $y=z=0$. If there are no strings at $y>0$ and $m$ strings at $z>0$, then by a similar argument from the no-string case there are two states (for fixed string positions) if $m$ is odd and one state if $m$ is even, with the same result for $y \leftrightarrow z$.

If there are $m>0$ strings at $z>0$ and $m'>0$ strings at $y>0$, we naively have eight states which are labeled by the parity of the number of layer-1 strings at $z>0$, the same parity for $y>0$, and the color of the string at $y=z=0$. Obviously some of these states are invariant under global SWAPs, but also both the $y>0$ and $z>0$ sets of strings can be moved around the torus with the $y=z=0$ string. This causes a parity label to flip along with the color of the $y=z=0$ string. Exhaustively counting the configurations shows that there are two states if $m,m'$ have opposite parity and one state otherwise.

In total, this sector has
\begin{widetext}
\begin{equation}
1 + \sum_{m=1}^{L_y-1}{L_y-1 \choose m}\left(\frac{3-(-1)^m}{2}\right)+\sum_{m=1}^{L_z-1}{L_z-1 \choose m}\left(\frac{3-(-1)^m}{2}\right) + \sum_{m=1}^{L_y-1}\sum_{m'=1}^{L_z-1}{L_y-1 \choose m}{L_z-1 \choose m'}\times \begin{cases}
2 & m,m' \text{ opposite parity}\\
1 & m,m' \text{ same parity}
\end{cases}
\label{eqn:cornerString}
\end{equation}
\end{widetext}

Adding Eqs. \eqref{eqn:noCornerString} and \eqref{eqn:cornerString} and evaluating the sums, we obtain $2^{L_y+L_z}-3\times (2^{L_y-2}+2^{L_z-2})+2$. After multiplying by $2^{2L_x+L_y+L_z-2}$ for the $y$- and $z$-oriented string configurations, then summing over the different choices of twist defect orientations (i.e. summing on cyclic permutations of $x,y,z$), we obtain the GSD in this sector 
\begin{equation}
\text{GSD}_{\text{2-twist}} = 3 \times 2^{N+1}+\sum_{i=x,y,z}\left(2^{\frac{N+1}{2}+L_i}-3\times 2^{N-L_i}\right)
\end{equation}

\subsection{Triple-twist sector}
In the wavefunction picture, every string must either be absent or wrap its handle of the torus twice. A global SWAP acts trivially on such states, so the number of ground states in the triple-twist sector is simply
\begin{equation}
\text{GSD}_{\text{3-twist}} = 2^N
\end{equation}

Adding all of the different twist sectors leads to the final result, Eq.\eqref{eqn:gaugedGSD}.

\section{Twist strings in gauged bilayer Haah's code}
\label{app:HaahStringOp}

In this appendix, we describe the construction of the membrane operator $\mathcal{M}_{\sigma}$ which creates twist string excitations $\sigma$ in the gauged bilayer Haah's code model.

We decompose
\begin{equation}
\mathcal{M}_{\sigma} = W \mathcal{O}
\end{equation}
where $W$ is a membrane of $\tau_x$ operators (for concreteness in the $y=1/2$ plane), which would create a flux string in the 3+1D toric code and $\mathcal{O}$ is to be described shortly. The operator $W$ alone has surface tension because for any cube $c$ centered at a point in the $y=1/2$ plane, $W$ anticommutes with any term in $A_c$ and $B_c$ which has an odd number of SWAP-odd operators in the $y=0$ plane (equivalently the $y=1$ plane). The purpose of $\mathcal{O}$ is to remove this tension.

Consider a ``reference" operator
\begin{widetext}
\begin{equation}
R = \bigotimes_{y=0} (1+(X1)^{(1)}(X1)^{(2)})(1+(1X)^{(1)}(1X)^{(2)})(1+(Z1)^{(1)}(Z1)^{(2)})(1+(1Z)^{(1)}(1Z)^{(2)})
\end{equation}
\end{widetext}
where the tensor product is over all sites below the membrane of $\tau_x$ operators. Note first that all terms in the product commute with each other, with $C_s$, and with $D_p$. For a cube $c$ in the $y=-1/2$ plane, none of the $\tau_z$ operators in $A_c$ and $B_c$ intersect $W$, so $[W,A_c]=[W,B_c]=0$. Furthermore,
\begin{align}
\lbrace (Z1)^{(1)}(Z1)^{(2)}, (X1)^{(\pm)} \rbrace &= 0\\
\left[ (Z1)^{(1)}(Z1)^{(2)}, (1X)^{(\pm)} \right] &= 0\\
\lbrace (Z1)^{(1)}(Z1)^{(2)}, (XX)^{(\pm)} \rbrace &= 0
\end{align}
with similar identities involving $1Z,ZZ$, etc. Therefore, commuting $A_c$ (resp. $B_c$) past $R$ simply changes some of the plus signs before $Z$ operators (resp. $X$ operators) in $R$ to minus signs, which does not change its commutation relations with $C_s$ or $D_p$. For each such $A_c$ and $B_c$ in the $y=-1/2$ plane whose support intersects the support of $R$, then, we can choose whether or not to commute it past $R$; in this way, we obtain $\sim 2^{2\ell^2}$ operators which differ by these sign choices, where $\ell$ is the linear size of the $\tau_x$ membrane. Superposing all of these operators will produce an operator which commutes with all of the cube operators in the $y=-1/2$ plane.

The remarkable thing about this choice of $R$ is that the same procedure may also be used to ensure commutation with the Hamiltonian terms in the $y=1/2$ plane, where $[W,A_c]\neq 0$. For example, a straightforward computation shows that $R$ times any term in (e.g.) $A_c$ which anticommutes with $W$ is zero. Therefore $WR$ commutes with those terms within $A_c$ despite those terms anticommuting with $W$. However, $R$ times the remaining terms in $A_c$, which commute with $W$, changes a set of minus signs in $R$, much like what happened for the operators in the $y=-1/2$ plane, producing an operator $R'$. Therefore, $A_cR=R'A_c$ since each term within $A_c$ either multiplies with both $R$ and $R'$ to zero or produces $R'$ upon commutation.

\begin{figure}
\includegraphics[width=0.7\columnwidth]{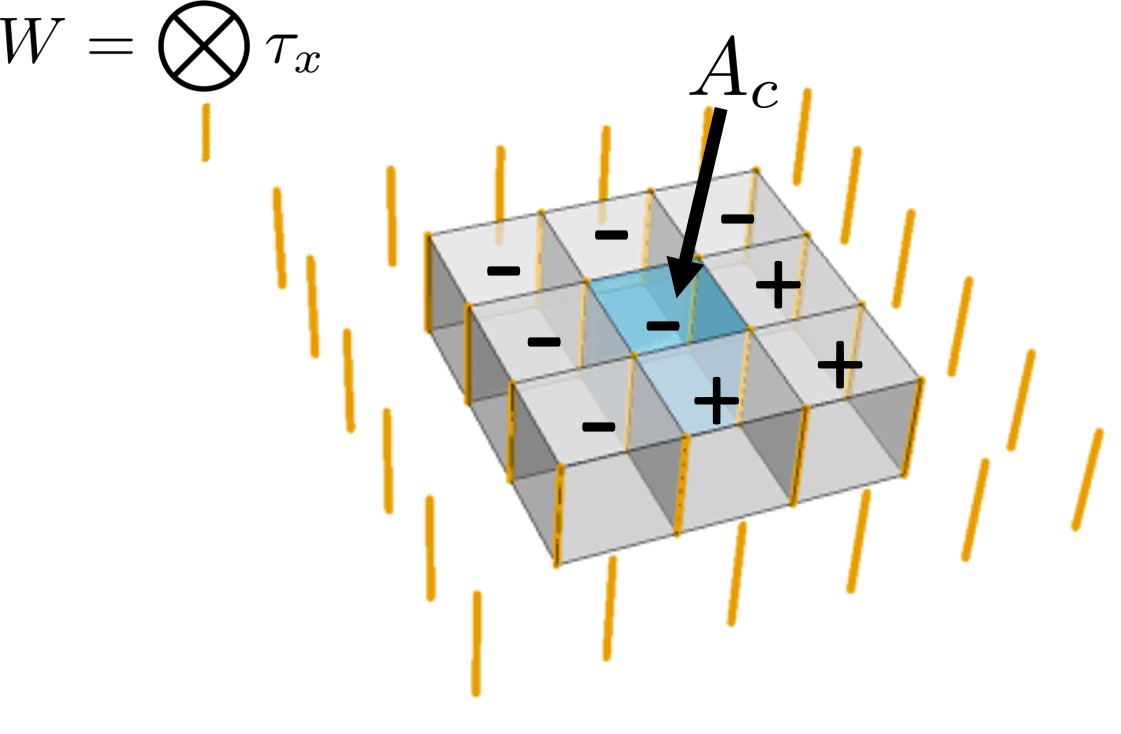}
\caption{Sign structure used in creating the operator $\mathcal{O}$ (see text) when $W$ is the product of $\tau_x$ over the orange links. If $A_c$, where $c$ is the blue cube in the operator, has been commuted past $R$ to produce $R'$, the operator $R''$ produced by commuting a $B_{c'}$ operator through $R'$ will enter $\mathcal{O}$ with a minus sign if $c'$ is a cube labeled $-$ in the figure.}
\label{fig:HaahSignStructure}
\end{figure}

In total, then, $\mathcal{O}$ is produced by starting with $R$, then superposing over all possible operators obtained from commuting $R$ through the product of any subset of the $\sim 4\ell^2$ Hamiltonian terms which intersect $R$. However, some terms in the superposition enter with an overall minus sign. For example, let $A_cR=R'A_c$ for a fixed $c$ in the $y=1/2$ plane. Then one can check that $R'$ times any term in $B_c$ (for the same cube $c$) which \textit{commutes} with $W$ is zero, while passing any term of $B_c$ which \textit{anticommutes} with $W$ produces another operator $R''$. Accordingly, the structure of $\mathcal{O}$ should be
\begin{equation}
\mathcal{O} = R + R' -R'' + \cdots
\end{equation}
so that $V\mathcal{O}$
One can check that the sign structure is consistent, in the sense that no additional terms in the superposition include $+R''$. In particular, for a given term $\tilde{R}$ in the superposition, i.e. a given subset of Hamiltonian terms, we count the number of times an $A_c$ and $B_{c'}$ in the set border each other in the relative positions shown in Fig.~\ref{fig:HaahSignStructure}. If the number of such borders is even (resp. odd), then $\tilde{R}$ enters the superposition with a plus (resp. minus) sign.

We do not know of a more compact way to describe the construction of this operator because the structure of the minus signs is highly nontrivial. We leave finding a ``nicer" description as an open problem.

\bibstyle{apsrev4-1} \bibliography{references}

\end{document}